\documentclass[12pt,a4paper]{article}
\usepackage[T2A]{fontenc}
\usepackage[cp1251]{inputenc}
\usepackage{cite}
\usepackage{mathtext}
\usepackage{amssymb,amsthm,amsmath}
\usepackage{array}
\usepackage[dvips]{epsfig}
\usepackage[russian, english]{babel}
\usepackage[title,titletoc]{appendix}

\begin{document}
\author{\bf Yu.A. Markov$^{1}\!\,$\thanks{e-mail:markov@icc.ru},\,
M.A. Markova$^{1}\!\,$\thanks{e-mail:markova@icc.ru}\, and
D.M. Gitman$^{\,2,\,3,\,4}\!\;$\thanks{e-mail:gitman@dfn.if.usp.br}}

\title{Unitary quantization and\\ para-Fermi statistics of order two}
%
%
\date{\it\normalsize
\begin{itemize}
\item[]$^{1}$Matrosov Institute for System Dynamics and Control Theory SB RAS
Irkutsk, Russia
\vspace{-0.3cm}
\item[]$^{2}$P.N. Lebedev Physical Institute, Moscow, Russia
\vspace{-0.3cm}
\item[]$^{3}$Department of Physics, Tomsk State University, Tomsk, Russia
\vspace{-0.3cm}
\item[]$^{4}$Institute of Physics, University of S\~ao Paulo, S\~ao Paulo, Brazil
\end{itemize}}
%
\thispagestyle{empty}
\maketitle{}
\def\theequation{\arabic{section}.\arabic{equation}}
\[
{\bf Abstract}
\]
{
\noindent
A connection between a unitary quantization scheme and para-Fermi statistics of order 2 is considered. An appropriate extension of Green's ansatz is suggested.
This extension allows one to transform bilinear and trilinear commutation relations for the annihilation and creation operators of two different para-Fermi fields $\phi_{a}$ and $\phi_{b}$ into identity. The way of incorporating para-Grassmann numbers $\xi_{k}$ into a general scheme of uniquantization is also offered. For parastatistics of order 2 a new fact is revealed, namely, the trilinear relations containing both the para-Grassmann variables $\xi_{k}$ and the field operators $a_{k}$, $b_{m}$ under a certain invertible mapping go over into the unitary equivalent relations, where commutators are replaced by anticommutators and vice versa. It is shown that the consequence of this circumstance is the existence of two alternative definitions of the coherent state for para-Fermi oscillators. The Klein transformation for Green's components of the operators $a_{k}$, $b_{m}$ is constructed in an explicit form that enables us to reduce the initial commutation rules for the components to the normal commutation relations of ordinary Fermi fields. A nontrivial connection between trilinear commutation relations of the unitary quantization scheme and so-called Lie-supertriple system is analysed. A brief discussion of the possibility of embedding the Duffin-Kemmer-Petiau theory into the unitary quantization scheme is provided.
}

{}


\newpage

\section{Introduction}
\setcounter{equation}{0}

The problem of quantization of finite-dimensional physical systems has attracted repeated attention of physicists. From the recent works in this line of investigations the Ref. \cite{assirati_2017} should be mentioned. However, in the present work, we would like to consider another line of research in quantization of finite-dimensional classical theories which places primary emphasis upon Lie algebraic aspects of the physical system under consideration.
In the papers by Govorkov \cite{govorkov_1979, govorkov_1980, govorkov_1983} and independently by Palev \cite{palev_1977, palev_1_1980, palev_2_1980}, Palev and Jeugt \cite{palev_2002} a formalism for the quantization of fields based on the relations of the Lie algebra of the unitary group $SU(2M+1)$ was set up\footnote{\,We point out however that some aspects of the special cases of the quantization based on $su(2)$ and $su(3)$ algebras were discussed in earlier papers by Govorkov \cite{govorkov_1968}, Ramakrishnan {\it et al.} \cite{ramakrishnan_1971}, Bracken and Green \cite{bracken_1973}.}. In \cite{govorkov_1979} the suggested scheme of quantization was called ``the uniquantization'', whereas in \cite{palev_1977} it was ``the A-quantization''. In this work we would like to investigate in more detail some properties of the relations obtained in \cite{govorkov_1979}, and in particular, to establish a connection between the unitary quantization and para-Fermi statistics of order 2. Below we will give a few basic formulae from \cite{govorkov_1979} to which we will repeatedly refer in the subsequent discussion. In Appendix~A, a brief scheme of deriving these formulae is given.\\
\indent
Let $a_k$ and $a_{k}^{\dagger}$ be the annihilation and creation operators obeyed the Green commutation relations \cite{green_1953} (we restrict our consideration only to the case of para-Fermi statistics)
{\setlength\abovedisplayskip{14pt}
\setlength\belowdisplayskip{14pt}
\begin{equation}
\hspace{0.7cm}
[\hspace{0.02cm}[\hspace{0.02cm}\hat{a}_{k},\hat{a}_{l}\hspace{0.02cm}],\hat{a}_{m}] =
2\hspace{0.02cm}\hat{\delta}_{lm}\hspace{0.02cm}\hat{a}_{k} - 2\hspace{0.02cm}\hat{\delta}_{km}\hspace{0.02cm}\hat{a}_{l},
\quad\;
\label{eq:1q}
\end{equation}
}
$\!\!$where $k,l,m = 1,2,\ldots,M$ and $[\,,]$ is commutator. The operator with hat above $\hat{a}_{k}$ stands for $a_k$ or $a_{k}^{\dagger}$ and $\hat{\delta}_{kl}=\delta_{kl}$ when $\hat{a}_{k}=a_{k}(a_{k}^{\dagger})$ and $\hat{a}_{l}=a_{l}^{\dagger}(a_{l})$, and $\hat{\delta}_{kl}=0$ otherwise. For uniquantization, in addition to the operators $\hat{a}_k$, another set of particle creation and annihilation operators $\hat{b}_k$ obeying the same commutation relations
{\setlength\abovedisplayskip{14pt}
\setlength\belowdisplayskip{14pt}
\begin{equation}
[\hspace{0.02cm}[\hspace{0.02cm}\hat{b}_{k},\hat{b}_{l}\hspace{0.02cm}],\hat{b}_{m}] =
2\hspace{0.02cm}\hat{\delta}_{lm}\hspace{0.02cm}\hat{b}_{k} - 2\hspace{0.02cm}\hat{\delta}_{km}\hspace{0.02cm}\hat{b}_{l}
\label{eq:1w}
\end{equation}
}
$\!\!$is introduced. In addition to (\ref{eq:1q}) and (\ref{eq:1w}), from the given scheme of quantization uniquely follow the mutual commutation relations of two types between the operators $\hat{a}_k$ and $\hat{b}_k$:
\vspace{-0.3cm}
\begin{flushleft}
1. trilinear relations
\end{flushleft}
\vspace{-0.7cm}
\begin{align}
&[\hspace{0.02cm}[\hspace{0.02cm}\hat{b}_{m},\hat{a}_{k}\hspace{0.02cm}],\hat{a}_{l}] =
4\hspace{0.02cm}\hat{\delta}_{km}\hspace{0.02cm}\hat{b}_{l}
+
2\hspace{0.02cm}\hat{\delta}_{lk}\hspace{0.02cm}\hat{b}_{m}
+
2\hspace{0.02cm}\hat{\delta}_{lm}\hspace{0.02cm}\hat{b}_{k},
\label{eq:1e}\\[0.8ex]
&[\hspace{0.02cm}[\hspace{0.02cm}\hat{a}_{m},\hat{b}_{k}\hspace{0.02cm}],\hat{b}_{l}] =
4\hspace{0.02cm}\hat{\delta}_{km}\hspace{0.02cm}\hat{a}_{l}
+
2\hspace{0.02cm}\hat{\delta}_{lk}\hspace{0.02cm}\hat{a}_{m}
+
2\hspace{0.02cm}\hat{\delta}_{lm}\hspace{0.02cm}\hat{a}_{k},
\label{eq:1r}
\end{align}
\vspace{-0.3cm}
\begin{flushleft}
2. bilinear relations
\end{flushleft}
\vspace{-0.9cm}
\begin{align}
&[\hspace{0.02cm}\hat{a}_{k},\hat{b}_{m}\hspace{0.02cm}]
=
[\hspace{0.02cm}\hat{a}_{m},\hat{b}_{k}\hspace{0.02cm}],
\label{eq:1t}\\[0.8ex]
&[\hspace{0.02cm}\hat{a}_{k},\hat{a}_{m}\hspace{0.02cm}]
=
[\hspace{0.02cm}\hat{b}_{k},\hat{b}_{m}\hspace{0.02cm}].
\label{eq:1y}
\end{align}
\indent
Thus we have a case of two para-Fermi fields of order $p$ between which there are not only trilinear relations but also bilinear ones. As is well known \cite{kamefuchi_1962, ryan_1963}, the commutation relations (\ref{eq:1q}) and (\ref{eq:1w}) generate an algebra which is isomorphic to the algebra of the orthogonal group $SO(2M+1)$. The other relations (\ref{eq:1e})\,--\,(\ref{eq:1y}) complete this algebra to the algebra of the unitary group $SU(2M+1)$. The particle-number operator
\begin{equation}
N =
\frac{1}{2}\,\sum^{M}_{k\hspace{0.02cm}=1}\,
\bigl(\hspace{0.03cm}[\hspace{0.02cm}a^{\dagger}_{k},
a^{\phantom{\dagger}\!}_{k}\hspace{0.02cm}]
\hspace{0.02cm}+\hspace{0.02cm}p\hspace{0.02cm}\bigr)
\Bigl(\equiv \frac{1}{2}\,\sum^{M}_{k\hspace{0.02cm}=1}\,
\bigl(\hspace{0.03cm}[\hspace{0.02cm}b^{\dagger}_{k},
b^{\phantom{\dagger}\!}_{k}\hspace{0.02cm}]
\hspace{0.02cm}+\hspace{0.02cm}p\hspace{0.02cm}\bigr) \Bigr)
\label{eq:1u}
\end{equation}
together with the algebra (\ref{eq:1q})\,--\,(\ref{eq:1y}) ultimately fixes the unitary quantization scheme.\\
\indent
In addition to the preceding, it should be noted that in Govorkov's construction \cite{govorkov_1979} for the group $SU(2M+1)$ there exists another important operator $\zeta_0$. This operator due to the relations (\ref{ap:A10}) and (\ref{ap:A18}) can be expressed in terms of the operators $\hat{a}_k$ and $\hat{b}_k$ as follows
\begin{equation}
\zeta_{0} = \frac{i}{2\hspace{0.02cm}(2\hspace{0.01cm}M + 1)}\,
\sum^{M}_{k\hspace{0.02cm}=1}\,\Bigl(
[\hspace{0.02cm}a^{\dagger}_{k},b^{\phantom{\dagger}\!}_{k}\hspace{0.02cm}] + [\hspace{0.02cm}b^{\dagger}_{k},a^{\phantom{\dagger}\!}_{k}\hspace{0.02cm}]\Bigr).
\label{eq:1i}
\end{equation}
It has the commutation properties
{\setlength\abovedisplayskip{14pt}
\setlength\belowdisplayskip{14pt}
\begin{equation}
[\hspace{0.02cm}\hat{a}_{k},\zeta_{0}] = 2\hspace{0.01cm}i\hspace{0.02cm}\hat{b}_{k},
\quad
[\hspace{0.02cm}\hat{b}_{k},\zeta_{0}] = -2\hspace{0.01cm}i\hspace{0.02cm}\hat{a}_{k}.
\label{eq:1o}
\end{equation}
}
\indent
$\!\!\!$The paper is organized as follows. In the section 2 a brief review of Greenberg and Messiah's work \cite{greenberg_1965} on several different parafields is presented. In sections 3 and 4 a generalization of the system of the commutation rules for the Green components of the annihilation and creation operators of two parafields $\phi_{a}$ and $\phi_{b}$ is given. The detailed proof that for parastatistics of order $p=2$ this system turns the Govorkov bilinear and trilinear relations into identity is presented. Section 5 is devoted to inclusion of the para-Grassmann numbers $\xi_{k}$ in the general scheme of unitary quantization. Section 6 is concerned with the construction of the commutation relations between the operators $a_{k}$, $b_{m}$, para-Grassmann numbers $\xi_{k}$ and the operator ${\rm e}^{\alpha\hspace{0.04cm}i\widetilde{N}}$, where $\alpha$ is an arbitrary real number. Two important special cases of the general relations, when $\alpha = \pm\hspace{0.02cm}\pi$ and $\alpha = \pm\hspace{0.02cm}\pi/2$ are considered. In the same section a certain invertible mapping of the trilinear relations that include both the para-Grassmann numbers and the field operators is considered. The nontrivial peculiarities of this mapping are revealed. In section 7 the acting of some operators, which arise within the unitary quantization scheme on the vacuum state is defined.\\
\indent
Section 8 is devoted to discussion of the properties of the coherent states. In particular, an interesting fact of the existence another state for para-Fermi statistics of order 2 that possesses the same properties as commonly used coherent states, is found. In section 9 the possibility of deriving Govorkov's trilinear relations from the requirement of invariance of the commutation relations between operators $a_{k},\,b_{m}$ and $\widetilde{N}$ under unitary transformation of $a_{k}$ and $b_{m}$ is analyzed. It is shown that, in contrast to the case with a single parafield, this requirement alone is not sufficient to recovery of all Govorkov's trilinear relations. In section 10 the construction of the so-called Klein transformation for the Green components of the parafield operators is given. In section 11 the question of a connection between the Govorkov trilinear relations and Lie-supertriple system is answered. Section 12 deals with the problem of embedding the Duffin-Kemmer-Petiau formalism into the unitary quantization scheme. It is shown that these two approaches, in the final analysis, are inconsistent. In concluding section 13 a possible connection between the unitary quantization scheme based on the Lie algebra of the unitary group $SU(2M)$ and para-Bose statistics is briefly discussed. In the same section some unusual properties unique to parastatistics of order 2 are sharply defined.\\
\indent
In Appendix A all of the basic relations of the Lie algebra of the unitary group $SU(2M + 1)$ are given. In the same Appendix some discrepancies noticed by us in Govorkov's works are mentioned. In Appendix B various operator identities, which we intensively use throughout the present work is written out. In Appendix C all basic commutation relations involving the operator
${\rm e}^{\alpha\hspace{0.04cm}i\widetilde{N}}$ are collected.

\section{\bf Review of Greenberg and Messiah's work}
\setcounter{equation}{0}

Let us write the general relation (\ref{eq:1q}) in more detail
\begin{align}
&[\hspace{0.02cm}[\hspace{0.02cm}a^{\dagger}_{k},
a^{\phantom{\dagger}\!}_{l}\hspace{0.02cm}],a^{\phantom{\dagger}\!}_{m}] = -2\hspace{0.02cm}\delta^{\phantom{\dagger}\!}_{km}\hspace{0.02cm}a^{\phantom{\dagger}\!}_{l},  \label{eq:2q}\\
&[\hspace{0.02cm}[\hspace{0.02cm}a_{k},a_{l}\hspace{0.02cm}],a_{m}] = 0 \label{eq:2w}
\end{align}
and as a consequence of Jacobi's identity (\ref{ap:B1}) we have
\begin{equation}
[\hspace{0.02cm}[\hspace{0.02cm}a^{\phantom{\dagger}\!}_{k},
a^{\phantom{\dagger}\!}_{l}\hspace{0.02cm}],a^{\dagger}_{m}] =
2\hspace{0.02cm}\delta^{\phantom{\dagger}\!}_{lm}\hspace{0.02cm}a^{\phantom{\dagger}\!}_{k} - 2\hspace{0.02cm}\delta^{\phantom{\dagger}\!}_{km}\hspace{0.02cm}a^{\phantom{\dagger}\!}_{l}.
\label{eq:2e}
\end{equation}
Greenberg and Messiah \cite{greenberg_1965} have suggested an extension of the relations (\ref{eq:2q})\,--\,(\ref{eq:2e}) to the case of several different parafields. In order for the relative commutation rules between different parafields to be defined, the authors have demanded the desired relations satisfy the following three requirements:
\begin{itemize}
\item[(i)] 	
the left-hand side must have the trilinear form\footnote{\,However, the authors didn't avoid the possibility of an existence of bilinear commutation or anticommutation relations between different parafields. Nevertheless, they have restricted the attention only to trilinear relations. In the scheme of unitary quantization the bilinear relations, Eqs.\,(\ref{eq:1t})\,--\,(\ref{eq:1y}), inevitably arise.}
\[
[\hspace{0.02cm}[\hspace{0.02cm}A,B\hspace{0.02cm}],C\hspace{0.02cm}]
\]
and the right-hand side must be linear;
\item[(ii)] when the internal pair $[\hspace{0.02cm}A,B]$ refers to the same field, it must
commute with $C$ if $C$ refers to another field;
\item[(iii)] these relations must be satisfied by ordinary Bose or Fermi fields.
\end{itemize}
In the case when these conditions apply to two para-Fermi fields $\phi_{a}$ and $\phi_{b}$, Greenberg and Messiah have obtained the following system of trilinear relations involving $\phi_{a}$ twice and $\phi_{b}$ once:
\begin{align}
&[\hspace{0.02cm}[\hspace{0.02cm}a^{\dagger}_{k},
a^{\phantom{\dagger}\!}_{l}\hspace{0.02cm}],b^{\phantom{\dagger}\!}_{m}] = 0, \label{eq:2r}\\
&[\hspace{0.02cm}[\hspace{0.02cm}a_{k},a_{l}\hspace{0.02cm}],b_{m}] = 0,
\label{eq:2t}\\
&[\hspace{0.02cm}[\hspace{0.02cm}a^{\dagger}_{k},a^{\dagger}_{l}\hspace{0.02cm}],
b^{\phantom{\dagger}\!}_{m}] = 0. \label{eq:2y}
\end{align}
By employing Jacobi's identity (\ref{ap:B1}) together with the conditions (i) and (iii) from (\ref{eq:2r}) two further trilinear relations follow
\begin{align}
&[\hspace{0.02cm}[\hspace{0.02cm}b^{\phantom{\dagger}\!}_{m},
a^{\dagger}_{k}\hspace{0.02cm}],a^{\phantom{\dagger}\!}_{l}] = 2\hspace{0.015cm}\delta^{\phantom{\dagger}\!}_{kl}\hspace{0.02cm}b^{\phantom{\dagger}\!}_{m},  \label{eq:2u}\\[0.5ex]
&[\hspace{0.02cm}[\hspace{0.02cm}a^{\phantom{\dagger}\!}_{l},
b^{\phantom{\dagger}\!}_{m}\hspace{0.02cm}],a^{\dagger}_{k}] = -2\hspace{0.015cm}\delta^{\phantom{\dagger}\!}_{kl}\hspace{0.02cm}b^{\phantom{\dagger}\!}_{m}.
\label{eq:2i}
\end{align}
In section 9 we will discuss more comprehensively a way of deriving (\ref{eq:2u}) and (\ref{eq:2i}). To the relations (\ref{eq:2r})\,--\,(\ref{eq:2i}) their Hermitian conjugation and also the 18 trilinear relations involving $\phi_{b}$ twice and $\phi_{a}$ once must be added.\\
\indent
Also in the paper \cite{greenberg_1965} a straightforward generalization of Green's ansatz \cite{green_1953} was suggested. Each field operator is expanded into the Green components
\begin{equation}
a_{k} = \sum^{p}_{\alpha\hspace{0.02cm}=\hspace{0.02cm}1}a^{(\alpha)}_{k}\,, \quad
b_{m} = \sum^{p}_{\alpha\hspace{0.02cm}=\hspace{0.02cm}1}b^{(\alpha)}_{m},
\label{eq:2o}
\end{equation}
where $p$ is the order of parastatistics. Each pair of components belonging to the same field satisfies the commutation relations
\begin{equation}
\begin{split}
&\{a^{(\alpha)}_{k}\!,a^{\dagger(\alpha)}_{l\!}\} = \delta_{kl}\hspace{0.02cm},\quad
\{a^{(\alpha)}_{k}\!,a^{(\alpha)}_{l}\} = 0,\\[1ex]
&\hspace{0.02cm}[\hspace{0.03cm}a^{(\alpha)}_{k}\!,a^{(\beta)}_{l}\hspace{0.02cm}] = [\hspace{0.03cm}a^{(\alpha)}_{k}\!,a^{\dagger(\beta)}_{l}\hspace{0.02cm}]
 = 0,
\quad \alpha\neq\beta,
\label{eq:2p}
\end{split}
\end{equation}
and similarly for the $\phi_{b}$ field. Here, $\{\,,\}$ is anticommutator. For each pair of Green's components of the different fields Greenberg and Messiah have postulated the following rules:
\begin{equation}
\begin{split}
&\{a^{\dagger(\alpha)}_{k}\!,b^{(\alpha)}_{m}\} = \{a^{(\alpha)}_{k}\!,b^{\dagger(\alpha)}_{m}\} = 0,\\[1ex]
&\{a^{(\alpha)}_{k}\!,b^{(\alpha)}_{m}\} = \{a^{\dagger(\alpha)}_{k}\!,b^{\dagger(\alpha)}_{m}\} = 0,
\label{eq:2a}
\end{split}
\hspace{1,6cm}
\end{equation}
and
\begin{equation}
\begin{split}
&[\hspace{0.03cm}a^{(\alpha)}_{k}\!,b^{(\beta)}_{m}\hspace{0.02cm}] =
[\hspace{0.03cm}a^{\dagger(\alpha)}_{k}\!,b^{\dagger(\beta)}_{m}\hspace{0.02cm}] = 0, \\[1ex]
&[\hspace{0.03cm}a^{\dagger(\alpha)}_{k}\!,b^{(\beta)}_{m}\hspace{0.02cm}] =
[\hspace{0.03cm}a^{(\alpha)}_{k}\!,b^{\dagger(\beta)}_{m}\hspace{0.02cm}] = 0,
\quad \alpha\neq\beta.
\label{eq:2s}
\end{split}
\end{equation}
The fields obeying the rules (\ref{eq:2a}) and (\ref{eq:2s}) verify the set of the trilinear relations (\ref{eq:2r})\,--\,(\ref{eq:2i}).\\
\indent
Finally, to the conditions of a unique vacuum state $|\hspace{0.02cm}0\rangle$:
\begin{equation}
a_{k} |\hspace{0.03cm}0\rangle = b_{k} |\hspace{0.02cm}0\rangle = 0, \quad
\mbox{for all}\; k
\label{eq:2d}
\end{equation}
and
\begin{equation}
\hspace{0.1cm}
\begin{split}
a^{\phantom{\dagger}\!}_{k} a^{\dagger}_{l}|\hspace{0.03cm}0\rangle &= p\,\delta^{\phantom{\dagger}\!}_{kl} |\hspace{0.03cm}0\rangle,
\quad\;\;
\mbox{for all}\; k,\,l \\[0.7ex]
b^{\phantom{\dagger}\!}_{m} b^{\dagger}_{n}|\hspace{0.03cm}0\rangle &= p\,\delta^{\phantom{\dagger}\!}_{mn} |\hspace{0.02cm}0\rangle, \quad
\mbox{for all}\; m,\,n
\label{eq:2f}
\end{split}
\end{equation}
Greenberg and Messiah have added two other conditions
\begin{equation}
\begin{split}
&b^{\phantom{\dagger}\!}_{m} a^{\dagger}_{k}|\hspace{0.03cm}0\rangle = 0,
\quad
\mbox{for all}\; m,\,k \\[0.7ex]
&a^{\phantom{\dagger}\!}_{k} b^{\dagger}_{m}|\hspace{0.03cm}0\rangle = 0.
\label{eq:2g}
\end{split}
\end{equation}
These conditions can be derived from the parafield commutation relations (\ref{eq:2r})\,--\,(\ref{eq:2i}) and from the uniqueness of the vacuum state $|\hspace{0.03cm}0\rangle$. This derivation will be considered in more detail in section 7 in the context of our problem.

\section{\bf Green's ansatz for the Govorkov relations}
\setcounter{equation}{0}

In Introduction, we have written out the trilinear and bilinear commutation relations, which arise within the framework of the scheme of Govorkov's unitary quantization. Our first step is to consider the trilinear relations for two different parafields. A particular consequence of the general formula (\ref{eq:1e}) is the following two relations
\begin{align}
&[\hspace{0.02cm}[\hspace{0.02cm}b^{\phantom{\dagger}\!}_{m},
a^{\dagger}_{k}\hspace{0.02cm}],a^{\phantom{\dagger}\!}_{l}] = 2\hspace{0.02cm}\delta^{\phantom{\dagger}\!}_{kl}\hspace{0.02cm}b^{\phantom{\dagger}\!}_{m} + 4\hspace{0.02cm}\delta^{\phantom{\dagger}\!}_{km}\hspace{0.02cm}b^{\phantom{\dagger}\!}_{l},
\label{eq:3q}\\[0.5ex]
&[\hspace{0.02cm}[\hspace{0.02cm}a^{\phantom{\dagger}\!}_{l},
b^{\phantom{\dagger}\!}_{m}\hspace{0.02cm}],a^{\dagger}_{k}] = -2\hspace{0.02cm}\delta^{\phantom{\dagger}\!}_{kl}\hspace{0.02cm}b^{\phantom{\dagger}\!}_{m} -2\hspace{0.02cm}\delta^{\phantom{\dagger}\!}_{km}\hspace{0.02cm}b^{\phantom{\dagger}\!}_{l}.
\label{eq:3w}
\end{align}
These relations differ from similar relations (\ref{eq:2u}) and (\ref{eq:2i}) of the Greenberg-Messiah scheme quantization by the presence of two last terms on the right-hand sides. Summing (\ref{eq:3q}) and (\ref{eq:3w}) and making use of the Jacobi identity, we obtain an analogue of the trilinear relation (\ref{eq:2r})
\begin{equation}
[\hspace{0.02cm}[\hspace{0.02cm}a^{\dagger}_{k},
a^{\phantom{\dagger}\!}_{l}\hspace{0.02cm}],b^{\phantom{\dagger}\!}_{m}] = -2\hspace{0.02cm}\delta^{\phantom{\dagger}\!}_{km}\hspace{0.02cm}b^{\phantom{\dagger}\!}_{l}.
\label{eq:3e}
\end{equation}
Here, we also observe an appearance of nonzero term on the right-hand side.\\
\indent
Let us present the $a$ and $b$ operators in the form of the Green expansions (\ref{eq:2o}). The question now arises as to what should be the commutations rules for the Green components $a_{k}^{(\alpha)}$ and $b_{m}^{(\beta)}$ so that the Govorkov trilinear relations (\ref{eq:3q}) and (\ref{eq:3e}) are identically fulfilled. It is clear that the commutation rules (\ref{eq:2p}) (and similarly for the $\phi_{b}$ field) must be true in this case also, since two sets of the $a$ and $b$ operators separately satisfy the standard trilinear relations for para-Fermi fields, Eqs.\,(\ref{eq:1q}) and (\ref{eq:1w}). Therefore, we need to generalize the relations (\ref{eq:2a}) and (\ref{eq:2s}) for the Green components of different fields. Note that these commutation rules are to a certain extent rather simple.\\
\indent
To be specific, let us consider the relation (\ref{eq:3e}). We write the left-hand side in terms of the Green components. For the commutator $[\hspace{0.02cm}a^{\dagger}_{k},a_{l}\hspace{0.02cm}]$, we have
\[
[\hspace{0.02cm}a^{\dagger}_{k},a^{\phantom{\dagger}\!}_{l}\hspace{0.02cm}] =
\sum^{p}_{\alpha\hspace{0.02cm}=\hspace{0.02cm}1}\,
[\hspace{0.02cm}a^{\dagger(\alpha)}_{k},a^{(\alpha)}_{l}\hspace{0.02cm}]
+
\sum_{\alpha\neq\beta}\,[\hspace{0.02cm}a^{\dagger(\alpha)}_{k},a^{(\beta)}_{l}\hspace{0.02cm}].
\]
By virtue of (\ref{eq:2p}), here the last term is equal to zero. For the double commutator, we get
\begin{equation}
[\hspace{0.02cm}[\hspace{0.02cm}a^{\dagger}_{k},
a^{\phantom{\dagger}\!}_{l}\hspace{0.02cm}],b^{\phantom{\dagger}\!}_{m}] =
\sum^{p}_{\alpha\hspace{0.02cm}=\hspace{0.02cm}1}\,[\hspace{0.02cm}
[\hspace{0.02cm}a^{\dagger(\alpha)}_{k},a^{(\alpha)}_{l}\hspace{0.02cm}],b^{(\alpha)}_{m}]
\hspace{0.02cm}+\hspace{0.02cm}
\sum_{\alpha\hspace{0.02cm}\neq\hspace{0.02cm}\beta}\,[\hspace{0.02cm}
[\hspace{0.02cm}a^{\dagger(\alpha)}_{k},a^{(\alpha)}_{l}\hspace{0.02cm}],b^{(\beta)}_{m}].
\label{eq:3r}
\end{equation}
The symbol $\sum_{\alpha\hspace{0.02cm}\neq\hspace{0.02cm}\beta}$  denotes a sum both over the $\alpha$ and $\beta$ indices with the only restriction $\alpha \neq \beta$. For the first expression under the sum sign on the right-hand side of (\ref{eq:3r}) we use the operator identity (\ref{ap:B2}), and for the second one we use usual Jacobi's identity (\ref{ap:B1})
\begin{align}
&[\hspace{0.02cm}
[\hspace{0.02cm}a^{\dagger(\alpha)}_{k},a^{(\alpha)}_{l}\hspace{0.02cm}],b^{(\alpha)}_{m}] =
\{a^{\dagger(\alpha)}_{k},\{b^{(\alpha)}_{m}, a^{(\alpha)}_{l}\}\!\hspace{0.02cm}\}
-
\{a^{(\alpha)}_{l},\{b^{(\alpha)}_{m}, a^{\dagger(\alpha)}_{k}\}\!\hspace{0.02cm}\},
\label{eq:3t}\\[0.8ex]
&[\hspace{0.02cm}
[\hspace{0.02cm}a^{\dagger(\alpha)}_{k},a^{(\alpha)}_{l}\hspace{0.02cm}],b^{(\beta)}_{m}] =
- [\hspace{0.02cm}
[\hspace{0.02cm}a^{(\alpha)}_{l},b^{(\beta)}_{m}\hspace{0.02cm}],a^{\dagger(\alpha)}_{k}]
- [\hspace{0.02cm}
[\hspace{0.02cm}b^{(\beta)}_{m},a^{\dagger(\alpha)}_{k}\hspace{0.02cm}],a^{(\alpha)}_{l}],
\quad \alpha\neq\beta.
\label{eq:3y}
\end{align}
By virtue of the Greenberg-Messiah commutation rules (\ref{eq:2a}) and (\ref{eq:2s}), both expressions vanish and thereby we arrive at (\ref{eq:2r}). Let us modify the first two relations in (\ref{eq:2a}) leaving the remaining ones unchanged (in this case, the second double commutator (\ref{eq:3y}) vanishes). For this purpose we introduce a new operator $\Omega$, as an additional algebraic element satisfying the relations
\begin{equation}
\begin{split}
&\{a^{\dagger(\alpha)}_{k}\!,b^{(\alpha)}_{m}\} = 2\hspace{0.015cm}\delta^{\phantom{\dagger}\!}_{mk}\hspace{0.02cm}\Omega, \quad\;
\{a^{(\alpha)}_{k}\!,\Omega\} = b^{(\alpha)}_{k},\\[1ex]
&\{b^{\dagger(\alpha)}_{m}\!,a^{(\alpha)}_{k}\} = 2\hspace{0.015cm}\delta^{\phantom{\dagger}\!}_{mk}\hspace{0.02cm}{\Omega}^{\dagger}, \quad
\{b^{(\alpha)}_{m}\!,{\Omega}\} = -\hspace{0.02cm}a^{(\alpha)}_{m}.
\label{eq:3u}
\end{split}
\end{equation}
In this case, as it can be easily seen, the expression (\ref{eq:3t}) results in
\[
[\hspace{0.02cm}
[\hspace{0.02cm}a^{\dagger(\alpha)}_{k},a^{(\alpha)}_{l}\hspace{0.02cm}],b^{(\alpha)}_{m}] =
-2\hspace{0.02cm}\delta_{mk}\hspace{0.02cm}b^{(\alpha)}_{l}
\]
and thus, by virtue of (\ref{eq:3r}), we reproduce (\ref{eq:3e}). However, if we try to apply the commutation rules (\ref{eq:3u}) to the trilinear relation (\ref{eq:3q}), we may see that the last term on the right-hand side is not reproduced. Here, we need a more radical modification of the relations (\ref{eq:2a}) and (\ref{eq:2s}). Below we shall postulate a new system of bilinear relations for Green's components $a_{k}^{(\alpha)}$ and $b_{m}^{(\beta)}$. Then, step by step, we verify that these commutation rules turn the Govorkov bilinear (\ref{eq:1t}), (\ref{eq:1y}) and trilinear (\ref{eq:3q}), (\ref{eq:3e}) relations into identity. However, it will take place only for a very special case of parastatistics, namely, for that of order 2.\\
\indent
Let us require that Green's components $a_k^{(\alpha)}$, $b_m^{(\beta)}$ and an additional operator $\Omega$ satisfy the following system of commutation rules
\begin{align}
&[\hspace{0.02cm}b^{(\alpha)}_{m}\!,a^{\dagger(\alpha)}_{k}\hspace{0.02cm}] = 2\hspace{0.02cm}\delta^{\phantom{\dagger}\!}_{mk}\hspace{0.02cm}\Omega, \quad
[\hspace{0.02cm}a^{(\alpha)}_{k}\!,b^{\dagger(\alpha)}_{m}\hspace{0.02cm}] = 2\hspace{0.02cm}\delta^{\phantom{\dagger}\!}_{mk}\hspace{0.02cm}{\Omega}^{\dagger},
\label{eq:3i}\\[0.8ex]
&[\hspace{0.02cm}a^{(\alpha)}_{k}\!,b^{(\alpha)}_{m}\hspace{0.02cm}] =
[\hspace{0.02cm}a^{\dagger(\alpha)}_{k}\!,b^{\dagger(\alpha)}_{m}\hspace{0.02cm}] = 0,
\label{eq:3o}\\[0.8ex]
&[\hspace{0.02cm}\Omega, a^{(\alpha)}_{k}\hspace{0.02cm}] = b^{(\alpha)}_{k},
\quad
[\hspace{0.03cm}\Omega, b^{(\alpha)}_{m}\hspace{0.02cm}] = -a^{(\alpha)}_{m},
\label{eq:3p}\\[0.8ex]
&\{a^{(\alpha)}_{k}\!,b^{(\beta)}_{m}\} = \{a^{\dagger(\alpha)}_{k}\!,b^{(\beta)}_{m}\} =
\{a^{(\alpha)}_{k}\!,b^{\dagger(\beta)}_{m}\} =
\{a^{\dagger(\alpha)}_{k}\!,b^{\dagger(\beta)}_{m}\} = 0, \quad \alpha\neq\beta.
\label{eq:3a}
\end{align}
It should be noted that not all relations (\ref{eq:3i})\,--\,(\ref{eq:3a}) are independent. As it will be shown at the end of this section, the relations (\ref{eq:3p}) are consequence of (\ref{eq:3i}), (\ref{eq:3a}) and of the bilinear relations (\ref{eq:1t}). Comparing (\ref{eq:3i}) and (\ref{eq:3p}) with the relations (\ref{eq:3u}), one sees that instead of anticommutators in (\ref{eq:3u}) here we have commutators. The same is true for the Greenberg-Messiah relations (\ref{eq:2s}), in which we replace commutators by anticommutators, Eqs.\,(\ref{eq:3a}).\\
\indent
First we consider the simplest relations from Govorkov's commutation rules, namely the bilinear relations (\ref{eq:1t}) and (\ref{eq:1y}). In particular, the relation (\ref{eq:1t}) implies
\[
[\hspace{0.02cm}a_{k},b_{m}\hspace{0.02cm}] = [\hspace{0.02cm}a_{m},b_{k}\hspace{0.02cm}].
\]
Substituting the decomposition (\ref{eq:2o}) into the left-hand side of the relation above and taking into account (\ref{eq:3o}),  (\ref{eq:3p}) and Jacoby's identity, we have the following chain of equalities:
\[
[\hspace{0.02cm}a_{k},b_{m}\hspace{0.02cm}] =
\sum_{\alpha\hspace{0.02cm}\neq\hspace{0.02cm}\beta}\,
[\hspace{0.02cm}a^{(\alpha)}_{k},b^{(\beta)}_{m}\hspace{0.02cm}] =
\sum_{\alpha\hspace{0.02cm}\neq\hspace{0.02cm}\beta}\,[\hspace{0.02cm}a^{(\alpha)}_{k},
[\hspace{0.02cm}\Omega, a^{(\beta)}_{m}\hspace{0.02cm}]\hspace{0.02cm}]
\]
\[
= -\sum_{\alpha\hspace{0.02cm}\neq\hspace{0.02cm}\beta}\,\Bigl(\hspace{0.02cm}
[\hspace{0.02cm}\Omega, [\hspace{0.02cm} a^{(\beta)}_{m}, a^{(\alpha)}_{k} \hspace{0.02cm}]\hspace{0.02cm}] +
[\hspace{0.02cm}a^{(\beta)}_{m},[\hspace{0.02cm} a^{(\alpha)}_{k}, \Omega\hspace{0.02cm}]\hspace{0.02cm}]\Bigr) =
\sum_{\alpha\hspace{0.02cm}\neq\hspace{0.02cm}\beta}\,
[\hspace{0.02cm}a^{(\beta)}_{m},b^{(\alpha)}_{k}\hspace{0.02cm}] \equiv
 [\hspace{0.02cm}a_{m},b_{k}\hspace{0.02cm}].
\]
In deriving the above relation we also have used the commutation rules for Green's components of the $\phi_{a}$ field,
Eq.\,(\ref{eq:2p}).\\
\indent
Further, we consider bilinear relation (\ref{eq:1y}), which in turns implies
\[
[\hspace{0.02cm}a_{k},a_{m}\hspace{0.02cm}] = [\hspace{0.02cm}b_{k},b_{m}\hspace{0.02cm}].
\]
Using the commutation rules (\ref{eq:2p}), the relations (\ref{eq:3o}), (\ref{eq:3p}) and Jacoby's identity (\ref{ap:B1}), here we have the chain of equalities
\[
[\hspace{0.02cm}a_{k},a_{m}\hspace{0.02cm}] =
\sum^{p}_{\alpha\hspace{0.02cm}=\hspace{0.02cm}1}\,
[\hspace{0.02cm}a^{(\alpha)}_{k},a^{(\alpha)}_{m}\hspace{0.02cm}] =
\sum^{p}_{\alpha\hspace{0.02cm}=\hspace{0.02cm}1}\,
[\hspace{0.02cm}a^{(\alpha)}_{k},[\hspace{0.03cm}\Omega, b^{(\alpha)}_{m}\hspace{0.02cm}]\hspace{0.02cm}]
\]
\[
= -\sum_{\alpha}\Bigl(\hspace{0.02cm}[\hspace{0.02cm}b^{(\alpha)}_{m}, [\hspace{0.02cm}a^{(\alpha)}_{k},\Omega\hspace{0.02cm}]\hspace{0.02cm}]
+
[\hspace{0.02cm}\Omega, [\hspace{0.02cm}b^{(\alpha)}_{m}, a^{(\alpha)}_{k}\hspace{0.02cm}]\hspace{0.02cm}]\Bigr)
= -\sum_{\alpha}\,[\hspace{0.02cm}b^{(\alpha)}_{m},b^{(\alpha)}_{k}\hspace{0.02cm}]
\equiv [\hspace{0.02cm}b_{k},b_{m}\hspace{0.02cm}].
\]
\indent
Let us return again to the bilinear relation (\ref{eq:1t}). We will analyze slightly more complicated case, when one operator is creation operator, and another is annihilation operator
\begin{equation}
[\hspace{0.02cm}a^{\dagger}_{k},b^{\phantom{\dagger}\!}_{m}\hspace{0.02cm}] = [\hspace{0.02cm}a^{\phantom{\dagger}\!}_{m},b^{\dagger}_{k}\hspace{0.02cm}].
\label{eq:3s}
\end{equation}
By using the relations (\ref{eq:3i}) and (\ref{eq:3p}), we find for the left-hand side of (\ref{eq:3s}):
\begin{equation}
[\hspace{0.02cm}a^{\dagger}_{k},b^{\phantom{\dagger}\!}_{m}\hspace{0.02cm}] =
\sum_{\alpha}\,[\hspace{0.02cm}a^{\dagger(\alpha)}_{k},b^{(\alpha)}_{m}\hspace{0.02cm}]
\hspace{0.04cm}+\hspace{0.02cm}
\sum_{\alpha\hspace{0.02cm}\neq\hspace{0.02cm}\beta}\,
[\hspace{0.02cm}a^{\dagger(\alpha)}_{k},b^{(\beta)}_{m}\hspace{0.02cm}]
= -2p\hspace{0.03cm}\delta^{\phantom{\dagger}\!}_{km}\hspace{0.01cm}\Omega
\hspace{0.04cm}+\hspace{0.02cm}
\sum_{\alpha\hspace{0.02cm}\neq\hspace{0.02cm}\beta}\,
[\hspace{0.02cm}a^{\dagger(\alpha)}_{k},[\hspace{0.03cm}\Omega, a^{(\beta)}_{m}\hspace{0.02cm}]\hspace{0.02cm}].
\label{eq:3d}
\end{equation}
Here, the expression under the sum sign in the last term, by virtue of Jacoby's identity and the commutation rules (\ref{eq:2p}), (\ref{eq:3p}), is equal to
\[
[\hspace{0.02cm}a^{\dagger(\alpha)}_{k},[\hspace{0.02cm}\Omega, a^{(\beta)}_{m}\hspace{0.02cm}]\hspace{0.02cm}] =
-[\hspace{0.02cm}\Omega, [\hspace{0.02cm}a^{(\beta)}_{m}, a^{\dagger(\alpha)}_{k} \hspace{0.02cm}]\hspace{0.02cm}]
-
[\hspace{0.02cm}a^{(\beta)}_{m},[\hspace{0.02cm} a^{\dagger(\alpha)}_{k}, \Omega\hspace{0.02cm}]\hspace{0.02cm}]
= [\hspace{0.02cm}a^{(\beta)}_{m},b^{\dagger(\alpha)}_{k}\hspace{0.02cm}].
\]
Adding and subtracting the sum $\sum_{\alpha}\,[\hspace{0.02cm}a^{(\alpha)}_{m},
b^{\dagger(\alpha)}_{k}\hspace{0.02cm}]\,(\equiv 2\hspace{0.02cm}p\hspace{0.03cm}\delta_{mk}\hspace{0.01cm}\Omega^{\dagger})$ to the right-hand side of (\ref{eq:3d}), we have
\[
[\hspace{0.01cm}a^{\dagger}_{k},b^{\phantom{\dagger}\!}_{m}\hspace{0.02cm}] = -2\hspace{0.015cm}p\hspace{0.04cm}\delta^{\phantom{\dagger}\!}_{km}\hspace{0.04cm}\Omega
-\sum_{\alpha}\,[\hspace{0.02cm}a^{(\alpha)}_{m},b^{\dagger(\alpha)}_{k}\hspace{0.02cm}]
+
\Bigl(\,\sum_{\alpha}\,[\hspace{0.02cm}a^{(\alpha)}_{m},b^{\dagger(\alpha)}_{k}\hspace{0.02cm}]
+
\sum_{\alpha\hspace{0.02cm}\neq\hspace{0.02cm}\beta}\,[\hspace{0.02cm}a^{(\beta)}_{m},
b^{\dagger(\alpha)}_{k}\hspace{0.02cm}]\Bigr)
\]
\[
\equiv -2\hspace{0.01cm}p\hspace{0.03cm}\delta^{\phantom{\dagger}\!}_{km}\hspace{0.02cm}
\bigl(\Omega + \Omega^{\dagger}\bigr)
+ [\hspace{0.02cm}a^{\phantom{\dagger}\!}_{m},b^{\dagger}_{k}\hspace{0.02cm}].
\]
Thus the bilinear relation (\ref{eq:3s}) will hold if the operator $\Omega$ satisfies the following condition:
\begin{equation}
\Omega + \Omega^{\dagger} = 0.
\label{eq:3f}
\end{equation}
The examples discussed are sufficient to state that the bilinear relations (\ref{eq:1t}), (\ref{eq:1y}) turn into identity by a system of the commutation rules (\ref{eq:2p}), (\ref{eq:3i})\,--\,(\ref{eq:3a}) and by an additional condition for the operator $\Omega$, Eq.\,(\ref{eq:3f}).\\
\indent
At the end of this section, we will show that for a particular case of parastatistics, namely for $p=2$, the commutation rules (\ref{eq:3p}) are consequence of (\ref{eq:3i}), (\ref{eq:3a}) and of the bilinear relation (\ref{eq:3s}). In other words, if we postulate the validity of (\ref{eq:3i}), (\ref{eq:3a}) and (\ref{eq:3s}), the relations (\ref{eq:3p}) will be their inevitable consequence. For this purpose we rewrite (\ref{eq:3s}) in terms of the Green components:
\begin{equation}
[\hspace{0.02cm}a^{\dagger(1)}_{k},b^{(2)}_{m}\hspace{0.02cm}]
+
[\hspace{0.02cm}a^{\dagger(2)}_{k},b^{(1)}_{m}\hspace{0.02cm}]
=
-\bigl(\hspace{0.02cm}[\hspace{0.02cm}b^{\dagger(1)}_{k},a^{(2)}_{m}\hspace{0.02cm}]
+
[\hspace{0.02cm}b^{\dagger(2)}_{k},a^{(1)}_{m}\hspace{0.02cm}]\hspace{0.02cm}\bigr).
\label{eq:3g}
\end{equation}
Now we calculate the commutator between (\ref{eq:3g}) and the operator $a_l^{(1)}$. Here we
have two commutators different from zero:
\begin{align}
&[\hspace{0.02cm}
[\hspace{0.02cm}a^{\dagger(1)}_{k},b^{(2)}_{m}\hspace{0.02cm}],a^{(1)}_{l}]
=
\{a^{\dagger(1)}_{k},\{a^{(1)}_{l},b^{(2)}_{m}\}\!\hspace{0.02cm}\}
-
\{b^{(2)}_{m}, \{a^{(1)}_{l}, a^{\dagger(1)}_{k}\}\!\hspace{0.02cm}\}
=
-2\hspace{0.015cm}\delta_{kl}\hspace{0.02cm}b^{(2)}_{m},
\label{eq:3h}\\[0.8ex]
&[\hspace{0.02cm}
[\hspace{0.02cm}b^{\dagger(1)}_{k},a^{(2)}_{m}\hspace{0.02cm}],a^{(1)}_{l}]
=
- [\hspace{0.02cm}
[\hspace{0.02cm}a^{(2)}_{m},a^{(1)}_{l}\hspace{0.02cm}],b^{\dagger(1)}_{k}]
- [\hspace{0.02cm}
[\hspace{0.02cm}a^{(1)}_{l},b^{\dagger(1)}_{k}\hspace{0.02cm}],a^{(2)}_{m}]
=
2\hspace{0.015cm}\delta_{kl}\hspace{0.02cm}[\hspace{0.02cm}\Omega, a^{(2)}_{m}\hspace{0.02cm}].
\notag
\end{align}
In the first case we have used the operator identity (\ref{ap:B2}). By doing so, the required commutator of the operator $a_l^{(1)}$ with  (\ref{eq:3g}) leads us to the relation $[\hspace{0.02cm}\Omega, a^{(2)}_{m}\hspace{0.02cm}] = b^{(2)}_{m}$, and a similar commutator with $a_l^{(2)}$ gives $[\hspace{0.02cm}\Omega, a^{(1)}_{m}\hspace{0.02cm}] = b^{(1)}_{m}$, and we reproduce the first relation in (\ref{eq:3p}). To provide the second relation it is necessary to take the commutator between $b_l^{(\alpha)}$ and (\ref{eq:3g}). For $\alpha = 1$ the commutators distinct from zero are
\begin{align}
&[\hspace{0.02cm}
[\hspace{0.02cm}a^{\dagger(1)}_{k},b^{(2)}_{m}\hspace{0.02cm}],b^{(1)}_{l}]
=
- [\hspace{0.02cm}
[\hspace{0.02cm}b^{(2)}_{m},b^{(1)}_{l}\hspace{0.02cm}],a^{\dagger(1)}_{k}]
- [\hspace{0.02cm}
[\hspace{0.02cm}b^{(1)}_{l},a^{\dagger(1)}_{k}\hspace{0.02cm}],b^{(2)}_{m}]
=
-2\hspace{0.015cm}\delta_{kl}\hspace{0.02cm}[\hspace{0.02cm}\Omega, b^{(2)}_{m}\hspace{0.02cm}],
\label{eq:3j}\\[0.8ex]
&[\hspace{0.02cm}
[\hspace{0.02cm}b^{\dagger(1)}_{k},a^{(2)}_{m}\hspace{0.02cm}],b^{(1)}_{l}]
=
\{b^{\dagger(1)}_{k},\{b^{(1)}_{l},a^{(2)}_{m}\}\!\hspace{0.02cm}\}
-
\{a^{(2)}_{m}, \{b^{(1)}_{l}, b^{\dagger(1)}_{k}\}\!\hspace{0.02cm}\}
=
-2\hspace{0.015cm}\delta_{kl}\hspace{0.02cm}a^{(2)}_{m}.
\notag
\end{align}
That gives us $[\hspace{0.02cm}\Omega, b^{(2)}_{m}\hspace{0.02cm}] = -\hspace{0.01cm}a^{(2)}_{m}$. The commutator containing $b_l^{(2)}$ results in a similar expression with the replacement of the Green index $2 \rightarrow 1$ and we reproduce thereby the second relation in (\ref{eq:3p}).

\section{\bf Govorkov's trilinear relations}
\setcounter{equation}{0}

Let us now turn immediately to analysis of the trilinear relations (\ref{eq:1e}) and (\ref{eq:1r}). Here, it will suffice to consider the special cases (\ref{eq:3q})\,--\,(\ref{eq:3e}). We have already analyzed the relation (\ref{eq:3e}) in the previous section, but now we will do this somewhat differently. For the first expression on the right-hand side of (\ref{eq:3r}) we use usual Jacobi's identity, and for the second expression we use the identity (\ref{ap:B2}). Then, by virtue of the commutation rules (\ref{eq:3i})\,--\,(\ref{eq:3a}), instead of (\ref{eq:3t}) and (\ref{eq:3y}), we get
\begin{align}
&\sum^{p}_{\alpha\hspace{0.02cm}=\hspace{0.02cm}1}\,[\hspace{0.02cm}
[\hspace{0.02cm}a^{\dagger(\alpha)}_{k},a^{(\alpha)}_{l}\hspace{0.02cm}],b^{(\alpha)}_{m}]
=
- \sum_{\alpha}\,\Bigl([\hspace{0.02cm}
[\hspace{0.02cm}a^{(\alpha)}_{l},b^{(\alpha)}_{m}\hspace{0.02cm}],a^{\dagger(\alpha)}_{k}]
+
\hspace{0.02cm}[\hspace{0.02cm}
[\hspace{0.02cm}b^{(\alpha)}_{m},a^{\dagger(\alpha)}_{k}\hspace{0.02cm}],a^{(\alpha)}_{l}]\Bigr)
=
\notag\\[0.8ex]
&= -2\hspace{0.01cm}\delta_{km}\hspace{0.01cm}
\sum_{\alpha}\,[\hspace{0.02cm}\Omega, a^{(\alpha)}_{l}\hspace{0.02cm}]
=
-2\hspace{0.01cm}\delta_{km}\hspace{0.01cm}b_{l},
\notag\\[0.8ex]
&\sum_{\alpha\hspace{0.02cm}\neq\hspace{0.02cm}\beta}\,[\hspace{0.02cm}
[\hspace{0.02cm}a^{\dagger(\alpha)}_{k},a^{(\alpha)}_{l}\hspace{0.02cm}],b^{(\beta)}_{m}]
= \sum_{\alpha\hspace{0.02cm}\neq\hspace{0.02cm}\beta}\,\Bigl(
\{a^{\dagger(\alpha)}_{k},\{b^{(\beta)}_{m}, a^{(\alpha)}_{l}\}\!\hspace{0.02cm}\}
-
\{a^{(\alpha)}_{l},\{b^{(\beta)}_{m}, a^{(\dagger\alpha)}_{k}\}\!\hspace{0.02cm}\}\Bigr)
= 0.
\notag
\end{align}
Here, we also reproduce the relation (\ref{eq:3e}), as it took place for the rules (\ref{eq:3u}).\\
\indent
Further, let us consider more complicated trilinear relation (\ref{eq:3q}), which for convenience we write out once again
\begin{equation}
[\hspace{0.02cm}[\hspace{0.02cm}b^{\phantom{\dagger}\!}_{m},
a^{\dagger}_{k}\hspace{0.02cm}],a^{\phantom{\dagger}\!}_{l}]
=
2\hspace{0.015cm}\delta^{\phantom{\dagger}\!}_{kl}\hspace{0.02cm}b^{\phantom{\dagger}\!}_{m} + 4\hspace{0.02cm}\delta^{\phantom{\dagger}\!}_{km}\hspace{0.02cm}b^{\phantom{\dagger}\!}_{l}.
\label{eq:4q}
\end{equation}
For the ``internal'' commutator we can use the result (\ref{eq:3d}):
\[
[\hspace{0.02cm}b^{\phantom{\dagger}\!}_{m},a^{\dagger}_{k}\hspace{0.02cm}] = 2p\hspace{0.025cm}\delta^{\phantom{\dagger}\!}_{km}\hspace{0.025cm}\Omega
\hspace{0.04cm}+\hspace{0.02cm}
\sum_{\alpha\hspace{0.02cm}\neq\hspace{0.02cm}\beta}\,
[\hspace{0.02cm}b^{(\alpha)}_{m},a^{\dagger(\beta)}_{k}\hspace{0.02cm}].
\]
Then the starting expression for an analysis of the left-hand side of (\ref{eq:4q}) takes the form
\begin{equation}
[\hspace{0.02cm}[\hspace{0.02cm}b^{\phantom{\dagger}\!}_{m},
a^{\dagger}_{k}\hspace{0.02cm}],a^{\phantom{\dagger}\!}_{l}]
=
2\hspace{0.01cm}p\hspace{0.025cm}\delta^{\phantom{\dagger}\!}_{mk}\hspace{0.035cm}
[\hspace{0.02cm}\Omega,a^{\phantom{\dagger}\!}_{l}\hspace{0.02cm}]
\hspace{0.04cm}+\hspace{0.02cm}
\sum_{\alpha\hspace{0.02cm}\neq\hspace{0.02cm}\beta}\,\sum_{\gamma}\,[\hspace{0.02cm}
[\hspace{0.02cm}b^{(\alpha)}_{m},a^{\dagger(\beta)}_{k}\hspace{0.02cm}], a^{(\gamma)}_{l}\hspace{0.02cm}].
\label{eq:4w}
\end{equation}
By means of the identity (\ref{ap:B2}), we write the double commutator under the sum sign
\[
[\hspace{0.02cm}[\hspace{0.02cm}b^{(\alpha)}_{m},a^{\dagger(\beta)}_{k}\hspace{0.02cm}], a^{(\gamma)}_{l}\hspace{0.02cm}]
=
\{b^{(\alpha)}_{m},\{a^{(\gamma)}_{l}, a^{\dagger(\beta)}_{k}\}\!\hspace{0.02cm}\}
-
\{a^{\dagger(\beta)}_{k},\{a^{(\gamma)}_{l}, b^{(\alpha)}_{m}\}\!\hspace{0.02cm}\},
\]
and present the triple sum as follows:
\begin{equation}
\sum_{\alpha\hspace{0.02cm}\neq\hspace{0.02cm}\beta}\,\sum_{\gamma}\hspace{0.02cm}
\hspace{0.02cm}=
\sum_{\alpha = \gamma\neq\beta} +
\sum_{\alpha\neq\beta = \gamma} +
\sum_{\alpha\neq\beta\neq\gamma}.
\label{eq:4e}
\end{equation}
\indent
Taking into account the relations (\ref{eq:2p}) and (\ref{eq:3a}), we have the following terms different from zero
\[
\sum_{\alpha\hspace{0.02cm}\neq\hspace{0.02cm}\beta}\,\sum_{\gamma}\,[\hspace{0.02cm}
[\hspace{0.02cm}b^{(\alpha)}_{m},a^{\dagger(\beta)}_{k}\hspace{0.02cm}], a^{(\gamma)}_{l}\hspace{0.02cm}]
\]
\[
=
2(p - 1)\hspace{0.03cm}\delta^{\phantom{\dagger}\!}_{lk}\,b^{\phantom{\dagger}\!}_{m}
+\!\!\sum_{\alpha = \gamma\neq\beta}\!\Bigl(
\{b^{(\alpha)}_{m},\{a^{(\alpha)}_{l}, a^{\dagger(\beta)}_{k}\}\!\hspace{0.02cm}\}
-
\{a^{\dagger(\beta)}_{k},\{a^{(\alpha)}_{l}, b^{(\alpha)}_{m}\}\!\hspace{0.02cm}\}\!\Bigr)
+\!\!\sum_{\alpha\neq\beta\neq\gamma}\!
\{b^{(\alpha)}_{m},\{a^{(\gamma)}_{l}, a^{\dagger(\beta)}_{k}\}\!\hspace{0.02cm}\}.
\]
The second contribution on the right-hand side with the help of the identities (\ref{ap:B2}) and (\ref{ap:B1}) can be presented as
\[
\sum_{\alpha = \gamma\neq\beta}\!\!
\Bigl(\{b^{(\alpha)}_{m},\{a^{(\alpha)}_{l}, a^{\dagger(\beta)}_{k}\}\!\hspace{0.02cm}\}
-
\{a^{\dagger(\beta)}_{k},\{a^{(\alpha)}_{l}, b^{(\alpha)}_{m}\}\!\hspace{0.02cm}\}\!\Bigr)
\!= \!-\!\!\!\sum_{\alpha = \gamma\neq\beta}\!\!\Bigl(
[\hspace{0.02cm}[\hspace{0.02cm}a^{\dagger(\beta)}_{m},a^{(\alpha)}_{l}\hspace{0.02cm}], b^{(\alpha)}_{m}\hspace{0.02cm}]
+
[\hspace{0.02cm}[\hspace{0.02cm}a^{(\alpha)}_{l},b^{(\alpha)}_{m}\hspace{0.02cm}], a^{\dagger(\beta)}_{k}\hspace{0.02cm}]\Bigr).
\]
It equals zero, by virtue of (\ref{eq:2p}) and (\ref{eq:3o}). We finally obtain, instead of (\ref{eq:4w}),
\begin{equation}
[\hspace{0.02cm}[\hspace{0.02cm}b^{\phantom{\dagger}\!}_{m},
a^{\dagger}_{k}\hspace{0.02cm}],a^{\phantom{\dagger}\!}_{l}]
=
2\hspace{0.01cm}(p - 1)\hspace{0.02cm}\delta^{\phantom{\dagger}\!}_{lk}\hspace{0.03cm}
b^{\phantom{\dagger}\!}_{m}
+
2\hspace{0.01cm}p\hspace{0.03cm}\delta^{\phantom{\dagger}\!}_{mk}\hspace{0.03cm}
b^{\phantom{\dagger}\!}_{l}
\hspace{0.04cm}+\!\sum_{\alpha\neq\beta\neq\gamma}\!
\{b^{(\alpha)}_{m},\{a^{(\gamma)}_{l}, a^{\dagger(\beta)}_{k}\}\!\hspace{0.02cm}\}.
\label{eq:4r}
\end{equation}
We see that this expression reproduces (\ref{eq:4q}) in the only case, when $p = 2$. In this special case the last term on the right-hand side of (\ref{eq:4r}) is simply absent and the numerical coefficients in the other terms take correct values.\\
\indent
The trilinear relation (\ref{eq:3w}) is automatically fulfilled on the strength of Jacoby's identity. Nevertheless, it is instructively to see this directly. By virtue of (\ref{eq:3o}), the equality holds
\[
[\hspace{0.02cm}a^{\phantom{\dagger}\!}_{l},b^{\phantom{\dagger}\!}_{m}\hspace{0.02cm}] =
\sum_{\alpha\hspace{0.02cm}\neq\hspace{0.02cm}\beta}\;
[\hspace{0.02cm}a^{(\alpha)}_{l},b^{(\beta)}_{m}\hspace{0.02cm}]
\]
and therefore,
\begin{equation}
[\hspace{0.02cm}[\hspace{0.02cm}a^{\phantom{\dagger}\!}_{l},b^{\phantom{\dagger}\!}_{m}
\hspace{0.02cm}],a^{\dagger}_{k}]
= \sum_{\alpha\hspace{0.02cm}\neq\hspace{0.02cm}\beta}\,\sum_{\gamma}\;
[\hspace{0.02cm}[\hspace{0.02cm}a^{(\alpha)}_{l},b^{(\beta)}_{m}\hspace{0.02cm}],
a^{\dagger(\gamma)}_{k}\hspace{0.02cm}]
\label{eq:4t}
\end{equation}
\[
=
\sum_{\alpha\hspace{0.02cm}\neq\hspace{0.02cm}\beta}\,\sum_{\gamma}\hspace{0.02cm}\Bigl(
\{a^{(\alpha)}_{l},\{a^{\dagger(\gamma)}_{k}, b^{(\beta)}_{m}\}\!\hspace{0.02cm}\}
-
\{b^{(\beta)}_{m},\{a^{\dagger(\gamma)}_{k}, a^{(\alpha)}_{l}\}\!\hspace{0.02cm}\}\!\hspace{0.02cm}\Bigr).
\]
We present again the triple sum on the right-hand side of (\ref{eq:4t}) in the form of the decomposition (\ref{eq:4e}). Then, with allowance for the commutation rules (\ref{eq:3a}) and (\ref{eq:2p}), the expression (\ref{eq:4t}) takes the form
\begin{align}
[\hspace{0.02cm}[\hspace{0.02cm}a^{\phantom{\dagger}\!}_{l},
b^{\phantom{\dagger}\!}_{m}\hspace{0.02cm}],a^{\dagger}_{k}]
=
-2\hspace{0.015cm}(p - 1)\hspace{0.01cm}\delta^{\phantom{\dagger}\!}_{kl}\hspace{0.02cm}
b^{\phantom{\dagger}\!}_{m}
&+\!
\sum_{\alpha\neq\beta = \gamma}\Bigl(
\{a^{(\alpha)}_{l},\{a^{\dagger(\beta)}_{k}, b^{(\beta)}_{m}\}\!\hspace{0.02cm}\}
-
\{b^{(\beta)}_{m},\{a^{\dagger(\beta)}_{k}, a^{(\alpha)}_{l}\}\!\hspace{0.02cm}\}\!\hspace{0.02cm}\Bigr)
\notag\\[0.8ex]
&-\!\sum_{\alpha\neq\beta\neq\gamma}\!
\{b^{(\beta)}_{m},\{a^{\dagger(\gamma)}_{k}, a^{(\alpha)}_{l}\}\!\hspace{0.02cm}\}.
\notag
\end{align}
For the second contribution on the right-hand side we have the chain of equalities
\begin{align}
\sum_{\alpha\neq\beta = \gamma}\!\!\Bigl(
\{a^{(\alpha)}_{l},\{a^{\dagger(\beta)}_{k}, b^{(\beta)}_{m}\}\!\hspace{0.02cm}\}
&-
\{b^{(\beta)}_{m},\{a^{\dagger(\beta)}_{k}, a^{(\alpha)}_{l}\}\!\hspace{0.03cm}\}\!\Bigr)
=
-\!\!\!\sum_{\alpha\neq\beta = \gamma}\!\!\Bigl(\hspace{0.02cm}
[\hspace{0.02cm}[\hspace{0.02cm}b^{(\beta)}_{m},a^{\dagger(\beta)}_{k}\hspace{0.02cm}], a^{(\alpha)}_{l}\hspace{0.02cm}]
+
[\hspace{0.02cm}[\hspace{0.02cm}a^{\dagger(\beta)}_{k},a^{(\alpha)}_{l}\hspace{0.02cm}], b^{(\beta)}_{m}\hspace{0.02cm}]\Bigr)
\notag\\[0.8ex]
&= 2\hspace{0.01cm}\delta_{mk}\!\!\sum_{\alpha\neq\beta = \gamma}[\hspace{0.02cm}\Omega,a^{(\alpha)}_{l}\hspace{0.02cm}]
=
2\hspace{0.015cm}(p - 1)\hspace{0.015cm}\delta_{mk}\hspace{0.03cm}b_{l}.
\notag
\end{align}
Then, instead of (\ref{eq:4t}), we have now
\[
[\hspace{0.02cm}[\hspace{0.02cm}a^{\phantom{\dagger}\!}_{l},
b^{\phantom{\dagger}\!}_{m}\hspace{0.02cm}],a^{\dagger}_{k}]
=
-2\hspace{0.01cm}(p - 1)\hspace{0.015cm}\delta^{\phantom{\dagger}\!}_{kl}\hspace{0.02cm}
b^{\phantom{\dagger}\!}_{m}
-
2\hspace{0.01cm}(p - 1)\hspace{0.015cm}\delta^{\phantom{\dagger}\!}_{mk}\hspace{0.03cm}
b^{\phantom{\dagger}\!}_{l}
\hspace{0.04cm}
-\!\!\sum_{\alpha\neq\beta\neq\gamma}\!
\{b^{(\beta)}_{m},\{a^{\dagger(\gamma)}_{k}, a^{(\alpha)}_{l}\}\!\hspace{0.02cm}\}.
\]
We see that this expression reproduces the trilinear relation (\ref{eq:3w}) only for $p = 2$.

\section{Inclusion of para-Grassmann numbers}
\setcounter{equation}{0}

In this section we want to include in the general scheme of uniquantization the para-Grassmann numbers, which we designate by $\xi_k,\,k = 1,\ldots,M$. Now our task is to derive the commutation rules involving simultaneously the $\xi_k$ and the operators $a_k$, $b_{m}$. In the case of a single para-Fermi field, for instance $\phi_{a}$, such commutation rules were suggested in the paper by Omote and Kamefuchi \cite{omote_1979}:
\begin{equation}
\begin{split}
&[\hspace{0.03cm}a^{\phantom{\dagger}\!}_{k}, [\hspace{0.02cm}a^{\phantom{\dagger}\!}_{l},
\xi^{\phantom{\dagger}\!}_{m}\hspace{0.02cm}]\hspace{0.02cm}] = 0,
\quad
[\hspace{0.03cm}a^{\phantom{\dagger}\!}_{k}, [\hspace{0.02cm}a^{\dagger}_{l},\xi^{\phantom{\dagger}\!}_{m}\hspace{0.02cm}]\hspace{0.02cm}]
=
2\hspace{0.02cm}\delta^{\phantom{\dagger}\!}_{kl}\hspace{0.02cm}\xi^{\phantom{\dagger}\!}_{m}, \\[0.8ex]
&[\hspace{0.03cm}\xi_{k}, [\hspace{0.02cm}\xi_{l},a_{m}\hspace{0.02cm}]\hspace{0.02cm}] = 0,
\quad\hspace{0.03cm}
[\hspace{0.03cm}\xi_{k}, [\hspace{0.02cm}\xi_{l},\xi_{m}\hspace{0.02cm}]\hspace{0.02cm}] = 0.
\label{eq:5q}
\end{split}
\end{equation}
For the special case of parastatistics $p = 2$ instead of the last relation in (\ref{eq:5q}) we can use more simple expression
\[
\xi_{k}\hspace{0.02cm}\xi_{l}\hspace{0.02cm}\xi_{m} \hspace{0.02cm}+\hspace{0.02cm} \xi_{m}\hspace{0.02cm}\xi_{l}\hspace{0.02cm}\xi_{k} = 0.
\]
The remaining relations follows from (\ref{eq:5q}) by a Hermitian conjugation. We believe that similar commutation rules take place for the second $\phi_{b}$ field also. For the para-Grassmann numbers $\xi_k$ the Green representation
\[
\xi^{\phantom{\dagger}\!}_{k} = \sum^{p}_{\alpha\hspace{0.02cm}=\hspace{0.02cm}1}\hspace{0.02cm}\xi^{(\alpha)}_{k}
\]
is also true. The bilinear commutation relations for the Green components $a_{k}^{(\alpha)}$, $b^{(\alpha)}_{m}$, and $\xi_l^{(\alpha)}$ were given in \cite{ohnuki_1980}:
\begin{equation}
\begin{split}
&\{a^{(\alpha)}_{k}\!,\hspace{0.02cm}\xi^{(\alpha)}_{l\!}\} = 0,
\quad
\{b^{(\alpha)}_{m},\hspace{0.02cm}\xi^{(\alpha)}_{l\!}\} = 0,
\quad
\{\xi^{(\alpha)}_{k}\!,\hspace{0.02cm}\xi^{(\alpha)}_{l\!}\} = 0, \\[1.2ex]
&[\hspace{0.03cm}a^{(\alpha)}_{k}\!,\hspace{0.02cm}\xi^{(\beta)}_{l}\hspace{0.02cm}] = 0,
\;\;\quad
[\hspace{0.03cm}b^{(\alpha)}_{m},\hspace{0.02cm}\xi^{(\beta)}_{l}\hspace{0.02cm}] = 0,
\quad\;
[\hspace{0.03cm}\xi^{(\alpha)}_{k}\!,\hspace{0.02cm}\xi^{(\beta)}_{l}\hspace{0.02cm}] = 0,
\quad \alpha\neq\beta
\label{eq:5w}
\end{split}
\end{equation}
plus Hermitian conjugation. They turn (\ref{eq:5q}) into identity.\\
\indent
As mentioned above, we are interested in the trilinear relations, including simultaneously the operators $a_{k}$, $b_{m}$, and the para-Grassmann numbers $\xi_k$. It might seem natural to start our investigation with the following expressions:
\[
[\hspace{0.03cm}b^{\phantom{\dagger}\!}_{m}, [\hspace{0.02cm}a^{\dagger}_{k},\xi^{\phantom{\dagger}\!}_{l}\hspace{0.02cm}]\hspace{0.02cm}],
\quad
[\hspace{0.03cm}a^{\phantom{\dagger}\!}_{k}, [\hspace{0.02cm}b^{\dagger}_{m},\xi^{\phantom{\dagger}\!}_{l}\hspace{0.02cm}]\hspace{0.02cm}].
\]
However, a preliminary analysis of these relations, by using (\ref{eq:3i})\,--\,(\ref{eq:3a}) and (\ref{eq:5w}), has shown that these double commutators eventually result in somewhat tangled expressions. Therefore, keeping in mind this analysis, we consider slightly different trilinear relations, namely,
\begin{equation}
\{\hspace{0.01cm}b^{\phantom{\dagger}\!}_{m}, [\hspace{0.02cm}a^{\dagger}_{k},\xi^{\phantom{\dagger}\!}_{l}\hspace{0.02cm}]\},
\quad
\{\hspace{0.01cm}a^{\phantom{\dagger}\!}_{k}, [\hspace{0.02cm}b^{\dagger}_{m},\xi^{\phantom{\dagger}\!}_{l}\hspace{0.02cm}]\hspace{0.005cm}\}.
\label{eq:5e}
\end{equation}
In terms of the Green components with regard to (\ref{eq:5w}), the first expression here, takes the form
\begin{equation}
\{\hspace{0.01cm}b^{\phantom{\dagger}\!}_{m}, [\hspace{0.02cm}a^{\dagger}_{k},\xi^{\phantom{\dagger}\!}_{l}\hspace{0.02cm}]\}
=
\sum_{\alpha}\,\{\hspace{0.01cm}b^{(\alpha)}_{m}, [\hspace{0.02cm}a^{\dagger(\alpha)}_{k},\xi^{(\alpha)}_{l}\hspace{0.02cm}]\}
\hspace{0.02cm}+\hspace{0.02cm}
\sum_{\alpha\hspace{0.02cm}\neq\hspace{0.02cm}\beta}\,\{\hspace{0.01cm}b^{(\beta)}_{m}, [\hspace{0.02cm}a^{\dagger(\alpha)}_{k},\xi^{(\alpha)}_{l}\hspace{0.02cm}]\}.
\label{eq:5r}
\end{equation}
For the first term on the right-hand side by using the identity (\ref{ap:B3}) and the rules (\ref{eq:3i}), (\ref{eq:5w}), we have
\[
\{\hspace{0.01cm}b^{(\alpha)}_{m}, [\hspace{0.02cm}a^{\dagger(\alpha)}_{k},\xi^{(\alpha)}_{l}\hspace{0.02cm}]\}
=
\{\hspace{0.01cm}\xi^{(\alpha)}_{l}, [\hspace{0.02cm}b^{(\alpha)}_{m},a^{\dagger(\alpha)}_{k}\hspace{0.02cm}]\}
+
[\hspace{0.02cm}a^{\dagger(\alpha)}_{k}, \{\hspace{0.01cm}\xi^{(\alpha)}_{l},b^{(\alpha)}_{m}\}]
=
2\hspace{0.015cm}\delta_{mk}\hspace{0.02cm}\{\hspace{0.01cm}\xi^{(\alpha)}_{l},\Omega\}.
\]
It is easily seen with the use of the same identity, that the second term in (\ref{eq:5r}) vanishes
and thus, instead of (\ref{eq:5r}), we obtain
\begin{equation}
\{\hspace{0.01cm}b^{\phantom{\dagger}\!}_{m}, [\hspace{0.02cm}a^{\dagger}_{k},\xi^{\phantom{\dagger}\!}_{l}\hspace{0.02cm}]\}
=
2\hspace{0.015cm}\delta^{\phantom{\dagger}\!}_{mk}\hspace{0.02cm}\{\hspace{0.01cm}
\xi^{\phantom{\dagger}\!}_{l}\hspace{0.02cm},\Omega\}.
\label{eq:5t}
\end{equation}
Similar reasoning for the second expression in (\ref{eq:5e}) leads to
\begin{equation}
\{\hspace{0.01cm}a_{k}, [\hspace{0.02cm}b^{\dagger}_{m},\xi_{l}\hspace{0.02cm}]\}
=
-2\hspace{0.015cm}\delta_{mk}\hspace{0.02cm}\{\hspace{0.01cm}\xi_{l}\hspace{0.02cm},\Omega\}.
\label{eq:5y}
\end{equation}
The following equalities
\[
[\hspace{0.02cm}\xi^{\phantom{\dagger}\!}_{l}, \{\hspace{0.01cm}b^{\phantom{\dagger}\!}_{m},a^{\dagger}_{k}\}\hspace{0.01cm}] = 0,
\quad
[\hspace{0.03cm}\xi^{\phantom{\dagger}\!}_{l}, \{\hspace{0.01cm}a^{\phantom{\dagger}\!}_{k},b^{\dagger}_{m}\}\hspace{0.005cm}] = 0
\]
are an immediate corollary of the relations (\ref{eq:5t}), (\ref{eq:5y}) and identity (\ref{ap:B3}).
It is to be noted that the relations (\ref{eq:5t}) and (\ref{eq:5y}) are merely a direct consequence of the commutation rules (\ref{eq:3i})\,--\,(\ref{eq:3a}) and (\ref{eq:5w}) for the Green components and do not contain any new information. However, here we can make a step forward and postulate the following relation:
\begin{equation}
\{\hspace{0.01cm}\xi_{l}\hspace{0.02cm},\Omega\} = \Lambda\hspace{0.03cm}\xi_{l},
\label{eq:5u}
\end{equation}
where $\Lambda$ is some constant satisfying (by virtue of (\ref{eq:3f})) the condition
\[
\Lambda = - \Lambda^{\!\hspace{0.01cm}\ast},
\]
where the asterisk denotes the complex conjugation. Thus, instead of (\ref{eq:5t}) and (\ref{eq:5y}), now we have
\begin{equation}
\begin{split}
&\{\hspace{0.01cm}b^{\phantom{\dagger}\!}_{m}, [\hspace{0.02cm}a^{\dagger}_{k},\xi^{\phantom{\dagger}\!}_{l}\hspace{0.02cm}]\}
=
2\hspace{0.01cm}\Lambda\hspace{0.03cm}
\delta^{\phantom{\dagger}\!}_{mk}\hspace{0.02cm}\xi_{l}, \quad
\\[1ex]
&\{\hspace{0.01cm}a^{\phantom{\dagger}\!}_{k}, [\hspace{0.02cm}b^{\dagger}_{m},\xi^{\phantom{\dagger}\!}_{l}\hspace{0.02cm}]\}
=
2\hspace{0.01cm}\Lambda^{\!\hspace{0.01cm}\ast}\hspace{0.01cm}
\delta^{\phantom{\dagger}\!}_{mk}\hspace{0.02cm}\xi^{\phantom{\dagger}\!}_{l}.
\label{eq:5i}
\end{split}
\end{equation}
\indent
In the paper \cite{govorkov_1979} Govorkov has introduced an important operator $\widetilde{N}$:
\begin{equation}
\widetilde{N}
=
\frac{i}{2\hspace{0.015cm}(2\hspace{0.005cm}M + 1)}
\sum^{M}_{k\hspace{0.02cm}=1}\,
\bigl(\hspace{0.03cm}[\hspace{0.02cm}a^{\dagger}_{k},
b^{\phantom{\dagger}\!}_{k}\hspace{0.02cm}] \hspace{0.02cm}+ \lambda\hspace{0.02cm}\bigr),
\label{eq:5o}
\end{equation}
where $\lambda$ is real constant different from zero\footnote{\,Note that neither in the paper \cite{govorkov_1979} nor in the review \cite{govorkov_1983} the number $\lambda$ was fixed.}. In terms of the operator $\zeta_0$, Eq.\,(\ref{eq:1i}), with the use of (\ref{eq:3s}) the expression (\ref{eq:5o}) can be also presented in the form
\begin{equation}
\widetilde{N}
= \frac{1}{2}\,\zeta_{0} \hspace{0.03cm}+\hspace{0.03cm}
\frac{i\hspace{0.02cm}M}{2\hspace{0.015cm}(2\hspace{0.005cm}M + 1)}\,\lambda.
\label{eq:5p}
\end{equation}
The operator $\widetilde{N}$ possesses the following properties:
\begin{equation}
[\hspace{0.02cm}i\widetilde{N}, a_{k}\hspace{0.02cm}] = b_{k},
\quad
[\hspace{0.03cm}i\widetilde{N}, b_{k}\hspace{0.02cm}] = -a_{k}.
\label{eq:5a}
\end{equation}
These relations are exactly the same as those given by Eq.\,(\ref{eq:3p}) for Green's components
$a^{(\alpha)}_{k}$ and $b^{(\alpha)}_{k}$. Nevertheless, the operator $\Omega$ cannot be literally identified with the operator $i \widetilde{N}$ since this will lead to a contradiction in the subsequent analysis. One can see a certain connection between $i \widetilde{N}$ and $\Omega$, if we find a relation of the type (\ref{eq:5u}) for the operator $i \widetilde{N}$. For this purpose, let us examine the following anticommutator:
\[
\{\hspace{0.01cm}\xi^{\phantom{\dagger}\!}_{l}\hspace{0.02cm}, \frac{1}{2}\,\zeta^{\phantom{\dagger}\!}_{0}\} =
\frac{i}{2\hspace{0.015cm}(2\hspace{0.005cm}M + 1)}
\sum^{M}_{k\hspace{0.02cm}=1}\,
\{\xi^{\phantom{\dagger}\!}_{l}\hspace{0.02cm},
[\hspace{0.02cm}a^{\dagger}_{k},b^{\phantom{\dagger}\!}_{k}\hspace{0.02cm}]\}
=
\frac{i}{2\hspace{0.015cm}(2\hspace{0.005cm}M + 1)}
\sum^{M}_{k\hspace{0.02cm}=1}\,\Bigl(\{\hspace{0.01cm}b^{\phantom{\dagger}\!}_{k}, [\hspace{0.02cm}\xi^{\phantom{\dagger}\!}_{l}\hspace{0.02cm},a^{\dagger}_{k}\hspace{0.02cm}]\}
+
[\hspace{0.02cm}a^{\dagger}_{k}, \{\hspace{0.01cm}b^{\phantom{\dagger}\!}_{k},
\xi^{\phantom{\dagger}\!}_{l}\}\hspace{0.01cm}]\hspace{0.02cm}\Bigr).
\]
The first term under the sum sign is defined by the first relation in (\ref{eq:5i}), and the second one, as will be shown in the next section, is equal to
\[
[\hspace{0.02cm}a^{\dagger}_{k}, \{\hspace{0.01cm}b^{\phantom{\dagger}\!}_{k},\xi^{\phantom{\dagger}\!}_{l}\}\hspace{0.01cm}]
=
2\hspace{0.01cm}\Lambda^{\!\hspace{0.01cm}\ast}\hspace{0.02cm}\xi^{\phantom{\dagger}\!}_{l}.
\]
Taking into account the aforesaid, we have
\[
\{\hspace{0.01cm}\xi_{l}\hspace{0.02cm}, \frac{1}{2}\,\zeta_{0}\} =
-\frac{i\hspace{0.02cm}M}{2\hspace{0.01cm}M + 1}\,(\Lambda - \Lambda^{\!\hspace{0.01cm}\ast})\hspace{0.03cm}\xi_{l}
\]
and then it follows from the expression (\ref{eq:5p}) that
\begin{equation}
\{\hspace{0.01cm}\xi_{l}\hspace{0.02cm},i\hspace{0.005cm}\widetilde{N}\} = \widetilde{\Lambda}\hspace{0.03cm}\xi_{l},
\label{eq:5s}
\end{equation}
where
\begin{equation}
\tilde{\Lambda} = \frac{M}{2\hspace{0.005cm}M + 1}\,(2\Lambda  - \lambda).
\label{eq:5d}
\end{equation}
\indent
We can derive the explicit expressions for the commutators between the operator $i \widetilde{N}$ and  Green components $a_{k}^{(\alpha)}$ and $b_{k}^{(\alpha)}$. To this end, we substitute (\ref{eq:3d}) for $p = 2$ into the right-hand side of (\ref{eq:5o}). Then, we obtain
\begin{equation}
i\widetilde{N}
=
\frac{2M}{2\hspace{0.01cm}M + 1}\,\Omega
\hspace{0.02cm}-\hspace{0.02cm}
\frac{1}{2\hspace{0.015cm}(2\hspace{0.005cm}M + 1)}
\sum^{M}_{k\hspace{0.02cm}=1}\,
\bigl(\hspace{0.02cm}[\hspace{0.02cm}a^{\dagger(1)}_{k}\!,b^{(2)}_{k}\hspace{0.02cm}]
+
[\hspace{0.02cm}a^{\dagger(2)}_{k}\!,b^{(1)}_{k}\hspace{0.02cm}]\hspace{0.02cm}\bigr)
-
\frac{M}{2\hspace{0.015cm}(2\hspace{0.005cm}M + 1)}\,\lambda.
\label{eq:5f}
\end{equation}
By using the relations (\ref{eq:3p}) and equality (\ref{eq:3h}), we get
\begin{equation}
\begin{split}
&[\hspace{0.02cm}i\widetilde{N}, a^{(1)}_{k}\hspace{0.02cm}]
=
\frac{1}{2\hspace{0.01cm}M + 1}\,\bigl(2\hspace{0.005cm}M\hspace{0.015cm}b^{(1)}_{k}\!
+\hspace{0.02cm} b^{(2)}_{k}\bigr),
\\[1ex]
&[\hspace{0.02cm}i\widetilde{N}, a^{(2)}_{k}\hspace{0.02cm}]
=
\frac{1}{2\hspace{0.01cm}M + 1}\,\bigl(2\hspace{0.005cm}M\hspace{0.015cm}b^{(2)}_{k}\!
+\hspace{0.02cm} b^{(1)}_{k}\bigr).
\label{eq:5g}
\end{split}
\hspace{0.4cm}
\end{equation}
Similar calculations for commutators with the Green components $b_{m}^{(\alpha)}$, by making use of (\ref{eq:3p}) and (\ref{eq:3j}), lead us to
\begin{equation}
\begin{split}
&[\hspace{0.02cm}i\hspace{0.01cm}\widetilde{N}, b^{(1)}_{m}\hspace{0.02cm}]
=
-\frac{1}{2\hspace{0.01cm}M + 1}\,\bigl(2\hspace{0.005cm}M\hspace{0.015cm}a^{(1)}_{m}\!
+\hspace{0.02cm} a^{(2)}_{m}\bigr),
\\[1ex]
&[\hspace{0.02cm}i\hspace{0.01cm}\widetilde{N}, b^{(2)}_{m}\hspace{0.02cm}]
=
-\frac{1}{2\hspace{0.01cm}M + 1}\,\bigl(2\hspace{0.005cm}M\hspace{0.015cm}a^{(2)}_{m}\!
+\hspace{0.02cm} a^{(1)}_{m}\bigr).
\label{eq:5h}
\end{split}
\end{equation}
In spite of somewhat unusual form of the commutation rules (\ref{eq:5g}) and (\ref{eq:5h}), they correctly reproduce the relations (\ref{eq:5a}), as it is easy to verify by summing two relations in (\ref{eq:5g}) and in (\ref{eq:5h}) and taking into account
\[
a^{\phantom{\dagger}\!}_{k} = a^{(1)}_{k}\!+\hspace{0.02cm}a^{(2)}_{k},
\quad
b^{\phantom{\dagger}\!}_{m} = b^{(1)}_{m}\!+\hspace{0.02cm}b^{(2)}_{m}.
\]

\section{Commutation relations with the operator
${\rm e}^{\alpha\hspace{0.04cm}i\widetilde{N}}$}
\setcounter{equation}{0}

Now we derive a set of the commutation relations between the operator
${\rm e}^{\alpha\hspace{0.03cm}i\widetilde{N}}$ and the operators $a_{k}$, $b_{m}$, and para-Grassmann numbers $\xi_{l}$. Here, $\alpha$ is an arbitrary real number. For this purpose, above all we note that for the operator $a_{k}$ the following equality holds
\begin{equation}
{\rm e}^{\alpha\hspace{0.02cm}i\widetilde{N}\!} a_{k}\hspace{0.02cm}
{\rm e}^{-\alpha\hspace{0.02cm}i\widetilde{N}}
=
a_{k} \hspace{0.02cm}+\hspace{0.02cm} \alpha\hspace{0.03cm}[\hspace{0.02cm}i\hspace{0.015cm}\widetilde{N}, a_{k}\hspace{0.02cm}]
\hspace{0.02cm}+\hspace{0.02cm}
\frac{\!1}{2!}\,\alpha^{2}\hspace{0.03cm}[\hspace{0.02cm}i\hspace{0.015cm}\widetilde{N}, [\hspace{0.02cm}i\hspace{0.015cm}\widetilde{N}, a_{k}\hspace{0.02cm}]\hspace{0.02cm}]
\hspace{0.02cm}+\hspace{0.02cm} \dots
\label{eq:6q}
\end{equation}
\[
= a_{k}\hspace{0.01cm}\Bigl(1 - \frac{\!1}{2!}\,\alpha^{2} + \frac{\!1}{4!}\,\alpha^{4} - \ldots\Bigr)
\hspace{0.02cm}+\hspace{0.02cm}
b_{k}\hspace{0.01cm}\Bigl(\alpha - \frac{\!1}{3!}\,\alpha^{3} + \ldots\Bigr)
\equiv
a_{k}\hspace{0.01cm}\cos\alpha + b_{k}\hspace{0.01cm}\sin\alpha.
\]
In deriving this relation we have taken into account the identity (\ref{ap:B7}) and relations (\ref{eq:5a}). A similar expression can be obtained and for the operator $b_{m}$. By using (\ref{eq:6q}), we write out the basic relations determining a rule of rearrangement between
${\rm e}^{\alpha\hspace{0.03cm}i\widetilde{N}}$ and $a_{k}$, $b_{m}$:
\begin{equation}
\begin{split}
&{\rm e}^{\alpha\hspace{0.02cm}i\widetilde{N}} a_{k}
=
(a_{k}\cos\alpha + b_{k}\sin\alpha)\hspace{0.03cm}{\rm e}^{\alpha\hspace{0.02cm}i\widetilde{N}},
\\[1ex]
&{\rm e}^{\alpha\hspace{0.03cm}i\widetilde{N}} b_{m}
=
(b_{m}\cos\alpha - a_{m}\sin\alpha)\hspace{0.03cm}{\rm e}^{\alpha\hspace{0.02cm}i\widetilde{N}}.
\label{eq:6w}
\end{split}
\end{equation}
Here we are mainly interested in two particular cases of the general formulae:
\vspace{-0.3cm}
\begin{flushleft}
1. in the case when $\alpha = \pm\hspace{0.015cm}\pi$, we have
\end{flushleft}
\vspace{-0.7cm}
\begin{equation}
\{\hspace{0.02cm}{\rm e}^{\pm\hspace{0.015cm}\pi\hspace{0.02cm}i\widetilde{N}}\!, a_{k}\} = 0,
\quad
\{\hspace{0.02cm}{\rm e}^{\pm\hspace{0.015cm}\pi\hspace{0.02cm}i\widetilde{N}}\!, b_{m}\} = 0;
\label{eq:6e}
\end{equation}
\vspace{-0.7cm}
\begin{flushleft}
2. in the case when $\alpha = \pm\hspace{0.015cm}\pi/2$, we have
\end{flushleft}
\vspace{-1.1cm}
\begin{align}
&{\rm e}^{\pm\hspace{0.015cm}\pi\hspace{0.01cm}i\widetilde{N}/2}\hspace{0,015cm} a_{k} =
\pm\hspace{0.04cm} b_{k}\hspace{0.03cm}{\rm e}^{\pm\hspace{0.015cm}\pi\hspace{0.01cm}i\widetilde{N}/2}\hspace{0,015cm},
\label{eq:6r} \\[0.5ex]
&{\rm e}^{\pm\hspace{0.015cm}\pi\hspace{0.01cm}i\widetilde{N}/2}\hspace{0,015cm} b_{m} =
\mp\hspace{0.04cm} a_{m}\hspace{0.03cm}{\rm e}^{\pm\hspace{0.015cm}\pi\hspace{0.01cm}i\widetilde{N}/2}\hspace{0,015cm}.
\label{eq:6t}
\end{align}
It is worthy of special emphasis that the relations (\ref{eq:6r}) and (\ref{eq:6t}) tell us about the possibility of two equivalent ``mapping'' the operator $b_k$ into the operator $a_{k}$:
\begin{equation}
\begin{split}
&a_{k} =
\pm\hspace{0.04cm}{\rm e}^{\mp\hspace{0.015cm}\pi\hspace{0.01cm}i\widetilde{N}/2}\hspace{0,015cm}
b_{k}\hspace{0.03cm}{\rm e}^{\pm\hspace{0.015cm}\pi\hspace{0.01cm}i\widetilde{N}/2}\hspace{0,015cm},
\\[1ex]
&a_{k} =
\mp\hspace{0.04cm}{\rm e}^{\pm\hspace{0.015cm}\pi\hspace{0.01cm}i\widetilde{N}/2}\hspace{0,015cm}
b_{k}\hspace{0.03cm}{\rm e}^{\mp\hspace{0.015cm}\pi\hspace{0.01cm}i\widetilde{N}/2}\hspace{0,015cm}.
\label{eq:6y}
\end{split}
\end{equation}
This circumstance is particularly convenient in an analysis of the concrete expressions. Anticommutation relations (\ref{eq:6e}) coincide with analogous relations for the operator
$(-1)^{N} = {\rm e}^{\pm\hspace{0.015cm}\pi\hspace{0.02cm}iN}$:
\begin{equation}
\{\hspace{0.02cm}{\rm e}^{\pm\hspace{0.015cm}\pi\hspace{0.02cm}iN}\!, a_{k}\} = 0,
\quad
\{\hspace{0.02cm}{\rm e}^{\pm\hspace{0.015cm}\pi\hspace{0.02cm}iN}\!, b_{m}\} = 0,
\label{eq:6u}
\end{equation}
where $N$ is the particle-number operator (\ref{eq:1u}). The relations (\ref{eq:6u}) are true by virtue of
\begin{equation}
\begin{split}
&[\hspace{0.02cm}a^{\phantom{\dagger}\!}_{k}, N\hspace{0.02cm}] = a^{\phantom{\dagger}\!}_{k},
\quad
[\hspace{0.02cm}a^{\dagger}_{k}, N\hspace{0.02cm}] = -\hspace{0.015cm}a^{\dagger}_{k},
\\[0.8ex]
&[\hspace{0.02cm}b^{\phantom{\dagger}\!}_{m}, N\hspace{0.02cm}] = b^{\phantom{\dagger}\!}_{m},
\quad
[\hspace{0.02cm}b^{\dagger}_{m}, N\hspace{0.02cm}] = -\hspace{0.015cm}b^{\dagger}_{m}.
\end{split}
\label{eq:6o}
\end{equation}
With regard to the relations (\ref{eq:6r}) and (\ref{eq:6t}), here, we can mention an interesting formal connection with the papers by Schwinger \cite{schwinger_1_1960, schwinger_2_1960} (see also \cite{schwinger_1970}) devoted to the construction and analysis of so-called ``the unitary operator bases''. The relation from \cite{schwinger_1_1960, schwinger_2_1960}
\[
\widehat{X}(\alpha)\hspace{0.02cm}\widehat{U} = \widehat{U}\hspace{0.05cm}\widehat{Y}(\alpha)
\]
will be analogue of (\ref{eq:6r}), (\ref{eq:6t}). Here, $\widehat{X}(\alpha)$ and $\widehat{Y}(\alpha)$ are two orthonormal operator bases in a given space and the operator
$\widehat{U}=(\widehat{U}_{ab})$ is an unitary operator:
\[
\widehat{U}_{ab} = \sum^{N}_{k\hspace{0.02cm}=\hspace{0.02cm}1}
|\hspace{0.04cm}a_{k}\rangle\langle\hspace{0.04cm}b_{k}|,
\]
where $|\hspace{0.04cm}a_{k}\rangle,\, |\hspace{0.04cm}b_{k}\rangle$ and their adjoints are the two ordered sets of vectors. In our case, the operator ${\rm e}^{\pm\hspace{0.015cm}\pi\hspace{0.01cm}i\widetilde{N}/2}\hspace{0,015cm}$ plays the role of the operator $\widehat{U}$. One can point out a number of the other close coincidences between two formalisms, but we will not go into a detailed analysis of this connection.\\
\indent
Let us introduce the para-Grassmann numbers $\xi_k$. Since now we have the anticommutation relation (\ref{eq:5u}), then, instead of (\ref{eq:6q}), we need to consider the following expression:
\begin{align}
{\rm e}^{\alpha\hspace{0.04cm}\Omega}\xi_{k}\hspace{0.03cm}
{\rm e}^{\alpha\hspace{0.03cm}\Omega}
&=
\xi_{k} \hspace{0.02cm}+\hspace{0.02cm} \alpha\hspace{0.03cm}
\{\hspace{0.02cm}\Omega, \xi_{k}\}
+
\frac{\!1}{2!}\hspace{0.03cm}\alpha^{2}\hspace{0.03cm}\{\hspace{0.02cm}\Omega, \{\hspace{0.02cm}\Omega, \xi_{k}\}\!\hspace{0.02cm}\} + \dots\, \notag \\[0.8ex]
&=
\xi_{k}\Bigl(1 + \Lambda\hspace{0.02cm}\alpha + \frac{\!1}{2!}\hspace{0.03cm}(\Lambda\alpha)^{2}
+ \dots\Bigr)
\equiv \xi_{k}\hspace{0.03cm}{\rm e}^{\alpha\hspace{0.03cm}\Lambda}.
\notag
\end{align}
Here, we have used the identity (\ref{ap:B8}). Thus we have the following commutation rule of the operator ${\rm e}^{\alpha\hspace{0.04cm}\Omega}$ with the para-Grassmann numbers $\xi_k$:
\[
{\rm e}^{\alpha\hspace{0.04cm}\Omega}\xi_{k} =
\xi_{k}\hspace{0.03cm}{\rm e}^{\alpha\hspace{0.02cm}\Lambda}
{\rm e}^{-\alpha\hspace{0.03cm}\Omega}.
\]
Similarly, for the operator ${\rm e}^{\alpha\hspace{0.03cm}i\widetilde{N}}$, by using (\ref{eq:5s}), we get
\begin{equation}
{\rm e}^{\alpha\hspace{0.03cm}i\widetilde{N}}\xi_{k} =
\xi_{k}\hspace{0.03cm}{\rm e}^{\alpha\hspace{0.02cm}\widetilde{\Lambda}}
{\rm e}^{-\alpha\hspace{0.02cm}i\widetilde{N}}.
\label{eq:6p}
\end{equation}
Having in hand the formulae (\ref{eq:6r}), (\ref{eq:6t}) and (\ref{eq:6p}), the question may be asked in which forms the various commutation trilinear relations turn under the mapping (\ref{eq:6y}). The answer is somewhat unexpected: not in all cases this mapping reduces only to a trivial replacement $a_{k}\rightleftharpoons b_{k}$.\\
\indent
Let us consider the mapping of the most simple trilinear relation from (\ref{eq:5q}) including the operator $a_{m}$ once:
\begin{equation}
[\hspace{0.03cm}\xi_{k}, [\hspace{0.02cm}\xi_{l},a_{m}\hspace{0.02cm}]\hspace{0.02cm}] = 0.
\label{eq:6a}
\end{equation}
First we consider the commutator $[\hspace{0.02cm}\xi_{l},a_{m}\hspace{0.02cm}]$. Making use of (\ref{eq:6y}) and (\ref{eq:6p}), we result in the following expression:
\begin{equation}
[\hspace{0.02cm}\xi_{l},a_{m}\hspace{0.02cm}] =
\mp\hspace{0.04cm}{\rm e}^{\pm\hspace{0.015cm}\pi\widetilde{\Lambda}/2}\hspace{0.015cm}
{\rm e}^{\mp\hspace{0.015cm}\pi\hspace{0.01cm}i\widetilde{N}/2}\hspace{0,015cm}
\{\hspace{0.02cm}\xi_{l},b_{m}\}\hspace{0.03cm}
{\rm e}^{\mp\hspace{0.015cm}\pi\hspace{0.01cm}i\widetilde{N}/2}\hspace{0,015cm}.
\label{eq:6s}
\end{equation}
We call attention to the fact that on the right-hand side {\it anticommutator} arises. Substitution of the preceding expression in (\ref{eq:6a}) leads to
\begin{equation}
\xi_{k}\{\hspace{0.02cm}\xi_{l},b_{m}\} -
{\rm e}^{\mp\hspace{0.015cm}\pi\hspace{0.02cm}i\widetilde{N}}
\{\hspace{0.02cm}\xi_{l},b_{m}\}\hspace{0.03cm}
\xi_{k}\hspace{0.03cm}{\rm e}^{\pm\hspace{0.015cm}\pi\hspace{0.02cm}i\widetilde{N}} =0.
\label{eq:6d}
\end{equation}
Further, by using the relations (\ref{eq:6e}) and (\ref{eq:6p}), we have for the last term in (\ref{eq:6d})
\[
{\rm e}^{\mp\hspace{0.015cm}\pi\hspace{0.02cm}i\widetilde{N}}
\{\hspace{0.02cm}\xi_{l},b_{m}\}\hspace{0.03cm}
\xi_{k}\hspace{0.03cm}{\rm e}^{\pm\hspace{0.015cm}\pi\hspace{0.02cm}i\widetilde{N}}
=
-\hspace{0.03cm}
{\rm e}^{\mp\hspace{0.015cm}\pi\hspace{0.01cm}\widetilde{\Lambda}}
\{\hspace{0.02cm}\xi_{l},b_{m}\}\hspace{0.02cm}
\Bigl({\rm e}^{\pm\hspace{0.015cm}\pi\hspace{0.02cm}i\widetilde{N}}\xi_{k}\hspace{0.03cm}
{\rm e}^{\pm\hspace{0.015cm}\pi\hspace{0.02cm}i\widetilde{N}}\Bigr)
=
-\hspace{0.03cm}\{\hspace{0.02cm}\xi_{l},b_{m}\},
\]
and thus we derive, instead of (\ref{eq:6d}),
\begin{equation}
\{\hspace{0.02cm}\xi_{k}, \{\hspace{0.02cm}\xi_{l}, b_{m}\}\} = 0.
\label{eq:6f}
\end{equation}
Contrary to the expectation, under the mapping (\ref{eq:6y}) the relation (\ref{eq:6a}) does not turn into a similar relation with the only replacement $a_{m} \rightarrow b_{m}$. We see that in addition to the replacement, all commutators are replaced by anticommutators. This circumstance can take place exceptionally for parastatistics of order two.\\
\indent
We are able to verify a validity of the trilinear relation (\ref{eq:6f}) with the help of the Green ansatz. Really, by virtue of the commutation rules (\ref{eq:5w}), the equality
\[
\{\hspace{0.02cm}\xi^{\phantom{\dagger}\!}_{l}, b^{\phantom{\dagger}\!}_{m}\} = \sum_{\alpha\hspace{0.02cm}\neq\hspace{0.02cm}\beta}\,
\{\hspace{0.02cm}\xi^{(\alpha)}_{l}, b^{(\beta)}_{m}\}
\]
takes place. Then, by making use of the decomposition of the triple sum (\ref{eq:4e}), we have
\begin{equation}
\{\hspace{0.02cm}\xi^{\phantom{\dagger}\!}_{k}, \{\hspace{0.02cm}\xi^{\phantom{\dagger}\!}_{l}, b^{\phantom{\dagger}\!}_{m}\}\!\hspace{0.02cm}\}
=\!\!\!
\sum_{\alpha = \gamma\neq\beta}\!
\{\hspace{0.02cm}\xi^{(\alpha)}_{k}, \{\hspace{0.02cm}\xi^{(\alpha)}_{l}, b^{(\beta)}_{m}\}\hspace{0.02cm}\!\}
\hspace{0.02cm}+\!\!
\sum_{\alpha\neq\beta = \gamma}\!
\{\hspace{0.02cm}\xi^{(\beta)}_{k}, \{\hspace{0.02cm}\xi^{(\alpha)}_{l}, b^{(\beta)}_{m}\}\hspace{0.02cm}\!\}
\hspace{0.02cm}+\!\!
\sum_{\alpha\neq\beta\neq\gamma}\!
\{\hspace{0.02cm}\xi^{(\gamma)}_{k}, \{\hspace{0.02cm}\xi^{(\alpha)}_{l}, b^{(\beta)}_{m}\}\hspace{0.02cm}\!\}.
\label{eq:6g}
\end{equation}
By using the identity (\ref{ap:B2}) and relations (\ref{eq:5w}), we find for the first two terms on the right-hand side of (\ref{eq:6g})
\[
\sum_{\alpha = \gamma\neq\beta}\!
\{\hspace{0.02cm}\xi^{(\alpha)}_{k}, \{\hspace{0.02cm}\xi^{(\alpha)}_{l}, b^{(\beta)}_{m}\}\hspace{0.02cm}\!\}
\hspace{0.02cm}=\!
\sum_{\alpha = \gamma\neq\beta}\!\Bigl(
\{\hspace{0.02cm}b^{(\beta)}_{m},\{\hspace{0.02cm}\xi^{(\alpha)}_{k}, \xi^{(\alpha)}_{l}\}\hspace{0.02cm}\!\}
+
[\hspace{0.03cm}\xi^{(\alpha)}_{l}, [\hspace{0.02cm}b^{(\beta)}_{m}, \xi^{(\alpha)}_{k}\hspace{0.02cm}]\hspace{0.02cm}]\Bigr) = 0,
\]
\[
\sum_{\alpha\neq\beta = \gamma}\!
\{\hspace{0.02cm}\xi^{(\beta)}_{k}, \{\hspace{0.02cm}b^{(\beta)}_{m},\xi^{(\alpha)}_{l}\}\!\hspace{0.02cm}\}
\hspace{0.02cm}=\!
\sum_{\alpha\neq\beta = \gamma}\!
\Bigl(\hspace{0.03cm}[\hspace{0.03cm}b^{(\beta)}_{m}, [\hspace{0.02cm}\xi^{(\alpha)}_{l},\xi^{(\beta)}_{k}\hspace{0.02cm}]\hspace{0.02cm}]
+
\{\hspace{0.02cm}\xi^{(\alpha)}_{l}, \{\hspace{0.02cm}b^{(\beta)}_{m},\xi^{(\beta)}_{k}\}\!\hspace{0.02cm}\}
\!\Bigr) = 0.
\]
The third term in (\ref{eq:6g}) is absent for $p = 2$.\\
\indent
Let us consider a mapping of more nontrivial trilinear relations (\ref{eq:5i}). To be specific, we shall deal with the second of them:
\begin{equation}
\{\hspace{0.01cm}a^{\phantom{\dagger}\!}_{k}, [\hspace{0.02cm}\xi^{\phantom{\dagger}\!}_{l},b^{\dagger}_{m}\hspace{0.02cm}]\}
=
2\hspace{0.01cm}\Lambda\hspace{0.015cm}\delta^{\phantom{\dagger}\!}_{mk}\hspace{0.02cm}
\xi^{\phantom{\dagger}\!}_{l}.
\label{eq:6h}
\end{equation}
By using the second formula in (\ref{eq:6y}), we have the starting expression for analysis
\[
\{\hspace{0.01cm}a^{\phantom{\dagger}\!}_{k}, [\hspace{0.02cm}\xi^{\phantom{\dagger}\!}_{l},b^{\dagger}_{m}\hspace{0.02cm}]\}
=
\mp\Bigl({\rm e}^{\pm\hspace{0.015cm}\pi\hspace{0.01cm}i\widetilde{N}/2}\hspace{0,015cm}
b^{\phantom{\dagger}\!}_{k}\hspace{0.03cm}{\rm e}^{\mp\hspace{0.015cm}\pi\hspace{0.01cm}i\widetilde{N}/2}\hspace{0,015cm} [\hspace{0.02cm}\xi^{\phantom{\dagger}\!}_{l},b^{\dagger}_{m}\hspace{0.02cm}]
\hspace{0.02cm}+\hspace{0.02cm}
[\hspace{0.02cm}\xi^{\phantom{\dagger}\!}_{l},b^{\dagger}_{m}\hspace{0.02cm}]\hspace{0.03cm}
{\rm e}^{\pm\hspace{0.015cm}\pi\hspace{0.01cm}i\widetilde{N}/2}\hspace{0,015cm}
b^{\phantom{\dagger}\!}_{k}\hspace{0.03cm}{\rm e}^{\mp\hspace{0.015cm}\pi\hspace{0.01cm}i\widetilde{N}/2}\hspace{0,015cm}\Bigr).
\]
Further, for the commutator on the right-hand side we make use of the expression, which is similar to (\ref{eq:6s})
\[
[\hspace{0.03cm}\xi^{\phantom{\dagger}\!}_{l},b^{\dagger}_{m}\hspace{0.02cm}]
=
\pm\hspace{0.04cm}{\rm e}^{\pm\hspace{0.015cm}\pi\widetilde{\Lambda}/2}\hspace{0.015cm}
{\rm e}^{\mp\hspace{0.015cm}\pi\hspace{0.01cm}i\widetilde{N}/2}\hspace{0,015cm}
\{\hspace{0.02cm}\xi^{\phantom{\dagger}\!}_{l},a^{\dagger}_{m}\}
\hspace{0.03cm}{\rm e}^{\mp\hspace{0.015cm}\pi\hspace{0.01cm}i\widetilde{N}/2}\hspace{0,015cm}.
\]
Let us multiply both sides of the relation (\ref{eq:6h}) by the operator
${\rm e}^{\pm\hspace{0.015cm}\pi\hspace{0.01cm}i\widetilde{N}/2}$. Then, taking into account the preceding, we get
\[
{\rm e}^{\mp\hspace{0.015cm}\pi\widetilde{\Lambda}/2}\hspace{0.015cm}
\Bigl({\rm e}^{\pm\hspace{0.015cm}\pi\hspace{0.02cm}i\widetilde{N}}
b^{\phantom{\dagger}\!}_{k}\hspace{0.02cm}
\{\hspace{0.02cm}\xi^{\phantom{\dagger}\!}_{l},a^{\dagger}_{m}\}
\hspace{0.03cm}{\rm e}^{\pm\hspace{0.015cm}\pi\hspace{0.02cm}i\widetilde{N}}\Bigr)
-
{\rm e}^{\pm\hspace{0.015cm}\pi\widetilde{\Lambda}/2}\hspace{0.015cm}
\{\hspace{0.02cm}\xi^{\phantom{\dagger}\!}_{l},a^{\dagger}_{m}\}\hspace{0.02cm}
b^{\phantom{\dagger}\!}_{k}
=
2\Lambda\hspace{0.01cm}\delta^{\phantom{\dagger}\!}_{mk}
\Bigl({\rm e}^{\pm\hspace{0.015cm}\pi\hspace{0.01cm}i\widetilde{N}/2}\hspace{0,015cm}
\xi^{\phantom{\dagger}\!}_{k}\hspace{0.03cm}
{\rm e}^{\pm\hspace{0.015cm}\pi\hspace{0.01cm}i\widetilde{N}/2}\hspace{0,015cm}\Bigr).
\]
The expression in parentheses on the left-hand side equals
${\rm e}^{\pm\hspace{0.015cm}\pi\hspace{0.01cm}\widetilde{\Lambda}}
b^{\phantom{\dagger}\!}_{k}\hspace{0.02cm}
\{\hspace{0.02cm}\xi^{\phantom{\dagger}\!}_{l}, a^{\dagger}_{m}\}$ and on the right-hand side does ${\rm e}^{\pm\hspace{0.015cm}\pi\widetilde{\Lambda}/2}\hspace{0.015cm}\xi_{k}$. The general factor ${\rm e}^{\pm\hspace{0.015cm}\pi\widetilde{\Lambda}/2}$ can be cancelled from the left- and right-hand sides and, finally, we arrive at
\begin{equation}
[\hspace{0.03cm}b^{\phantom{\dagger}\!}_{k}, \{\hspace{0.02cm}\xi^{\phantom{\dagger}\!}_{l},a^{\dagger}_{m}\hspace{0.02cm}\}]
=
2\hspace{0.01cm}\Lambda\hspace{0.02cm}\delta^{\phantom{\dagger}\!}_{mk}\hspace{0.02cm}
\xi^{\phantom{\dagger}\!}_{l}.
\label{eq:6j}
\end{equation}
Here, we see once again that under the mapping (\ref{eq:6y}) in the trilinear relation (\ref{eq:6h}) not only the replacement of operators $a\rightleftharpoons b$ occurs, but commutator is replaced by anticommutator and vice versa. Similarly to the previous case (\ref{eq:6f}), we can verify (\ref{eq:6j}) with the help of the Green representation for the operators and para-Grassmann numbers.\\
\indent
Let us consider a mapping of the trilinear relation from (\ref{eq:5q}) containing the operators $a^{\phantom{\dagger}\!}_{k}$ and $a_{l}^{\dagger}$ simultaneously:
\[
[\hspace{0.03cm}a^{\phantom{\dagger}\!}_{k}, [\hspace{0.02cm}a^{\dagger}_{l},\xi^{\phantom{\dagger}\!}_{m}\hspace{0.02cm}]\hspace{0.02cm}]
=
2\hspace{0.01cm}\delta^{\phantom{\dagger}\!}_{kl}\hspace{0.02cm}\xi^{\phantom{\dagger}\!}_{m}.
\]
By arguments that are completely similar to the previous cases leads to the following relation:
\begin{equation}
\{\hspace{0.02cm}b^{\phantom{\dagger}\!}_{k}, \{\hspace{0.02cm}b^{\dagger}_{l},\xi^{\phantom{\dagger}\!}_{m}\}\!\hspace{0.02cm}\}
=
2\hspace{0.015cm}\delta^{\phantom{\dagger}\!}_{kl}\hspace{0.02cm}\xi^{\phantom{\dagger}\!}_{m},
\label{eq:6k}
\end{equation}
which may be verified by means of the Green ansatz.\\
\indent
The peculiarity of all examples considered above is that the para-Grassmann number $\xi_{k}$
{\it always} enters into the commutator or anticommutator along with the operator $a_{k}$ or $b_{m}$ (or with their Hermitian conjugation). Let us discuss a mapping of the relations, where this circumstance doesn't take place, for example, the mapping of the relations of the form
\begin{equation}
\begin{split}
&\{\hspace{0.01cm}\xi^{\phantom{\dagger}\!}_{l}, [\hspace{0.02cm}b^{\dagger}_{m},a^{\phantom{\dagger}\!}_{k}\hspace{0.02cm}]\}
=
2\hspace{0.01cm}\bigl(\Lambda - \Lambda^{\ast}\bigr)\hspace{0.0cm}
\delta^{\phantom{\dagger}\!}_{mk}\hspace{0.02cm}\xi^{\phantom{\dagger}\!}_{l}, \\[0.7ex]
&[\hspace{0.03cm}\xi^{\phantom{\dagger}\!}_{l}, \{\hspace{0.02cm}a^{\dagger}_{k},b^{\phantom{\dagger}\!}_{m}\hspace{0.02cm}\}]
= 0.
\end{split}
\label{eq:6kk}
\end{equation}
It is evident that under the mapping (\ref{eq:6y}) these two relations can never go over into each other, since their right-hand sides are different. By repeating the above arguments, we obtain
\begin{align}
&\{\hspace{0.01cm}\xi^{\phantom{\dagger}\!}_{l}, [\hspace{0.02cm}a^{\dagger}_{m},b^{\phantom{\dagger}\!}_{k}\hspace{0.02cm}]\}
=
2\hspace{0.01cm}\bigl(\Lambda^{\ast} - \Lambda\bigr)\hspace{0.01cm}
\delta^{\phantom{\dagger}\!}_{mk}\hspace{0.02cm}\xi^{\phantom{\dagger}\!}_{l}, \notag \\[0.7ex]
&[\hspace{0.03cm}\xi^{\phantom{\dagger}\!}_{l}, \{\hspace{0.02cm}b^{\dagger}_{k},a^{\phantom{\dagger}\!}_{m}\hspace{0.02cm}\}]
= 0, \notag
\end{align}
i.e the structure of the trilinear relations remains unchanged and in the given case they merely represent Hermitian conjugation of (\ref{eq:6kk}). The same is true for the trilinear relations, which don't contain the variable $\xi_{k}$ at all, for example,
\[
[\hspace{0.02cm}[\hspace{0.02cm}a^{\dagger}_{k},
a^{\phantom{\dagger}\!}_{l}\hspace{0.02cm}],b^{\phantom{\dagger}\!}_{m}] = -2\hspace{0.015cm}\delta^{\phantom{\dagger}\!}_{km}\hspace{0.02cm}b^{\phantom{\dagger}\!}_{l}.
\]
Under the mapping (\ref{eq:6y}) this relation passes into
\[
[\hspace{0.02cm}[\hspace{0.02cm}b^{\dagger}_{k},b^{\phantom{\dagger}\!}_{l}\hspace{0.02cm}],
a^{\phantom{\dagger}\!}_{m}] = -2\hspace{0.015cm}\delta^{\phantom{\dagger}\!}_{km}\hspace{0.02cm}a^{\phantom{\dagger}\!}_{l}.
\]
Here, the structure of the relation is completely conserved. Thus, all trilinear commutation relations break up into two sets, one of which changes its structure under the mapping (\ref{eq:6y}), and the other conserves it. Everything depends on how the para-Grassmann variable $\xi_{k}$ enters into the specific trilinear relation. It is clear that all reasoning above is true for the mapping that is the inverse of (\ref{eq:6y}), i.e.
\begin{equation}
\left\{
\begin{array}{l}
b_{m} =
\pm\hspace{0.04cm}{\rm e}^{\pm\hspace{0.015cm}\pi\hspace{0.01cm}i\widetilde{N}/2}\hspace{0,015cm}
a_{m}\hspace{0.03cm}{\rm e}^{\mp\hspace{0.015cm}\pi\hspace{0.01cm}i\widetilde{N}/2}\hspace{0,015cm},  \\[1ex]
b_{m} =
\mp\hspace{0.04cm}{\rm e}^{\mp\hspace{0.015cm}\pi\hspace{0.01cm}i\widetilde{N}/2}\hspace{0,015cm}
a_{m}\hspace{0.03cm}{\rm e}^{\pm\hspace{0.015cm}\pi\hspace{0.01cm}i\widetilde{N}/2}\hspace{0,015cm}.
\end{array}
\right.
\label{eq:6l}
\end{equation}
It is necessary only in the initial formulae (\ref{eq:6a}), (\ref{eq:6h}), $\ldots$ to replace the operators $a_{k}$ by $b_{k}$ (and vice versa) and to do the same in final formulae (\ref{eq:6f}), (\ref{eq:6j}), $\ldots$\;.

\section{Action of the operators $\Omega$ and $i\widetilde{N}$ on the vacuum state}
\setcounter{equation}{0}

Let us consider the problem of acting the operators $\Omega$ and $\widetilde{N}$ on the vacuum state $|\hspace{0.04cm}0\rangle$. For the operator $\widetilde{N}$, Eq.\,(\ref{eq:5p}), we have
\begin{equation}
\widetilde{N}|\hspace{0.03cm}0\rangle = \frac{1}{2}\,\zeta_{0}|\hspace{0.04cm}0\rangle
+
\frac{i\hspace{0.01cm}M}{2\hspace{0.02cm}(2\hspace{0.01cm}M + 1)}\,\lambda\hspace{0.02cm} |\hspace{0.03cm}0\rangle.
\label{eq:7q}
\end{equation}
Further, by virtue of the definition of the operator $\zeta_{0}$, Eq.\,(\ref{eq:1i}), with regard for (\ref{eq:2d}), we derive
\begin{equation}
\zeta_{0}|\hspace{0.04cm}0\rangle
=
-\frac{i}{2\hspace{0.01cm}M + 1}\,\sum^{M}_{k\hspace{0.02cm}=1}
b^{\phantom{\dagger}\!}_{k}\hspace{0.015cm}a^{\dagger}_{k}|\hspace{0.04cm}0\rangle.
\label{eq:7w}
\end{equation}
If one uses an additional condition of Greenberg and Messiah, Eq.\,(\ref{eq:2g}), then we get
\[
\zeta_{0}|\hspace{0.04cm}0\rangle = 0,
\]
and so it follows from (\ref{eq:7q}) that
\[
\widetilde{N}|\hspace{0.03cm}0\rangle =
\lambda\,\frac{i\hspace{0.01cm}M}{2\hspace{0.02cm}(2\hspace{0.01cm}M + 1)}\, |\hspace{0.03cm}0\rangle.
\]
Hence, if we wanted to demand the fulfillment of the condition
\begin{equation}
\widetilde{N}|\hspace{0.03cm}0\rangle = 0
\label{eq:7e}
\end{equation}
by analogy with a similar condition for the particle-number operator
\[
N|\hspace{0.03cm}0\rangle = 0,
\]
we would lead to the trivial requirement: $\lambda = 0$. However, the latter actually results in the degeneration of the theory under consideration. The only way to avoid this is to give up the condition (\ref{eq:2g}).\\
\indent
To understand how (\ref{eq:2g}) should be changed, let us consider in detail deriving the condition (\ref{eq:2g}), as presented in the paper \cite{greenberg_1965}. But now we shall proceed from Govorkov's trilinear relations. The first step is to act by the relation (\ref{eq:4q}) on the vacuum
state
\[
a^{\phantom{\dagger}\!}_{l}\hspace{0.02cm}(b^{\phantom{\dagger}\!}_{m}
a^{\dagger}_{k})|\hspace{0.03cm}0\rangle = 0,\quad \mbox{for all } k,\,l,\,m.
\]
The uniqueness of $|\hspace{0.03cm}0\rangle$ implies
\begin{equation}
b^{\phantom{\dagger}\!}_{m}a^{\dagger}_{k}|\hspace{0.03cm}0\rangle = c^{\phantom{\dagger}\!}_{mk}|\hspace{0.03cm}0\rangle,
\label{eq:7r}
\end{equation}
where $c_{mk}$ is a number. Note that at this point of consideration, the additional term $4\hspace{0.02cm}\delta_{km}\hspace{0.02cm}b_{l}$ on the right-hand side of (\ref{eq:4q}) plays
no role. Further, we consider commutator in the form
\[
[\hspace{0.02cm}[\hspace{0.02cm}b^{\phantom{\dagger}\!}_{l}, b^{\dagger}_{m}\hspace{0.02cm}],b^{\phantom{\dagger}\!}_{m}a^{\dagger}_{k}]
=
[\hspace{0.02cm}[\hspace{0.02cm}b^{\phantom{\dagger}\!}_{l}, b^{\dagger}_{m}\hspace{0.02cm}],b^{\phantom{\dagger}\!}_{m}]\hspace{0.02cm}a^{\dagger}_{k}
\hspace{0.02cm}+\hspace{0.02cm}
b^{\phantom{\dagger}\!}_{m}[\hspace{0.02cm}[\hspace{0.02cm}b^{\phantom{\dagger}\!}_{l}, b^{\dagger}_{m}\hspace{0.02cm}],a^{\dagger}_{k}]
\quad (\mbox{no summation!}).
\]
Using the trilinear relations (\ref{eq:2q}) and (\ref{eq:3e}) with $a$ and $b$ interchanged, we find
\[
[\hspace{0.02cm}[\hspace{0.02cm}b^{\phantom{\dagger}\!}_{l}, b^{\dagger}_{m}\hspace{0.02cm}],b^{\phantom{\dagger}\!}_{m}a^{\dagger}_{k}]
=
2\hspace{0.015cm}\delta^{\phantom{\dagger}\!}_{mm}\hspace{0.02cm}
b^{\phantom{\dagger}\!}_{l}\hspace{0.02cm}a^{\dagger}_{k}
-
2\hspace{0.015cm}\delta^{\phantom{\dagger}\!}_{kl}\hspace{0.02cm}
b^{\phantom{\dagger}\!}_{m}\hspace{0.02cm}a^{\dagger}_{m}.
\]
In comparison with the case of Greenberg and Messiah, on the right-hand side a new term arises. Acting on the vacuum by the above expression and using Eqs\,(\ref{eq:2f}) and (\ref{eq:7r}), we obtain
\[
0 = 2\hspace{0.015cm}b^{\phantom{\dagger}\!}_{l}\hspace{0.02cm}
a^{\dagger}_{k}|\hspace{0.03cm}0\rangle
-
2\hspace{0.015cm}\delta^{\phantom{\dagger}\!}_{kl}\hspace{0.02cm}
b^{\phantom{\dagger}\!}_{m}\hspace{0.02cm}
a^{\dagger}_{m}|\hspace{0.03cm}0\rangle
=
2\hspace{0.015cm}c^{\phantom{\dagger}\!}_{lk}|\hspace{0.03cm}0\rangle
-
2\hspace{0.015cm}\delta^{\phantom{\dagger}\!}_{lk}\hspace{0.02cm}
c^{\phantom{\dagger}\!}_{mm}|\hspace{0.03cm}0\rangle
\]
or
\[
c_{lk} = \delta_{lk}\hspace{0.03cm}c_{mm}.
\]
We set
\[
c_{mm}\equiv c \quad \mbox{for all } m,
\]
where $c$ is some, generally speaking, complex constant. Thus, within the framework of the unitary quantization we lead to the following additional conditions, instead of (\ref{eq:2g}):
\begin{equation}
\begin{split}
&b^{\phantom{\dagger}\!}_{m} a^{\dagger}_{k}|\hspace{0.03cm}0\rangle = c\hspace{0.04cm}\delta^{\phantom{\dagger}\!}_{mk}|\hspace{0.03cm}0\rangle,
\\[0.8ex]
&a^{\phantom{\dagger}\!}_{k} b^{\dagger}_{m}|\hspace{0.03cm}0\rangle = c^{\ast}\hspace{0.02cm}\delta^{\phantom{\dagger}\!}_{mk}|\hspace{0.03cm}0\rangle.
\label{eq:7t}
\end{split}
\end{equation}
\indent
In this case, it follows from (\ref{eq:7w}) that
\begin{equation}
\zeta_{0}|\hspace{0.03cm}0\rangle = -\hspace{0.02cm}c\,\frac{i\hspace{0.01cm}M}{2\hspace{0.01cm}M + 1}\,
|\hspace{0.03cm}0\rangle
\label{eq:7tt}
\end{equation}
and therefore
\begin{equation}
\widetilde{N}|\hspace{0.03cm}0\rangle =
-\hspace{0.02cm}(c - \lambda)\,
\frac{i\hspace{0.01cm}M}{2\hspace{0.02cm}(2\hspace{0.01cm}M + 1)}\,|\hspace{0.03cm}0\rangle.
\label{eq:7y}
\end{equation}
Further, if we operate on the vacuum by (\ref{eq:5s}), then in view of the rule \cite{ohnuki_1980}
\[
\xi_{l}|\hspace{0.04cm}0\rangle = |\hspace{0.04cm}0\rangle\hspace{0.04cm}\xi_{l}
\]
and Eq.\,(\ref{eq:7y}), we get the equation relating the constants $\Lambda$ and $c$
\[
(c - \lambda)\,\frac{M}{2\hspace{0.01cm}M + 1} =
(2\hspace{0.02cm}\Lambda - \lambda)\,\frac{M}{2\hspace{0.01cm}M + 1}
\]
or
\begin{equation}
\Lambda = \frac{1}{2}\,c.
\label{eq:7u}
\end{equation}
As an immediate corollary of (\ref{eq:7u}) and (\ref{eq:5u}), we have a rule of acting the operator $\Omega$ on the vacuum:
\[
\Omega\hspace{0.02cm}|\hspace{0.04cm}0\rangle =
\frac{1}{4}\,c\hspace{0.03cm}|\hspace{0.04cm}0\rangle.
\]
\indent
If we wanted to require the fulfillment of the condition (\ref{eq:7e}), then Eqs.\,(\ref{eq:7y}) and (\ref{eq:7u}) would result in an unique fixing of the constants $\Lambda$ and $c$ in terms of the parameter $\lambda$:
\begin{equation}
c = \lambda, \qquad \Lambda = \frac{1}{2}\,\lambda.
\label{eq:7i}
\end{equation}
It is the only parameter which remains undefined in the theory in question. Note that in the case of (\ref{eq:7i}) the constant $\tilde{\Lambda}$ vanishes and, instead of (\ref{eq:5s}), we have now
\[
\{\hspace{0.01cm}\xi_{l}\hspace{0.02cm},\widetilde{N}\} = 0.
\]

\section{Coherent states}
\setcounter{equation}{0}

Omote and Kamefuchi \cite{omote_1979} have made a detailed study of the coherent states of para-Fermi operators on the basis of the para-Grassmann algebra. These authors defined the coherent state of a system of para-Fermi oscillators $a_{k}$ in the following form:
\begin{equation}
|\hspace{0.02cm}(\xi)_{p}\,;a\hspace{0.02cm}\rangle = \exp\Bigl(-\frac{1}{2}\sum^{M}_{l\hspace{0.02cm}=\hspace{0.02cm}1}\,
[\hspace{0.02cm}\xi^{\phantom{\dagger}\!}_{l}, a^{\dagger}_{l}\hspace{0.02cm}]\Bigr)|\hspace{0.03cm}0\rangle,
\label{eq:8q}
\end{equation}
in so doing
\begin{equation}
a_{k}|\hspace{0.04cm}(\xi)_{p}\,;a\hspace{0.02cm}\rangle = \xi_{k}|\hspace{0.04cm}(\xi)_{p}\,;a\hspace{0.02cm}\rangle.
\label{eq:8w}
\end{equation}
In notation of the coherent state $|\hspace{0.04cm}(\xi)_{\!\hspace{0.02cm}p}\hspace{0.04cm}\rangle$ accepted in \cite{omote_1979}, we have inserted an additional symbol $a$ to emphasize that this state is associated with the field $\phi_{a}$. The formula (\ref{eq:8w}) is a consequence of the operator equality
\begin{equation}
a_{k}\hspace{0.04cm}{\rm e}^{\textstyle-\frac{1}{2}\sum_{l}\hspace{0.03cm}
[\hspace{0.02cm}\xi^{\phantom{\dagger}\!}_{l}, a^{\dagger}_{l}\hspace{0.02cm}]}
=
{\rm e}^{\textstyle-\frac{1}{2}\sum_{l}\hspace{0.03cm}
[\hspace{0.02cm}\xi^{\phantom{\dagger}\!}_{l}, a^{\dagger}_{l}\hspace{0.02cm}]}a_{k}
+\hspace{0.02cm}
\xi_{k}\hspace{0.04cm}{\rm e}^{\textstyle-\frac{1}{2}\sum_{l}\hspace{0.03cm}
[\hspace{0.02cm}\xi^{\phantom{\dagger}\!}_{l}, a^{\dagger}_{l}\hspace{0.02cm}]},
\label{eq:8e}
\end{equation}
which can be easily obtained by using the identity (\ref{ap:B7}) and the trilinear relations (\ref{eq:5q}). Here, we only note that a simple form of the second term on the right-hand side of (\ref{eq:8e}) is conditioned by precise cutting off a series
\[
{\rm e}^{-\hspace{0.02cm}\textstyle\frac{1}{2}\hspace{0.03cm}
[\hspace{0.02cm}\xi^{\phantom{\dagger}\!}_{l}, a^{\dagger}_{l}\hspace{0.02cm}]}\,a_{k}\,
{\rm e}^{\hspace{0.02cm}\textstyle\frac{1}{2}\hspace{0.03cm}
[\hspace{0.02cm}\xi^{\phantom{\dagger}\!}_{l^{\prime}}\hspace{0.02cm}, a^{\dagger}_{l^{\prime}}\hspace{0.02cm}]}
=
a_{k} + \biggl(\!-\frac{1}{2}\biggr)\hspace{0.03cm}[\hspace{0.02cm}[\hspace{0.02cm}
\xi^{\phantom{\dagger}\!}_{l}\hspace{0.02cm}, a^{\dagger}_{l}\hspace{0.02cm}], a^{\phantom{\dagger}\!}_{k}\hspace{0.02cm}]
+
\frac{\!1}{2\hspace{0.02cm}!}\hspace{0.03cm}\biggl(\!-\frac{1}{2}\biggr)^{\!\!2\!}
[\hspace{0.02cm}[\hspace{0.02cm}\xi^{\phantom{\dagger}\!}_{l}\hspace{0.02cm}, a^{\dagger}_{l}\hspace{0.02cm}], [\hspace{0.02cm}[\hspace{0.02cm}\xi^{\phantom{\dagger}\!}_{l^{\prime}}, a^{\dagger}_{l^{\prime}}\hspace{0.02cm}], a^{\phantom{\dagger}\!}_{k}\hspace{0.02cm}]\hspace{0.02cm}]
+\, \ldots
= a_{k} - \xi_{k}
\]
after the second term of the expansion.\\
\indent
In a similar way, we can define a coherent state for a system of para-Fermi oscillators $b_{k}$:
\begin{equation}
|\hspace{0.02cm}(\xi)_{p}\,;b\hspace{0.02cm}\rangle = \exp\Bigl(-\frac{1}{2}\sum^{M}_{l=1}\,
[\hspace{0.02cm}\xi^{\phantom{\dagger}\!}_{l}, b^{\dagger}_{l}\hspace{0.02cm}]\Bigr)|\hspace{0.03cm}0\rangle,
\label{eq:8r}
\end{equation}
so that
\begin{equation}
b_{m}|\hspace{0.03cm}(\xi)_{p}\,;b\hspace{0.02cm}\rangle = \xi_{m}|\hspace{0.03cm}(\xi)_{p}\,;b\hspace{0.02cm}\rangle.
\label{eq:8t}
\end{equation}
Of course, in the general case the coherent state (\ref{eq:8r}) for the $b$\hspace{0.02cm}-\hspace{0.02cm}operators will never be the coherent one for the $a$\hspace{0.02cm}-\hspace{0.02cm}operators. However, for parastatistics of order 2, within the framework of uniquantization the situation is somewhat different. Really, let us consider the following operator identity:
\begin{equation}
a_{k}\hspace{0.04cm}{\rm e}^{\textstyle-\frac{1}{2}\hspace{0.03cm}
[\hspace{0.02cm}\xi^{\phantom{\dagger}\!}_{l}, b^{\dagger}_{l}\hspace{0.02cm}]}
\hspace{0.02cm}-\hspace{0.02cm}
{\rm e}^{\hspace{0.02cm}\textstyle\frac{1}{2}\hspace{0.03cm}
[\hspace{0.02cm}\xi^{\phantom{\dagger}\!}_{l}, b^{\dagger}_{l}\hspace{0.02cm}]}a_{k}
\equiv
\Bigl(a_{k} - {\rm e}^{\hspace{0.02cm}\textstyle\frac{1}{2}\hspace{0.03cm}
[\hspace{0.02cm}\xi^{\phantom{\dagger}\!}_{l}, b^{\dagger}_{l}\hspace{0.02cm}]}
\,a_{k}\,
{\rm e}^{\hspace{0.02cm}\textstyle\frac{1}{2}\hspace{0.03cm}
[\hspace{0.02cm}\xi^{\phantom{\dagger}\!}_{l^{\prime}}, b^{\dagger}_{l^{\prime}}\hspace{0.02cm}]}\Bigr)\hspace{0.02cm}
{\rm e}^{\textstyle-\frac{1}{2}\hspace{0.03cm}
[\hspace{0.02cm}\xi^{\phantom{\dagger}\!}_{l^{\prime\prime}}, b^{\dagger}_{l^{\prime\prime}}\hspace{0.02cm}]}.
\label{eq:8y}
\end{equation}
From here on for simplicity we adopt the usual summation convention over repeated indices. By virtue of the identity (\ref{ap:B8}) and the relations (\ref{eq:5i}), we have
\begin{equation}
{\rm e}^{\hspace{0.02cm}\textstyle\frac{1}{2}\hspace{0.03cm}
[\hspace{0.02cm}\xi^{\phantom{\dagger}\!}_{l}, b^{\dagger}_{l}\hspace{0.02cm}]}\,a_{k}\,
{\rm e}^{\hspace{0.02cm}\textstyle\frac{1}{2}\hspace{0.03cm}
[\hspace{0.02cm}\xi^{\phantom{\dagger}\!}_{l^{\prime}}\hspace{0.02cm}, b^{\dagger}_{l^{\prime}}\hspace{0.02cm}]}
=
a_{k} + \frac{1}{2}\hspace{0.03cm}\bigl\{[\hspace{0.02cm}\xi^{\phantom{\dagger}\!}_{l}\hspace{0.02cm}, b^{\dagger}_{l}\hspace{0.02cm}], a^{\phantom{\dagger}\!}_{k}\!\hspace{0.02cm}\bigr\}
+
\frac{\!1}{2\hspace{0.02cm}!}\hspace{0.03cm}\biggl(\frac{1}{2}\biggr)^{\!\!2}\bigl\{
[\hspace{0.02cm}\xi^{\phantom{\dagger}\!}_{l}\hspace{0.02cm}, b^{\dagger}_{l}\hspace{0.02cm}], \bigl\{ [\hspace{0.02cm}\xi^{\phantom{\dagger}\!}_{l^{\prime}}, b^{\dagger}_{l^{\prime}}\hspace{0.02cm}], a^{\phantom{\dagger}\!}_{k}\bigr\}\!
\bigr\}
+\, \ldots
\label{eq:8u}
\end{equation}
\[
= a_{k} + \Lambda\hspace{0.02cm}\xi_{k} +
\frac{\!1}{2\hspace{0.02cm}!}\hspace{0.03cm}\Lambda\hspace{0.03cm}
\xi\hspace{0.02cm}_{k}\hspace{0.03cm}
[\hspace{0.02cm}\xi^{\phantom{\dagger}\!}_{l}\hspace{0.02cm}, b^{\dagger}_{l}\hspace{0.02cm}] \hspace{0.02cm} \hspace{0.02cm}
+\hspace{0.04cm}\ldots
\hspace{1.6cm}
\]
and thus from (\ref{eq:8y}) the following operator relation follows:
\begin{equation}
a_{k}\hspace{0.04cm}{\rm e}^{\textstyle-\frac{1}{2}\hspace{0.03cm}
[\hspace{0.02cm}\xi^{\phantom{\dagger}\!}_{l}\hspace{0.02cm}, b^{\dagger}_{l}\hspace{0.02cm}]}
\hspace{0.02cm}=\hspace{0.02cm}
{\rm e}^{\hspace{0.02cm}\textstyle\frac{1}{2}\hspace{0.03cm}
[\hspace{0.02cm}\xi^{\phantom{\dagger}\!}_{l}\hspace{0.02cm}, b^{\dagger}_{l}\hspace{0.02cm}]}\hspace{0.02cm}a_{k}
-
\Lambda\hspace{0.04cm}\xi_{k}\Bigl(\hspace{0.04cm}\sum_{s}\hspace{0.02cm}\frac{1}{(s + 1)!}\;
[\hspace{0.02cm}\xi^{\phantom{\dagger}\!}_{m}\hspace{0.02cm}, b^{\dagger}_{m}\hspace{0.02cm}]^{\hspace{0.02cm}s}\Bigr)
{\rm e}^{\textstyle-\frac{1}{2}\hspace{0.03cm}
[\hspace{0.02cm}\xi^{\phantom{\dagger}\!}_{l}\hspace{0.02cm}, b^{\dagger}_{l}\hspace{0.02cm}]}.
\label{eq:8i}
\end{equation}
In contrast to (\ref{eq:8e}), in the first term on the right-hand side of (\ref{eq:8i}) the sign in the exponential function has changed. However, it does not play any role, since in acting (\ref{eq:8i}) on the vacuum state this term turns to zero. We see a much more serious change in the second term, it takes an additional factor as compared with (\ref{eq:8e})
\begin{equation}
\Lambda\sum_{s}\hspace{0.02cm}\frac{1}{(s + 1)!}\;
[\hspace{0.02cm}\xi^{\phantom{\dagger}\!}_{m}\hspace{0.02cm}, b^{\dagger}_{m}\hspace{0.02cm}]^{\hspace{0.02cm}s}
=
\Lambda\Bigl(1 + \frac{\!1}{2\hspace{0.02cm}!}\,[\hspace{0.02cm}\xi^{\phantom{\dagger}\!}_{m}\hspace{0.02cm}, b^{\dagger}_{m}\hspace{0.02cm}]
+
\frac{\!1}{3\hspace{0.02cm}!}\,[\hspace{0.02cm}\xi^{\phantom{\dagger}\!}_{m}\hspace{0.02cm}, b^{\dagger}_{m}\hspace{0.02cm}]^{2} \hspace{0.02cm}+\hspace{0.03cm} \ldots\,\Bigr).
\label{eq:8o}
\end{equation}
This circumstance is associated with the fact that a series in (\ref{eq:8u}) is not cut off exactly after the second term $\Lambda\hspace{0.02cm}\xi_{k}$ as this takes place in deriving (\ref{eq:8e}). The only consolation here is a finiteness of the series (\ref{eq:8o}). In particular, for a more important case from the physical point of view, when $M = 2$, this series contains only the terms written out in (\ref{eq:8o}). Acting by the operator relation (\ref{eq:8i}) on the vacuum, we find further
\begin{equation}
a_{k}|\hspace{0.025cm}(\xi)_{2}\,;b\hspace{0.02cm}\rangle =
\Lambda\hspace{0.03cm}\xi_{k}\Bigl(\hspace{0.02cm}\sum_{s}\hspace{0.02cm}\frac{1}{(s + 1)!}\;
[\hspace{0.02cm}\xi^{\phantom{\dagger}\!}_{m}\hspace{0.02cm}, b^{\dagger}_{m}\hspace{0.02cm}]^{\hspace{0.02cm}s}\Bigr)
|\hspace{0.025cm}(\xi)_{2}\,;b\hspace{0.02cm}\rangle.
\label{eq:8p}
\end{equation}
If we introduce the conjugate coherent state for the $\phi_{b}$ field
\[
\langle\hspace{0.02cm}(\bar{\xi}^{\,\prime})_{2};b\hspace{0.02cm}|
\equiv
\langle\hspace{0.03cm}0|\exp\Bigl(\hspace{0.02cm}\frac{1}{2}\sum^{M}_{l=1}\,
[\hspace{0.02cm}\bar{\xi}^{\,\prime}_{l}, b^{\phantom{\prime}}_{l}\hspace{0.02cm}]\Bigr),
\]
then for the special case $M = 2$ a matrix element of the operator $a_{k}$ in the basis of coherent states for the $b$\hspace{0.02cm}-\hspace{0.02cm}operators can be written as follows:
\[
\langle\hspace{0.03cm}(\bar{\xi}^{\,\prime})_{2};b\hspace{0.02cm}|\, a_{k} |\hspace{0.02cm}(\xi)_{2}\,;b\rangle
=
-\Lambda\Bigl(1 + \frac{1}{2\hspace{0.02cm}!}\,[\hspace{0.02cm}\xi^{\phantom{\prime}}_{l}\hspace{0.02cm}, \bar{\xi}^{\,\prime}_{l}\hspace{0.02cm}]
+
\frac{\!1}{3\hspace{0.02cm}!}\,[\hspace{0.02cm}\xi^{\phantom{\prime}}_{l}\hspace{0.02cm}, \bar{\xi}^{\,\prime}_{l}\hspace{0.02cm}]^{\,2}\Bigr)
\langle\hspace{0.02cm}(\bar{\xi}^{\,\prime})_{2};b\hspace{0.02cm}
|\hspace{0.02cm}(\xi)_{2}\,;b\hspace{0.02cm}\rangle,
\]
where the overlap function has the standard form \cite{omote_1979}
\begin{equation}
\langle\hspace{0.025cm}(\bar{\xi}^{\,\prime})_{2};b\hspace{0.02cm}
|\hspace{0.02cm}(\xi)_{2}\,;b\hspace{0.02cm}\rangle
=
{\rm e}^{\hspace{0.025cm}\textstyle\frac{1}{2}\hspace{0.03cm}
[\hspace{0.02cm}\bar{\xi}^{\,\prime}_{l}\hspace{0.02cm},\xi^{\phantom{\prime}}_{l}
\hspace{0.02cm}\hspace{0.02cm}]}.
\label{eq:8a}
\end{equation}
\indent
The complexity of the expression on the right-hand side of (\ref{eq:8p}) is inevitable in the approach under consideration. It is ultimately a consequence of  ``involving'' the coherent state with the opposite sign of the Grassmann variable $\xi_k$. Actually, if we act by the operator $[\hspace{0.02cm}\xi^{\phantom{\dagger}\!}_{l}\hspace{0.02cm}, b^{\dagger}_{l}\hspace{0.03cm}]$ on (\ref{eq:8p}), we will have
\[
[\hspace{0.02cm}\xi^{\phantom{\dagger}\!}_{l}\hspace{0.02cm}, b^{\dagger}_{l}\hspace{0.02cm}]\hspace{0.03cm}
a_{k}|\hspace{0.02cm}(\xi)_{2}\,;b\hspace{0.02cm}\rangle
=
\Lambda\hspace{0.03cm}\xi_{k}
\Bigl(|\hspace{0.02cm}(\xi)_{2}\,;b\hspace{0.02cm}\rangle
\hspace{0.02cm}+\hspace{0.02cm}
 |\hspace{0.02cm}(-\xi)_{2}\,;b\hspace{0.02cm}\rangle\!\hspace{0.01cm}\Bigr).
\]
By virtue of (\ref{eq:5i}), the preceding expression can be also casted as follows:
\[
a_{k}\hspace{0.02cm}[\hspace{0.02cm}\xi^{\phantom{\dagger}\!}_{l}\hspace{0.02cm}, b^{\dagger}_{l}\hspace{0.02cm}]\,
|\hspace{0.02cm}(\xi)_{2}\,;b\hspace{0.02cm}\rangle
=
\Lambda\hspace{0.03cm}\xi_{k}
\Bigl(|\hspace{0.02cm}(\xi)_{2}\,;b\hspace{0.02cm}\rangle
\hspace{0.02cm}-\hspace{0.02cm}
|\hspace{0.03cm}(-\xi)_{2}\,;b\hspace{0.02cm}\rangle\!\hspace{0.01cm}\Bigr)
\]
(recall that the summation is taken over repeated indices). The state $|\hspace{0.03cm}(-\xi)_{2}\,;b\hspace{0.02cm}\rangle$ in turn can be presented as a result of acting on the ordinary coherent state $|\hspace{0.02cm}(\xi)_{2}\,;b\hspace{0.02cm}\rangle$ by the parafermion number counter $(-1)^{N}$ (``$G$\hspace{0.02cm}-\hspace{0.02cm}parity operator''), with the particle-number operator (\ref{eq:1u}), i.e.
\[
(-1)^{N}\hspace{0.02cm}
|\hspace{0.02cm}(\xi)_{2}\,;b\hspace{0.02cm}\rangle
\hspace{0.02cm}=\hspace{0.02cm}
|\hspace{0.02cm}(-\xi)_{2}\,;b\hspace{0.02cm}\rangle.
\]
We mention that within the usual Fermi statistics such state was considered by D'Hoker and Gagn\'e \cite{hoker_1996} in the context of the construction of worldline path integral for the {\it imaginary} part of the effective action, i.e. the phase of the fermion functional determinant. It is also interesting to note that the number counter enters into the so-called deformed Heisenberg algebra (the Calogero\hspace{0.03cm}-Vasiliev oscillator) \cite{vasiliev_1990, vasiliev_1991, dodlov_1993} involving the reflection operator $R=(-1)^{N}$ and a deformation parameter $ \nu \in \mathbb{R}$. In the paper by Plyushchay \cite{plyushchay_1997} it was shown that the single-mode algebra has finite-dimensional representations of some deformed parafermion algebra which at the deformation parameter $\nu = -\hspace{0.02cm}3$ is reduced to the standard parafermionic algebra of order 2. This may point to the fact that there is a certain connection between Govorkov's unitary quantization and the deformed Heisenberg algebra. \\
\indent
In section\,6, we have examined the map of different trilinear commutation relations. Here, we should also like to consider a problem of mapping the coherent states (\ref{eq:8q}) and (\ref{eq:8r}). The interesting question now arises whether these coherent states relate between each other by the transformation of the type (\ref{eq:6y}) (or (\ref{eq:6l})). To be specific, as the starting expression we take (\ref{eq:8q}), and as the transformation connecting the operators $a_{k}$ and $b_{k}$ we take the following relation:
\begin{equation}
a_{k} =
-\hspace{0.04cm}{\rm e}^{\pi\hspace{0.01cm}i\widetilde{N}/2}\hspace{0,015cm}
b_{k}\hspace{0.03cm}{\rm e}^{-\pi\hspace{0.01cm}i\widetilde{N}/2}\hspace{0,015cm}.
\label{eq:8s}
\end{equation}
Although the final expression has a simple form, nevertheless the calculations are somewhat lengthly.\\
\indent
Taking into account (\ref{eq:8s}), the formula (\ref{eq:8w}) can be presented in the form (for $p =2 $):
\[
-\hspace{0.04cm}{\rm e}^{\pi\hspace{0.01cm}i\widetilde{N}/2}\hspace{0,015cm}
b_{k}\,{\rm e}^{-\hspace{0.02cm}\pi\hspace{0.01cm}i\widetilde{N}/2}\hspace{0,015cm}
{\rm e}^{\textstyle-\frac{1}{2}\hspace{0.03cm}
[\hspace{0.02cm}\xi^{\phantom{\dagger}\!}_{l}, a^{\dagger}_{l}\hspace{0.02cm}]}|\hspace{0.03cm}0\rangle
=
\xi_{k}\hspace{0.03cm}{\rm e}^{\textstyle-\frac{1}{2}\hspace{0.03cm}
[\hspace{0.02cm}\xi^{\phantom{\dagger}\!}_{l}, a^{\dagger}_{l}\hspace{0.02cm}]}|\hspace{0.03cm}0\rangle.
\]
Let us operate on the left by ${\rm e}^{-\pi\hspace{0.01cm}i\widetilde{N}/2}$. Further, in front of the vacuum vector $|\hspace{0.03cm}0\rangle$ we insert the identity operator
\[
 I \equiv {\rm e}^{\pm\pi\hspace{0.01cm}i\widetilde{N}/2}\hspace{0,02cm}
{\rm e}^{\mp\pi\hspace{0.01cm}i\widetilde{N}/2}\hspace{0,015cm}.
\]
Besides, we insert the identity operator also between $\xi_{k}$ and the exponential function. Then, we obtain
\begin{equation}
b_{k}\Bigl({\rm e}^{-\pi\hspace{0.01cm}i\widetilde{N}/2}\hspace{0,015cm}
{\rm e}^{\textstyle-\frac{1}{2}\hspace{0.03cm}[\hspace{0.02cm}\xi^{\phantom{\dagger}\!}_{l}, a^{\dagger}_{l}\hspace{0.02cm}]}\,
{\rm e}^{\pi\hspace{0.01cm}i\widetilde{N}/2}\hspace{0,015cm}\Bigr)
{\rm e}^{-\pi\hspace{0.01cm}i\widetilde{N}/2}\hspace{0,015cm}|\hspace{0.03cm}0\rangle
\label{eq:8d}
\end{equation}
\[
=
-\Bigl({\rm e}^{-\pi\hspace{0.01cm}i\widetilde{N}/2}\hspace{0,015cm}\xi_{k}\,
{\rm e}^{-\pi\hspace{0.01cm}i\widetilde{N}/2}\hspace{0,015cm}\Bigr)\!
\Bigl({\rm e}^{\pi\hspace{0.01cm}i\widetilde{N}/2}\hspace{0,015cm}
{\rm e}^{\textstyle-\frac{1}{2}\hspace{0.03cm}[\hspace{0.02cm}\xi^{\phantom{\dagger}\!}_{l}, a^{\dagger}_{l}\hspace{0.02cm}]}\,
{\rm e}^{-\pi\hspace{0.01cm}i\widetilde{N}/2}\hspace{0,015cm}\Bigr)
{\rm e}^{\pi\hspace{0.01cm}i\widetilde{N}/2}\hspace{0,015cm}|\hspace{0.03cm}0\rangle.
\]
By using the operator identity (\ref{ap:B9}), we derive
\begin{align}
{\rm e}^{\pm\pi\hspace{0.01cm}i\widetilde{N}/2}\hspace{0,015cm}
{\rm e}^{\textstyle-\frac{1}{2}\hspace{0.03cm}[\hspace{0.02cm}
\xi^{\phantom{\dagger}\!}_{l}\hspace{0.02cm}, a^{\dagger}_{l}\hspace{0.02cm}]}\,
{\rm e}^{\mp\pi\hspace{0.01cm}i\widetilde{N}/2}
&=
\exp\Bigr[\!-\!\frac{1}{2}\hspace{0.02cm}\Bigl({\rm e}^{\pm\pi\hspace{0.01cm}i\widetilde{N}/2}\hspace{0,015cm}
[\hspace{0.02cm}\xi^{\phantom{\dagger}\!}_{l}\hspace{0.02cm}, a^{\dagger}_{l}\hspace{0.02cm}]\,
{\rm e}^{\pm\pi\hspace{0.01cm}i\widetilde{N}/2}\hspace{0,015cm}\Bigr)
{\rm e}^{\mp\pi\hspace{0.02cm}i\widetilde{N}}\hspace{0.03cm}\Bigr] \notag \\[0.7ex]
&=
\exp\Bigr(\!\pm\!\frac{1}{2}\hspace{0.04cm}{\rm e}^{\pm\pi\widetilde{\Lambda}/2}\hspace{0.015cm}
\bigl\{\xi^{\phantom{\dagger}\!}_{l}\hspace{0.02cm}, b^{\dagger}_{l}\bigr\}\,
{\rm e}^{\mp\pi\hspace{0.02cm}i\widetilde{N}}\hspace{0.03cm}\Bigr).\notag
\end{align}
At the last step we have used the relation (\ref{eq:6s}). Further, by virtue of (\ref{eq:6p}), the expression in the first parentheses on the right-hand side of (\ref{eq:8d}) equals
\[
{\rm e}^{-\pi\hspace{0.01cm}i\widetilde{N}/2}\hspace{0,015cm}\xi_{k}\,
{\rm e}^{-\pi\hspace{0.01cm}i\widetilde{N}/2}
= {\rm e}^{-\pi\widetilde{\Lambda}/2}\hspace{0.015cm}\xi_{k}.
\]
Taking into account the above-mentioned reasoning, we get, instead of (\ref{eq:8d}),
\begin{align}
b_{k}\exp\Bigr(\!-\!\frac{1}{2}\;{\rm e}^{-\pi\widetilde{\Lambda}/2}\hspace{0.015cm}
\bigl\{\xi^{\phantom{\dagger}\!}_{l}\hspace{0.02cm}, b^{\dagger}_{l}\bigr\}\,
{\rm e}^{\pi\hspace{0.02cm}i\widetilde{N}}\hspace{0.03cm}\Bigr)
{\rm e}^{-\pi\hspace{0.01cm}i\widetilde{N}/2}\hspace{0,015cm}&|\hspace{0.03cm}0\rangle
\label{eq:8f}\\[0.7ex]
=
-\hspace{0.03cm}{\rm e}^{-\pi\widetilde{\Lambda}/2}\hspace{0.015cm}\xi_{k}
\exp\Bigr(\hspace{0.03cm}\frac{1}{2}\;{\rm e}^{\pi\widetilde{\Lambda}/2}\hspace{0.015cm}
\bigl\{\xi^{\phantom{\dagger}\!}_{l}\hspace{0.02cm}, b^{\dagger}_{l}\bigr\}\,
{\rm e}^{-\pi\hspace{0.02cm}i\widetilde{N}}\hspace{0.03cm}\Bigr)
{\rm e}^{\pi\hspace{0.01cm}i\widetilde{N}/2}\hspace{0,015cm}&|\hspace{0.03cm}0\rangle.
\notag
\end{align}
In the previous section we have obtained the rule of acting the operator $i \widetilde{N}$ on the vacuum
\begin{equation}
i\widetilde{N}|\hspace{0.03cm}0\rangle =
\frac{1}{2}\hspace{0.03cm}\tilde{\Lambda}\hspace{0.03cm}|\hspace{0.03cm}0\rangle.
\label{eq:8g}
\end{equation}
Using this, we find
\[
{\rm e}^{\mp\pi\hspace{0.01cm}i\widetilde{N}/2}\hspace{0,015cm}|\hspace{0.03cm}0\rangle
=
{\rm e}^{\mp\pi\widetilde{\Lambda}/2}\hspace{0.015cm}|\hspace{0.03cm}0\rangle.
\]
\indent
It remains to analyze the exponential operator
\begin{equation}
\exp\Bigr(\!\mp\!\frac{1}{2}\;{\rm e}^{\mp\pi\widetilde{\Lambda}/2}\hspace{0.015cm}
\bigl\{\xi^{\phantom{\dagger}\!}_{l}\hspace{0.02cm}, b^{\dagger}_{l}\bigr\}\,
{\rm e}^{\pm\pi\hspace{0.02cm}i\widetilde{N}}\hspace{0.03cm}\Bigr)
\equiv
{\rm e}^{A}.
\label{eq:8h}
\end{equation}
Let us consider the following expansion:
\begin{equation}
{\rm e}^{A} = \cosh A + \sinh A = \sum^{\infty}_{s\hspace{0.02cm}=\hspace{0.02cm}0}
\frac{A^{2s + 1}}{(2s + 1)!}
\;+\, \sum^{\infty}_{s\hspace{0.02cm}=\hspace{0.02cm}0}\frac{A^{2\hspace{0.01cm}s}}{(2s)!}
\hspace{0.03cm}.
\label{eq:8j}
\end{equation}
At first we define an explicit form of the operator $A$ squared
\[
A^{2} = \biggr(\!\!\mp\!\frac{1}{2}\biggr)^{\!\!2}{\rm e}^{\mp\pi\hspace{0.02cm}\widetilde{\Lambda}}
\bigl\{\xi^{\phantom{\dagger}\!}_{l}\hspace{0.02cm}, b^{\dagger}_{l}\bigr\}
\Bigl({\rm e}^{\pm\pi\hspace{0.02cm}i\widetilde{N}}
\bigl\{\xi^{\phantom{\dagger}\!}_{l^{\prime}}\hspace{0.02cm}, b^{\dagger}_{l^{\prime}}\bigr\}\hspace{0.02cm}
{\rm e}^{\pm\pi\hspace{0.02cm}i\widetilde{N}}\hspace{0.03cm}\Bigr)
=
(-1)\hspace{0.03cm}\biggr(\!\!\mp\!\frac{1}{2}\biggr)^{\!\!2}
\bigl\{\xi^{\phantom{\dagger}\!}_{l}\hspace{0.02cm}, b^{\dagger}_{l}\bigr\}^{2}.
\]
In deriving this expression we have taken into account that by virtue (\ref{eq:6e}) and (\ref{eq:6p}), the following equality holds
\[
{\rm e}^{\pm\pi\hspace{0.02cm}i\widetilde{N}}
\bigl\{\xi^{\phantom{\dagger}\!}_{l^{\prime}}\hspace{0.02cm}, b^{\dagger}_{l^{\prime}}\bigr\}\hspace{0.02cm}
{\rm e}^{\pm\pi\hspace{0.02cm}i\widetilde{N}}
=
(-1)\hspace{0.03cm}{\rm e}^{\pm\pi\hspace{0.02cm}\widetilde{\Lambda}}
\bigl\{\xi^{\phantom{\dagger}\!}_{l^{\prime}}\hspace{0.02cm}, b^{\dagger}_{l^{\prime}}\bigr\}.
\]
Then
\[
A^{2\hspace{0.01cm}s} = (-1)^{s}\hspace{0.03cm}\biggr(\!\!\mp\!\frac{1}{2}\biggr)^{\!\!2\hspace{0.01cm}s\!}
\bigl\{\xi^{\phantom{\dagger}\!}_{l}\hspace{0.02cm}, b^{\dagger}_{l}\bigr\}^{2\hspace{0.01cm}s}
\]
and
\[
A^{2\hspace{0.01cm}s + 1} = (-1)^{s}\hspace{0.03cm}\biggr(\!\!\mp\!\frac{1}{2}\biggr)^{\!\!2\hspace{0.01cm}s + 1\!}
\bigl\{\xi^{\phantom{\dagger}\!}_{l}\hspace{0.02cm}, b^{\dagger}_{l}\bigr\}^{2\hspace{0.01cm}s
+ 1}
\hspace{0.02cm}{\rm e}^{\mp\pi\widetilde{\Lambda}/2}\hspace{0.015cm}
\hspace{0.02cm}{\rm e}^{\pm\pi\hspace{0.02cm}i\widetilde{N}}.
\]
Substituting the derived expressions into (\ref{eq:8j}), we obtain that the exponential operator (\ref{eq:8h}) can be presented as follows
\[
\exp\Bigr(\!\mp\!\frac{1}{2}\;{\rm e}^{\mp\pi\widetilde{\Lambda}/2}\hspace{0.015cm}
\bigl\{\xi^{\phantom{\dagger}\!}_{l}\hspace{0.02cm}, b^{\dagger}_{l}\bigr\}\,
{\rm e}^{\pm\pi\hspace{0.02cm}i\widetilde{N}}\hspace{0.03cm}\Bigr)
=
\cos\Bigl(\frac{1}{2}\hspace{0.02cm}\bigl\{\xi^{\phantom{\dagger}\!}_{l}\hspace{0.02cm}, b^{\dagger}_{l}\bigr\}\hspace{0.02cm}\!\Bigr)
\hspace{0.01cm}\mp\,
\sin\Bigl(\frac{1}{2}\hspace{0.02cm}\bigl\{\xi^{\phantom{\dagger}\!}_{l}\hspace{0.02cm}, b^{\dagger}_{l}\bigr\}\!\Bigr)
\hspace{0.02cm}{\rm e}^{\mp\pi\widetilde{\Lambda}/2}\hspace{0.015cm}
\hspace{0.02cm}{\rm e}^{\pm\pi\hspace{0.02cm}i\widetilde{N}}.
\]
Here, on the right-hand side the act of the operator
${\rm e}^{\pm\pi\hspace{0.02cm}i\widetilde{N}}$ on the vacuum is defined by formula (\ref{eq:8g}). In view of the above, our main expression (\ref{eq:8f}) takes the form
\begin{equation}
\begin{split}
&b_{k}\Bigl[\hspace{0.03cm}\cos\Bigl(\frac{1}{2}\hspace{0.02cm}
\bigl\{\xi^{\phantom{\dagger}\!}_{l}\hspace{0.02cm}, b^{\dagger}_{l}\bigr\}\hspace{0.02cm}\!\Bigr)
\hspace{0.01cm}-\,
\sin\Bigl(\frac{1}{2}\hspace{0.02cm}\bigl\{\xi^{\phantom{\dagger}\!}_{l}\hspace{0.02cm}, b^{\dagger}_{l}\bigr\}\!\Bigr)
\Bigr]|\hspace{0.03cm}0\rangle = \\[1ex]
-\hspace{0.02cm}&\xi_{k}\Bigl[\hspace{0.03cm}\cos\Bigl(\frac{1}{2}\hspace{0.03cm}
\bigl\{\xi^{\phantom{\dagger}\!}_{l}\hspace{0.02cm}, b^{\dagger}_{l}\bigr\}\hspace{0.02cm}\!\Bigr)
\hspace{0.01cm}+\,
\sin\Bigl(\frac{1}{2}\hspace{0.02cm}\bigl\{\xi^{\phantom{\dagger}\!}_{l}\hspace{0.02cm}, b^{\dagger}_{l}\bigr\}\!\Bigr)
\Bigr]|\hspace{0.03cm}0\rangle.
\label{eq:8k}
\end{split}
\end{equation}
It is worthy of special emphasis that all exponential factors containing the constant $\widetilde{\Lambda}$ has exactly canceled out. This circumstance serves as an indirect proof of the correctness of our reasoning. A somewhat cumbersome expression (\ref{eq:8k}) turns into
identity, if the following relation holds\footnote{\,The relation of the type (\ref{eq:8l}), of course, is not unique. For example, the relation
\[
b_{k}\hspace{0.04cm}
{\rm e}^{\textstyle\pm\frac{1}{2}\hspace{0.03cm}i\hspace{0.02cm}
\{\hspace{0.02cm}\xi^{\phantom{\dagger}\!}_{l}\hspace{0.02cm}, b^{\dagger}_{l}\}\,}|\hspace{0.03cm}0\rangle
=
-\hspace{0.03cm}\xi_{k}\hspace{0.05cm}
{\rm e}^{\textstyle\mp\frac{1}{2}\hspace{0.03cm}i\hspace{0.02cm}
\{\hspace{0.02cm}\xi^{\phantom{\dagger}\!}_{l}\hspace{0.02cm}, b^{\dagger}_{l}\}\,}
|\hspace{0.03cm}0\rangle
\]
also turns (\ref{eq:8k}) into identity. However, here on the right-hand side the sign in the exponential function has changed and there is no the factor $i$ in front of $\xi_k$. Besides, a direct check shows that the relation just above is not satisfied in contrast to (\ref{eq:8l}).}:
\begin{equation}
b_{k}\hspace{0.04cm}
{\rm e}^{\textstyle\pm\frac{1}{2}\hspace{0.03cm}i\hspace{0.02cm}
\{\hspace{0.02cm}\xi^{\phantom{\dagger}\!}_{l}\hspace{0.02cm}, b^{\dagger}_{l}\hspace{0.02cm}\}}|\hspace{0.03cm}0\rangle
=
(\pm\hspace{0.02cm}i)\hspace{0.03cm}\xi_{k}\hspace{0.05cm}
{\rm e}^{\textstyle\pm\frac{1}{2}\hspace{0.03cm}i\hspace{0.02cm}
\{\hspace{0.02cm}\xi^{\phantom{\dagger}\!}_{l}\hspace{0.02cm}, b^{\dagger}_{l}\hspace{0.01cm}\}}
|\hspace{0.03cm}0\rangle.
\label{eq:8l}
\end{equation}
Thus in mapping the coherent state (\ref{eq:8q}) with (\ref{eq:8w}) we lead not to the coherent state (\ref{eq:8r}) with (\ref{eq:8t}), but to the expression (\ref{eq:8l}). This is consistent with the rule stated in section 6: if the para-Grassmann number $\xi_{k}$ enters into commutator (anticommutator) together with the operator $a_{k}$ or $b_{k}$, then in mapping (\ref{eq:6y}) or (\ref{eq:6l}) in addition to the replacement $a_{k} \rightleftharpoons b_{k}$, the commutator (anticommutator) has to be replaced by anticommutator (commutator). It is clear that the factor $(\pm i)$ in (\ref{eq:8l}) does not play any role there.\\
\indent
Now, by a straightforward calculation we prove the relation (\ref{eq:8l}). Omitting the factor $(\pm i)$ on the left- and right-hand sides, we rewrite the relation as follows:
\begin{equation}
b_{k}\hspace{0.04cm}
{\rm e}^{\hspace{0.02cm}\textstyle\frac{1}{2}\hspace{0.03cm}
\{\hspace{0.02cm}\xi^{\phantom{\dagger}\!}_{l}\hspace{0.02cm}, b^{\dagger}_{l}\hspace{0.02cm}\}}|\hspace{0.03cm}0\rangle
=
\xi_{k}\hspace{0.05cm}
{\rm e}^{\hspace{0.02cm}\textstyle\frac{1}{2}\hspace{0.03cm}
\{\hspace{0.02cm}\xi^{\phantom{\dagger}\!}_{l}\hspace{0.02cm}, b^{\dagger}_{l}\hspace{0.01cm}\}}
|\hspace{0.03cm}0\rangle.
\label{eq:8z}
\end{equation}
To prove the relation, it is sufficient to consider the following expression:
\begin{align}
&{\rm e}^{\textstyle-\frac{1}{2}\hspace{0.03cm}
\{\hspace{0.02cm}\xi^{\phantom{\dagger}\!}_{l}\hspace{0.02cm}, b^{\dagger}_{l}\hspace{0.02cm}\}}
b_{k}\hspace{0.04cm}
{\rm e}^{\textstyle-\frac{1}{2}\hspace{0.03cm}
\{\hspace{0.02cm}\xi^{\phantom{\dagger}\!}_{l^{\prime}}\hspace{0.02cm}, b^{\dagger}_{l^{\prime}}\hspace{0.02cm}\}}
\label{eq:8x} \\[1ex]
=\;
&b_{k} + \biggl(\!-\frac{1}{2}\biggr)\bigl\{\!\bigl\{\xi^{\phantom{\dagger}\!}_{l}\hspace{0.02cm}, b^{\dagger}_{l}\bigr\},b^{\phantom{\dagger}\!}_{k}\bigr\}
+
\frac{\!1}{2!}\hspace{0.02cm}
\biggl(\!-\frac{1}{2}\biggr)^{\!\!2\!}\bigl\{\!\bigl\{\xi^{\phantom{\dagger}\!}_{l}\hspace{0.02cm}, b^{\dagger}_{l}\bigr\},\bigl\{\!\bigl\{\xi^{\phantom{\dagger}\!}_{l^{\prime}}\hspace{0.02cm}, b^{\dagger}_{l^{\prime}}\bigr\},b^{\phantom{\dagger}\!}_{k}\bigr\}\!\bigr\} +\, \ldots \notag\\[1ex]
=\;
&b_{k} + \biggl(\!-\frac{1}{2}\biggr)2\hspace{0.02cm}\xi_{k}
+ \frac{1}{2!}\hspace{0.02cm}\biggl(\!-\frac{1}{2}\biggr)^{\!\!2\!}2\hspace{0.02cm}
\bigl\{\!\bigl\{\xi^{\phantom{\dagger}\!}_{l}\hspace{0.02cm}, b^{\dagger}_{l}\bigr\},\xi^{\phantom{\dagger}\!}_{k}\bigr\}
+\, \ldots\, = b_{k} - \xi_{k}.
\notag
\end{align}
Here, we have used the trilinear commutation rules (\ref{eq:6k}) and (\ref{eq:6f}), which are valid
for $p = 2$. The series in (\ref{eq:8x}) is exactly cut off as this took place in calculating the
operator relation (\ref{eq:8e}) for the standard definition of the coherent state (\ref{eq:8q}), (\ref{eq:8w}). Analogue of the formula (\ref{eq:8e}) is now the expression
\[
b_{k}\hspace{0.04cm}
{\rm e}^{\hspace{0.02cm}\textstyle\frac{1}{2}\hspace{0.03cm}
\{\hspace{0.02cm}\xi^{\phantom{\dagger}\!}_{l}\hspace{0.02cm}, b^{\dagger}_{l}\hspace{0.02cm}\}}
=
{\rm e}^{\textstyle-\frac{1}{2}\hspace{0.03cm}
\{\hspace{0.02cm}\xi^{\phantom{\dagger}\!}_{l}\hspace{0.02cm}, b^{\dagger}_{l}\hspace{0.02cm}\}} b_{k}
\,+\,
\xi_{k}\hspace{0.04cm}
{\rm e}^{\hspace{0.02cm}\textstyle\frac{1}{2}\hspace{0.03cm}
\{\hspace{0.02cm}\xi^{\phantom{\dagger}\!}_{l}\hspace{0.02cm}, b^{\dagger}_{l}\hspace{0.02cm}\}}.
\]
Acting on the vacuum state $|\hspace{0.03cm}0\rangle$ by the previous operator relation, we results in the formula (\ref{eq:8z}). Further, it can be shown that instead of (\ref{eq:8i}), we now have
\[
a_{k}\hspace{0.04cm}
{\rm e}^{\hspace{0.02cm}\hspace{0.02cm}\textstyle\frac{1}{2}\hspace{0.03cm}
\{\hspace{0.02cm}\xi^{\phantom{\dagger}\!}_{l}\hspace{0.02cm}, b^{\dagger}_{l}\hspace{0.02cm}\}}
=
{\rm e}^{\hspace{0.02cm}\hspace{0.02cm}\textstyle\frac{1}{2}\hspace{0.03cm}
\{\hspace{0.02cm}\xi^{\phantom{\dagger}\!}_{l}\hspace{0.02cm}, b^{\dagger}_{l}\hspace{0.02cm}\}} a_{k}
-
\Lambda\hspace{0.03cm}\xi_{k}\Bigl(\hspace{0.03cm}\sum_{s}\frac{(-1)^{s}}{(s + 1)!}\;
\bigl\{\xi^{\phantom{\dagger}\!}_{m}\hspace{0.02cm}, b^{\dagger}_{m}\bigr\}^{\hspace{0.02cm}s}\Bigr)\hspace{0.03cm}
{\rm e}^{\hspace{0.02cm}\hspace{0.02cm}\textstyle\frac{1}{2}\hspace{0.03cm}
\{\hspace{0.02cm}\xi^{\phantom{\dagger}\!}_{l}\hspace{0.02cm}, b^{\dagger}_{l}\hspace{0.02cm}\}}.
\]
Finally, we may derive the overlap function for the ``coherent'' state ${\rm e}^{\hspace{0.02cm}\textstyle\frac{1}{2}\hspace{0.03cm}
\{\hspace{0.02cm}\xi^{\phantom{\dagger}\!}_{l}\hspace{0.02cm}, b^{\dagger}_{l}\hspace{0.01cm}\}}
|\hspace{0.03cm}0\rangle$. A somewhat cumbersome calculation, which is omitted here, leads to
\[
\langle\hspace{0.03cm}0|\,
{\rm e}^{\hspace{0.02cm}\textstyle\frac{1}{2}\hspace{0.03cm}
\{\hspace{0.02cm}\bar{\xi}^{\prime}_{l}\hspace{0.02cm}, b^{\dagger}_{l}\hspace{0.01cm}\}}\,
{\rm e}^{\hspace{0.02cm}\textstyle\frac{1}{2}\hspace{0.03cm}
\{\hspace{0.02cm}\xi^{\phantom{\dagger}\!}_{l^{\prime}}\hspace{0.02cm}, b^{\dagger}_{l^{\prime}}\hspace{0.01cm}\}}
|\hspace{0.03cm}0\rangle
=
{\rm e}^{\textstyle\hspace{0.02cm}\frac{1}{2}\hspace{0.03cm}
[\hspace{0.02cm}
\bar{\xi}^{\,\prime}_{l}\hspace{0.02cm},\xi^{\phantom{\dagger}\!}_{l}\hspace{0.02cm}
\hspace{0.02cm}]}.
\]
This should be expected, since the right-hand side in (\ref{eq:8a}) is invariant in mapping (\ref{eq:6y}) (or (\ref{eq:6l})).

\section{Unitary transformations}
\setcounter{equation}{0}

Bialynicki-Berula \cite{bialynicki_1963} has shown that the trilinear relations (\ref{eq:2q})\,--\,(\ref{eq:2e}) for a single para-Fermi field $\phi_{a}$ can be obtained from the requirement that the equations
\begin{equation}
[\hspace{0.02cm}a_{k}, N\hspace{0.02cm}] = a_{k},
\quad
[\hspace{0.02cm}a^{\dagger}_{k}, N\hspace{0.02cm}] = -a^{\dagger}_{k},
\label{eq:9q}
\end{equation}
where $N$ is the particle-number operator, Eq.\,(\ref{eq:1u}), be invariant under unitary transformation of the $a_{k}$'s operators:
\[
a^{\prime}_{k} = a_{k} + \sum^{M}_{l\hspace{0.02cm}=\hspace{0.02cm}1}\hspace{0.02cm}\alpha_{kl\,}a_{l}.
\]
Here, the infinitesimal parameters $\alpha_{kl}$ are subject to the condition
\begin{equation}
\alpha_{kl} + \alpha^{\ast}_{lk} = 0.
\label{eq:9w}
\end{equation}
\indent
The following question suggests itself: can we obtain Govorkov's trilinear relations (\ref{eq:3q})\,--\,(\ref{eq:3e}) containing operators of two different para-Fermi fields $\phi_{a}$ and $\phi_{b}$ as a consequence of the requirement of invariance for the equations
\begin{equation}
[\hspace{0.02cm}i\widetilde{N}, a_{k}\hspace{0.02cm}] = b_{k},
\quad
[\hspace{0.03cm}i\widetilde{N}, b_{k}\hspace{0.02cm}] = -\hspace{0.02cm}a_{k},
\label{eq:9e}
\end{equation}
and its Hermitian conjugation, under an infinitesimal linear transformation of operators $a_{k}$ and $b_{k}$. For convenience of the further reference, we write out once more an explicit form of the operator
$i \widetilde{N}$:
\begin{equation}
i\widetilde{N}
=
\varrho\,
\sum^{M}_{k\hspace{0.02cm}=1}\,
\bigl(\hspace{0.02cm}
[\hspace{0.02cm}a^{\dagger}_{k},b^{\phantom{\dagger}\!}_{k}\hspace{0.02cm}] \hspace{0.02cm}+ \lambda\hspace{0.02cm}\bigr),
\quad
\varrho\equiv-\frac{1}{2\hspace{0.02cm}(2\hspace{0.01cm}M + 1)}.
\label{eq:9r}
\end{equation}
It is clear that the required transformation leaving (\ref{eq:9e}) unchanged, must ``mix up'' the operators $a_{k}$ and $b_{k}$:
\begin{equation}
\begin{array}{l}
a^{\prime}_{k} = a_{k} + \alpha_{kl\,}b_{l},  \\[1ex]
b^{\prime}_{k} = b_{k} + \beta_{kl\,}a_{l},
\end{array}
\qquad
\begin{array}{l}
a^{\prime\,\dagger}_{k} = a^{\dagger}_{k} - \alpha_{lk\,}b^{\dagger}_{l},  \\[1ex]
b^{\prime\,\dagger}_{k} = b^{\dagger}_{k} - \beta_{lk\,}a^{\dagger}_{l}.
\end{array}
\label{eq:9t}
\end{equation}
Here, for the sake of brevity, we have again used the summation convention over repeated indices and required that the infinitesimal parameters $\alpha_{kl}$ and $\beta_{kl}$ satisfy the condition (\ref{eq:9w}). For the commutator in (\ref{eq:9r}), we have
\[
[\hspace{0.02cm}a^{\prime\,\dagger}_{k},b^{\prime}_{k}\hspace{0.02cm}]
=
[\hspace{0.02cm}a^{\dagger}_{k},b^{\phantom{\dagger}\!}_{k}\hspace{0.02cm}]
-
\alpha_{lk\,}[\hspace{0.02cm}b^{\dagger}_{l},b^{\phantom{\dagger}\!}_{k}\hspace{0.02cm}]
+
\beta_{kl\,}[\hspace{0.02cm}a^{\dagger}_{k},a^{\phantom{\dagger}\!}_{l}\hspace{0.02cm}].
\]
The requirement of invariance for the first equation in (\ref{eq:9e}) leads to the following relation:
\[
\alpha_{ms}\hspace{0.02cm}[\hspace{0.03cm}b_{s}, i\widetilde{N}\hspace{0.02cm}]
-
\varrho\,\alpha_{lk}\hspace{0.02cm}
[\hspace{0.03cm}a_{m}, [\hspace{0.02cm}b^{\dagger}_{l},b^{\phantom{\dagger}\!}_{k}\hspace{0.02cm}]\hspace{0.02cm}]
+
\varrho\hspace{0.02cm}\beta^{\phantom{\dagger}\!}_{kl}\hspace{0.02cm}
[\hspace{0.03cm}a^{\phantom{\dagger}\!}_{m}, [\hspace{0.02cm}a^{\dagger}_{k},a^{\phantom{\dagger}\!}_{l}\hspace{0.02cm}]\hspace{0.02cm}]
=
\beta_{ms}\hspace{0.03cm}a_{s}.
\]
The commutator in the first term on the left-hand side, by virtue of (\ref{eq:9e}), is equal to
$-a_{s}$ and the double commutator in the third term in view of (\ref{eq:2q}) equals $2\hspace{0.02cm} \delta_{mk}\hspace{0.02cm}a_{l}$. Therefore, the above expression can be presented as
\[
- \varrho\,\alpha^{\phantom{\dagger}\!}_{lk}\hspace{0.02cm}
[\hspace{0.03cm}a^{\phantom{\dagger}\!}_{m}, [\hspace{0.02cm}b^{\dagger}_{l},b^{\phantom{\dagger}\!}_{k}\hspace{0.02cm}]\hspace{0.02cm}]
=
\delta_{ml}\bigl(\beta_{lk\,}(1 - 2\varrho) + \alpha_{lk}\bigr)a_{k}.
\]
Here, we take one step further: we require the fulfillment of an additional condition connecting the infinitesimal parameters $\alpha_{kl}$ and $\beta_{kl}$ among themselves
\[
\alpha_{kl\,} = -\beta_{kl}.
\]
Only in this special case the parameter $\varrho$ is exactly cancelled from the left- and right-hand sides and we lead to
\[[\hspace{0.03cm}a^{\phantom{\dagger}\!}_{m}, [\hspace{0.02cm}b^{\dagger}_{l},b^{\phantom{\dagger}\!}_{k}\hspace{0.02cm}]\hspace{0.02cm}]
=
-2\hspace{0.015cm}\delta_{ml}\hspace{0.02cm}a_{k}.
\]
The requirement of invariance for the second equation in (\ref{eq:9e}) results in a similar expression
\begin{equation}
[\hspace{0.02cm}[\hspace{0.02cm}a^{\dagger}_{k},a^{\phantom{\dagger}\!}_{l}\hspace{0.02cm}],
b^{\phantom{\dagger}\!}_{m}] = -2\hspace{0.01cm}\delta_{km}\hspace{0.02cm}b_{l}.
\label{eq:9y}
\end{equation}
We thereby reproduce the Govorkov relation (\ref{eq:3e}).\\
\indent
The appearance of the transformation (\ref{eq:9t}) is enhanced by using the matrix notations
\[
\left(\!
\begin{array}{l}
\tilde{a}  \\[0.7ex]
\tilde{b}
\end{array}
\!\right)
=
\left(\!
\begin{array}{cc}
I & 0  \\[0.7ex]
0 & I
\end{array}
\!\right)\!
\left(\!
\begin{array}{l}
a  \\[0.7ex]
b
\end{array}
\!\right)
+
\left(\!
\begin{array}{cc}
0 & \alpha  \\[0.7ex]
-\alpha & 0
\end{array}
\!\right)\!
\left(\!
\begin{array}{l}
a  \\[0.7ex]
b
\end{array}
\!\right),
\]
where $a = (a_{1},\ldots,a_{M})^{T},\,b = (b_{1},\ldots,b_{M})^{T}$ ($T$ is the sign of transposition) and $\alpha = (\alpha_{kl})$. The matrix
\[
X =
\left(\!
\begin{array}{cc}
0 & \alpha  \\[0.7ex]
-\alpha & 0
\end{array}
\!\right)\!
\]
satisfies the condition $X^{\dagger} = X$ and in so doing it belongs to algebra of the unimodular group $SU(2M)$. This is quite natural, since all Govorkov's relations was obtained within the framework of quantization of the fields based on the relations of Lie algebra of the group $SU(2M + 1)$.\\
\indent
However, here the question remains as to whether it might be possible to obtain from one trilinear relation (\ref{eq:9y}) the other relations (\ref{eq:3q}) or (\ref{eq:3w}). In the case of a single para-Fermi field $\phi_{a}$ the answer is positive \cite{bialynicki_1963}. Really, the requirement of invariance for (\ref{eq:9q}) leads to the relation (\ref{eq:2q}). The use of the Jacobi identity and relation (\ref{eq:2q}) is sufficient to restore the other trilinear relation (\ref{eq:2e}). In the case of (\ref{eq:9y}) the Jacobi identity gives
\begin{equation}
[\hspace{0.02cm}[\hspace{0.02cm}b^{\phantom{\dagger}\!}_{m},
a^{\dagger}_{k}\hspace{0.02cm}],a^{\phantom{\dagger}\!}_{l}]
+
[\hspace{0.02cm}[\hspace{0.02cm}a^{\phantom{\dagger}\!}_{l},
b^{\phantom{\dagger}\!}_{m}\hspace{0.02cm}],a^{\dagger}_{k}]
=
2\hspace{0.015cm}\delta_{km}\hspace{0.02cm}b_{l}
\label{eq:9u}
\end{equation}
and here, on the left-hand side, as opposed to the case of the single field, we have two different expressions and a priori it is not clear how they can be ``decoupled''.\\
\indent
Greenberg and Messiah \cite{greenberg_1965} suggested an approach to decoupling the relations of the (\ref{eq:9u}) type. We have already mentioned the fact in section\,2. Let us consider their method in more detail. The starting relation is (\ref{eq:2r}). The use of the Jacobi identity gives
\begin{equation}
[\hspace{0.02cm}[\hspace{0.02cm}b^{\phantom{\dagger}\!}_{m},
a^{\dagger}_{k}\hspace{0.02cm}],a^{\phantom{\dagger}\!}_{l}]
+
[\hspace{0.02cm}[\hspace{0.02cm}a^{\phantom{\dagger}\!}_{l},
b^{\phantom{\dagger}\!}_{m}\hspace{0.02cm}],a^{\dagger}_{k}]
= 0.
\hspace{0.7cm}
\label{eq:9i}
\end{equation}
In ``decoupling'' this relation, Greenberg and Messiah have used the conditions $(i)$ and $(iii)$, which was given in section 2. In particular, the condition $(iii)$ requires that the desired trilinear relations be satisfied by ordinary Bose and Fermi fields.\\
\indent
Let the operators $a_{k}$ and $b_{k}$ be the operators of Fermi oscillators, i.e. they satisfy usual anticommutation relations:
\begin{eqnarray}
&\{\hspace{0.02cm}a^{\dagger}_{k},a^{\phantom{\dagger}\!}_{l}\hspace{0.02cm}\} = \delta^{\phantom{\dagger}\!}_{kl},
\quad
&\{\hspace{0.02cm}b^{\dagger}_{m},b^{\phantom{\dagger}\!}_{n}\hspace{0.02cm}\} = \delta^{\phantom{\dagger}\!}_{mn},
\label{eq:9o}\\[0.5ex]
&\!\!\!\{\hspace{0.02cm}a_{k},a_{l}\hspace{0.02cm}\} = 0,
\quad
&\{\hspace{0.02cm}b_{m},b_{n}\hspace{0.02cm}\} = 0,\;
\ldots\;,
\label{eq:9p}\\[0.5ex]
&\{\hspace{0.02cm}a^{\dagger}_{k},b^{\phantom{\dagger}\!}_{m}\hspace{0.02cm}\} = 0,
\quad
&\{\hspace{0.02cm}a_{k},b_{m}\hspace{0.02cm}\} = 0,\;
\ldots\;.
\label{eq:9a}
\end{eqnarray}
Then for the first term on the left-hand side of (\ref{eq:9i}), by using the identity (\ref{ap:B2}), we have
\begin{equation}
[\hspace{0.02cm}[\hspace{0.02cm}b^{\phantom{\dagger}\!}_{m},
a^{\dagger}_{k}\hspace{0.02cm}],a^{\phantom{\dagger}\!}_{l}]
=
\{\hspace{0.02cm}b^{\phantom{\dagger}\!}_{m}, \{\hspace{0.02cm}a^{\phantom{\dagger}\!}_{l},a^{\dagger}_{k}\}\!\hspace{0.02cm}\}
-
\{\hspace{0.02cm}a^{\dagger}_{k},\{\hspace{0.02cm}a^{\phantom{\dagger}\!}_{l},
b^{\phantom{\dagger}\!}_{m}\}\!\hspace{0.02cm}\}
=
2\hspace{0.015cm}\delta_{lk}\hspace{0.02cm}b_{m},
\label{eq:9s}
\end{equation}
and for the second term, correspondingly, we get
\begin{equation}
[\hspace{0.02cm}[\hspace{0.02cm}a^{\phantom{\dagger}\!}_{l},
b^{\phantom{\dagger}\!}_{m}\hspace{0.02cm}],a^{\dagger}_{k}]
=
\{\hspace{0.02cm}a^{\phantom{\dagger}\!}_{l},
\{\hspace{0.02cm}a^{\dagger}_{k},b^{\phantom{\dagger}\!}_{m}\}\!\hspace{0.02cm}\}
-
\{\hspace{0.02cm}b^{\phantom{\dagger}\!}_{m}, \{\hspace{0.02cm}a^{\dagger}_{k},a^{\phantom{\dagger}\!}_{l}\}\!\hspace{0.02cm}\}
=
-2\hspace{0.015cm}\delta_{lk}\hspace{0.02cm}b_{m}.
\label{eq:9d}
\end{equation}
The obtained expressions on the right-hand sides of (\ref{eq:9s}) and (\ref{eq:9d}) are simply
{\it postulated} as the right-hand sides in the trilinear relations for the para-Fermi oscillators $a_{k}$ and $b_{k}$, as was indeed done in (\ref{eq:2u}) and (\ref{eq:2i}).\\
\indent
Let us apply this approach to the relation (\ref{eq:9u}). It is clear that the standard system of the commutation rules (\ref{eq:9o})\,--\,(\ref{eq:9a}) is not appropriate here. This system must somehow be modified to reproduce the right-hand sides in the trilinear relations (\ref{eq:3q}) and (\ref{eq:3w}). The relations (\ref{eq:9o}) and (\ref{eq:9p}) should be kept without any modification, but instead of the first relation in (\ref{eq:9a}), we should consider, for example, the expression
\[
\{\hspace{0.02cm}a^{\dagger}_{k},b^{\phantom{\dagger}\!}_{m}\hspace{0.02cm}\} =
-2\hspace{0.02cm}\delta_{km}\hspace{0.02cm}G.
\]
Here, we have introduced an additional algebraic quantity $G$ such that
\[
\{\hspace{0.02cm}a_{l},G\hspace{0.02cm}\} = b_{l}.
\]
In this case, instead of (\ref{eq:9d}), we obtain what we need
\[
[\hspace{0.02cm}[\hspace{0.02cm}a^{\phantom{\dagger}\!}_{l},
b^{\phantom{\dagger}\!}_{m}\hspace{0.02cm}],a^{\dagger}_{k}]
=
-2\hspace{0.02cm}\delta_{km}\hspace{0.02cm}b_{l}
-2\hspace{0.02cm}\delta_{kl}\hspace{0.02cm}b_{m}.
\]
However, the relation (\ref{eq:9s}) remains still unchanged. This fact can serve as a hint that a system of the Govorkov trilinear relations, in contrast to a system of the Greenberg-Messiah trilinear ones, in principle does not allow any reduction to more simple {\it bilinear} relations of the usual Fermi statistics even with a modification of the bilinear relations containing different fields. Thus, the question of a rule of decoupling the relation (\ref{eq:9u}) remains open.

\section{\bf Klein transformation}
\setcounter{equation}{0}

In the remarkable paper by Dr\"uhl, Haag, and Roberts \cite{druhl_1970} (see also \cite{cusson_1969}), a comprehensive construction of the so-called Klein transformation \cite{klein_1970, rosenfeld_1948, luders_1958, umezawa_1955, ohnuki_1984, ohnuki_1986} of Green's component of a para-Fermi field for an arbitrary order $p$ was given. This transformation has allowed to lead the initial relations (\ref{eq:2p}), which contain both commutators and anticommutators, to the normal commutation relations for $p$ ordinary Fermi fields. In particular, Dr\"uhl {\it et al.} have shown that to reduce the formalism to ordinary Fermi statistics, it is necessary to introduce $p\hspace{0.02cm}/2$ Klein operators $H_{2j},\,j=1,\ldots, p\hspace{0.02cm}/2$ for $p$ even, and $(p-1)/2$ those for $p$ odd. Thus, for parastatistics $p = 2, \hspace{0.02cm}3$ we need one Klein operator $H_{2}$, and for $p = 4$ we need already two Klein operators $H_{2}$ and $H_{4}$.\\
\indent
In the problem under consideration with two different para-Fermi fields $\phi_{a}$ and $\phi_{b}$ of order $p = 2$, it can be assumed that we require at least two Klein operators, which we designate as $H_{A}^{(2)}$ and $H_{B}^{(1)}$. The meaning of these notations will become clear just below. It is necessary to define the Klein transformation of Green's components $a_{k}^{(\alpha)}$ and $b_{m}^{(\alpha)}$ so that to reduce simultaneously to the normal form both the commutation relations (\ref{eq:2p}) separately for each set $\bigl\{a_{k}^{(\alpha)}\bigr\}$, $\bigl\{b_{m}^{(\alpha)}\bigr\}$ and the commutation relations (\ref{eq:3i})\,--\,(\ref{eq:3a}) of the mixed type. We state that the required Klein transformation has the form:
\begin{equation}
\begin{array}{lll}
&a^{(1)}_{k} = A^{(1)}_{k}H^{(2)}_{A}, \qquad\qquad &b^{(1)}_{m} = -i\hspace{0.01cm}B^{(1)}_{m}H^{(1)}_{B}, \\[1ex]
&a^{(2)}_{k} = i\hspace{0.01cm}A^{(2)}_{k}H^{(2)}_{A}, \qquad\qquad &b^{(2)}_{m} = B^{(2)}_{m}H^{(1)}_{B},
\end{array}
\label{eq:10q}
\end{equation}
where $A_k^{(\alpha)}$ and $B_m^{(\alpha)}$ are new Green's components satisfying the
following commutation rules with the Klein operators $(H_{A}^{(2)},\,H_{B}^{(1)})$:
\begin{equation}
\left\{\!\!\!\!\!
\begin{array}{lll}
&[\hspace{0.03cm} A^{(1)}_{k}\!,H^{(2)}_{A}\hspace{0.02cm}]  = 0, \qquad &\{A^{(1)}_{k}\!,H^{(1)}_{B}\} = 0,
\\[1.3ex]
&\{A^{(2)}_{k}\!,H^{(2)}_{A}\} = 0, \qquad
&[\hspace{0.03cm} A^{(2)}_{k}\!,H^{(1)}_{B}\hspace{0.02cm}]  = 0,
\end{array}
\right.
\qquad
\left\{\!\!\!\!\!
\begin{array}{lll}
&\{B^{(1)}_{m}\!,H^{(1)}_{B}\} = 0, \qquad &[\hspace{0.03cm} B^{(1)}_{m}\!,H^{(2)}_{A}\hspace{0.02cm}]  = 0,
\\[1.3ex]
&[\hspace{0.01cm} B^{(2)}_{m}\!,H^{(1)}_{B}\hspace{0.02cm}]  = 0, \qquad
&\{B^{(2)}_{m}\!,H^{(2)}_{A}\} = 0.
\end{array}
\right.
\label{eq:10w}
\end{equation}
At the same time the Klein operators themselves satisfy the conditions
\begin{equation}
\bigl(H^{(2)}_{A}\bigr)^{2} = \bigl(H^{(1)}_{B}\bigr)^{2} = I,\qquad
[\hspace{0.03cm} H^{(2)}_{A},H^{(1)}_{B}\hspace{0.02cm}] = 0.
\label{eq:10e}
\end{equation}
An explicit form of the Klein operators will be given below, and now, we shall simply show that the Klein transformation (\ref{eq:10q}) with the rules (\ref{eq:10w}) and (\ref{eq:10e}) actually furnishes the desired result.\\
\indent
It is easy to verify that the transformation (\ref{eq:10q}) leads to the normal form of the commutation relations (\ref{eq:2p}). This enables us to write them as follows:
\begin{equation}
\begin{array}{llll}
&\bigl\{A^{(\alpha)}_{k}\!,A^{\dagger(\alpha)}_{l\!}\bigr\} = \delta_{kl}\hspace{0.02cm},
\quad\quad
&\bigl\{A^{(\alpha)}_{k}\!,A^{\dagger(\beta)}_{l\!}\bigr\} = 0,
\quad\quad
&\bigl\{A^{(\alpha)}_{k}\!,A^{(\beta)}_{l\!}\bigr\} = 0,\quad\alpha\neq\beta
\\[1.3ex]
&\bigl\{B^{(\alpha)}_{m}\!,B^{\dagger(\alpha)}_{n\!}\bigr\} = \delta_{mn}\hspace{0.02cm},
\quad\quad
&\bigl\{B^{(\alpha)}_{m}\!,B^{\dagger(\beta)}_{n\!}\bigr\} = 0,
\quad\quad
&\bigl\{B^{(\alpha)}_{m}\!,B^{(\beta)}_{n\!}\bigr\} = 0.
\end{array}
\label{eq:10r}
\end{equation}
Therefore, we will restrict our attention to analysis of the system (\ref{eq:3i})\,--\,(\ref{eq:3a}). Let us consider the relation (\ref{eq:3i}) for $\alpha = 1$. By direct substitution of (\ref{eq:10q}) into (\ref{eq:3i}) with the help of (\ref{eq:10w}) and (\ref{eq:10e}), we get
\begin{equation}
[\hspace{0.02cm}b^{(1)}_{m},a^{\dagger(1)}_{k}\hspace{0.02cm}]
= i\hspace{0.01cm}H^{(2)}_{A}\bigl\{B^{(1)}_{m}\!,A^{\dagger(1)}_{k\!}\bigr\}H^{(1)}_{B}
= 2\hspace{0.02cm}\delta_{mk}\hspace{0.02cm}\Omega.
\label{eq:10t}
\end{equation}
The expression obtained suggests that the operator $\Omega$ must also be subjected to the Klein transformation. Thus, the Klein transformation (\ref{eq:10q}) need to be supplemented by the following rule:
\begin{equation}
\Omega = H^{(2)}_{A}\hspace{0.02cm}\widetilde{\Omega}\hspace{0.03cm}H^{(1)}_{B},
\label{eq:10y}
\end{equation}
where $\widetilde{\Omega}$ is a new operator. The relation similar to (\ref{eq:10t}) is valid for $\alpha = 2$ also. Therefore, instead of (\ref{eq:3i}), we now get
\begin{equation}
\bigl\{B^{(\alpha)}_{m}\!,A^{\dagger(\alpha)}_{k\!}\bigr\} =
\frac{2}{i}\,\delta_{mk}\hspace{0.03cm}\widetilde{\Omega}.
\label{eq:10u}
\end{equation}
Further, it can be easily verified that instead of (\ref{eq:3o}), we have
\begin{equation}
\bigl\{\!\hspace{0.03cm}A^{(\alpha)}_{k\!}, B^{(\alpha)}_{m\!}\bigr\} = 0,
\quad
\bigl\{\!\hspace{0.03cm}A^{\dagger(\alpha)}_{k\!\!}, B^{\dagger(\alpha)}_{m\!}\bigr\} = 0,
\label{eq:10i}
\end{equation}
and the relations (\ref{eq:3p}) with the use of (\ref{eq:10y}) transform to
\begin{equation}
\bigl\{\!\hspace{0.04cm}A^{(\alpha)}_{k\!}, \widetilde{\Omega}\hspace{0.02cm}\bigr\} = i\hspace{0.005cm}B^{(\alpha)}_{k},
\quad
\bigl\{\!\hspace{0.03cm}B^{(\alpha)}_{m\!}, \widetilde{\Omega}\hspace{0.02cm}\bigr\} = i\hspace{0.005cm}A^{(\alpha)}_{m}.
\label{eq:10o}
\end{equation}
Let us call attention to one point: on the right-hand sides of (\ref{eq:10o}) the signs are the same, in contrast to (\ref{eq:3p}). Finally, the anticommutation relations (\ref{eq:3a}) retain the form with the replacements $a_{k}^{(\alpha)}\rightarrow A_{k}^{(\alpha)}$ and $b_{m}^{(\alpha)} \rightarrow B_{m}^{(\alpha)}$.\\
\indent
Let us define now an explicit form of the Klein operators $H_{A}^{(2)}$ and $H_{B}^{(1)}$. For this purpose we rewrite the parafermion number operators, Eq.\,(\ref{eq:1u}), in terms of new Green's components $A_{k}^{(\alpha)}$ and $B_{m}^{(\alpha)}$. More specifically, we consider the particle-number operator for the $b$ para-Fermi oscillators. For $p = 2$ we have
\[
N_{b} =
\frac{1}{2}\hspace{0.03cm}\sum^{M}_{m\hspace{0.02cm}=1}
\bigl(\hspace{0.025cm}[\hspace{0.02cm}b^{\dagger(1)}_{m},b^{(1)}_{m}\hspace{0.02cm}]
\hspace{0.02cm}+\hspace{0.02cm}
[\hspace{0.02cm}b^{\dagger(2)}_{m},b^{(2)}_{m}\hspace{0.02cm}]\hspace{0.02cm}\bigr)
\hspace{0.02cm}+\hspace{0.02cm}M.
\]
Substituting the Klein transformation (\ref{eq:10q}) into the foregoing expression and taking into account (\ref{eq:10w})\,--\,(\ref{eq:10r}), we find that this parafermion number operator can be presented in the following form:
\[
N_{b} = N^{(1)}_{B} + N^{(2)}_{B},
\]
where
\begin{equation}
N^{(\alpha)}_{B} \equiv \frac{1}{2}\hspace{0.03cm}\sum^{M}_{m\hspace{0.02cm}=1}\;
[\hspace{0.02cm}B^{\dagger(\alpha)}_{m},B^{(\alpha)}_{m}\hspace{0.02cm}]
\hspace{0.02cm}+\hspace{0.02cm}\frac{1}{2}\hspace{0.03cm}M,
\quad\alpha = 1, 2
\label{eq:10p}
\end{equation}
is the particle-number operator of ordinary fermions. A similar representation is true for the $a$ para-Fermi oscillators also. An explicit form of the Klein operators $H_A^{(2)}, H_B^{(1)}$ is given by the following expressions:
\[
H^{(2)}_{A\!} = (-1)^{N^{(2)}_{A}},\quad H^{(1)}_{B\!} = (-1)^{N^{(1)}_{B}}.
\]
Most of the commutation relations in (\ref{eq:10w}) are apparently fulfilled. Only two of them require special consideration, namely,
\begin{equation}
\{A^{(1)}_{k}\!,H^{(1)}_{B}\} = 0\;\;\mbox{and}\;\; \{B^{(2)}_{m}\!,H^{(2)}_{A}\} = 0.
\label{eq:10a}
\end{equation}
Let us consider the first of them. Above all, we present the Klein operator $H_B^{(1)}$ in a more common way
\[
H^{(1)}_{B} = {\rm e}^{i\hspace{0.015cm}\pi N^{(1)}_{B}}.
\]
Then the relation under examination can be presented as
\begin{equation}
\{A^{(1)}_{k}\!,H^{(1)}_{B}\} =
\bigl({\rm e}^{i\hspace{0.015cm}\pi N^{(1)}_{B}\!} A^{(1)}_{k}
{\rm e}^{-i\hspace{0.015cm}\pi N^{(1)}_{B}}
+
A^{(1)}_{k}\bigr)\hspace{0.03cm}{\rm e}^{i\hspace{0.015cm}\pi N^{(1)}_{B}}.
\label{eq:10s}
\end{equation}
Further, making use of the identity (\ref{ap:B7}), we have
\begin{equation}
{\rm e}^{i\hspace{0.02cm}\pi N^{(1)}_{B}\!} A^{(1)}_{k}
{\rm e}^{-i\hspace{0.02cm}\pi N^{(1)}_{B}}
=
A^{(1)}_{k}
\,+\,
i\hspace{0.02cm}\pi\hspace{0.03cm}[\hspace{0.02cm}N^{(1)}_{B}, A^{(1)}_{k}\hspace{0.02cm}]
\,+\,
\frac{\!1}{2!}\,(i\hspace{0.02cm}\pi)^{2}\hspace{0.04cm}[\hspace{0.02cm}N^{(1)}_{B}, [\hspace{0.02cm}N^{(1)}_{B}, A^{(1)}_{k}\hspace{0.02cm}]\hspace{0.02cm}]
\,+\, \dots\,.
\label{eq:10d}
\end{equation}
By this means our proof has been limited to the calculation of the commutator $[\hspace{0.02cm}N^{(1)}_{B}, A^{(1)}_{k}\hspace{0.02cm}]$. Using the definition of the fermion number operator (\ref{eq:10p}) and identity (\ref{ap:B2}), we obtain
\[
[\hspace{0.02cm}N^{(1)}_{B}, A^{(1)}_{k}\hspace{0.02cm}]
=
\frac{1}{2}\hspace{0.02cm}\sum^{M}_{m\hspace{0.02cm}=1}\;
[\hspace{0.02cm}[\hspace{0.02cm}B^{\dagger(1)\!}_{m}, B^{(1)}_{m}\hspace{0.02cm}],
A^{(1)}_{k}\hspace{0.02cm}]
=
\frac{1}{2}\hspace{0.02cm}\sum^{M}_{m\hspace{0.02cm}=1}\left(\!\hspace{0.02cm}
\bigl\{\!\hspace{0.03cm}B^{\dagger(1)\!}_{m}, \bigl\{\!\hspace{0.03cm}A^{(1)\!}_{k}, B^{(1)}_{m}\bigr\}\!\bigr\}
-
\bigl\{\!\hspace{0.03cm}B^{(1)\!}_{m}, \bigl\{\!\hspace{0.03cm}A^{(1)\!}_{k}, B^{\dagger(1)}_{m}\bigr\}\!\hspace{0.01cm}\bigr\}\!\right).
\]
The first term on the right-hand side vanishes by virtue of (\ref{eq:10i}), and for the second term, in view of (\ref{eq:10u}) and (\ref{eq:10o}), we get
\[
\bigl\{\!\hspace{0.03cm}B^{(1)\!}_{m}, \bigl\{\!\hspace{0.03cm}A^{(1)\!}_{k}, B^{\dagger(1)}_{m}\bigr\}\!\hspace{0.01cm}\bigr\}
=
\frac{2}{i}\,\delta_{km}\hspace{0.02cm}
\bigl\{\!\hspace{0.03cm}B^{(1)}_{m\!}, \widetilde{\Omega}\hspace{0.02cm}\bigr\}
=
2\hspace{0.01cm}\delta_{km}\hspace{0.03cm}A^{(1)\!}_{m}.
\]
Thus, the desired commutator is equal to
\[
[\hspace{0.02cm}N^{(1)}_{B}, A^{(1)}_{k}\hspace{0.02cm}] = - A^{(1)}_{k}.
\]
Substituting the obtained expression into (\ref{eq:10d}), we further find
\[
{\rm e}^{i\hspace{0.015cm}\pi N^{(1)}_{B}\!} A^{(1)}_{k}
{\rm e}^{-i\hspace{0.015cm}\pi N^{(1)}_{B}}
=
A^{(1)}_{k}\Bigl( 1 - i\hspace{0.02cm}\pi \,+\,\frac{\!1}{2!}\hspace{0.03cm}(i\hspace{0.02cm}\pi)^{2}
- \frac{\!1}{3!}\hspace{0.03cm}(i\hspace{0.02cm}\pi)^{3} \,+\, \dots\,\Bigr)
\equiv
A^{(1)}_{k}{\rm e}^{-i\hspace{0.02cm}\pi} = -A^{(1)}_{k}.
\]
Thus, the right-hand side of the equality (\ref{eq:10s}) actually equals zero. The second relation in (\ref{eq:10a}) is proved by a similar way.

\section{\bf Lie-supertriple system}
\setcounter{equation}{0}

In this section we would like to discuss an interesting connection between the Govorkov trilinear relations (\ref{eq:3q})\,--\,(\ref{eq:3e}) and so-called Lie-supertriple system. The Lie-supertriple system, which is a generalization of the standard Lie-triple system \cite{jacobson_1949, lister_1952, sagle_1965, yamaguti_1969}, was studied in detail in the works of Okubo {\it et al.} \cite{okubo_1994, okubo_1997, kamiya_2000, kamiya_2016}. Our consideration will be based on the work \cite{okubo_1994}, in which the author has reformulated the parastatistics as a Lie-supertriple system. In \cite{okubo_1994} a number of examples of such a reformulation was presented. We are specially interested in the Example 3. The explicit form of this example will be given just bellow. A few more definitions are required first (in notations of the
paper \cite{okubo_1994}).\\
\indent
Let $V$ be a vector superspace, i.e. it represents a direct sum
\[
V = V_{B}\oplus V_{F}\hspace{0.02cm}.
\]
In this superspace the grade is entered by
\begin{equation}
\sigma(x) = \left\{\!\!\!\!\!\!\!
\begin{array}{lll}
&0, \qquad &\mbox{if}\; x\in V_{B}
\\[1ex]
&1, \qquad &\mbox{if}\; x\in V_{F}
\end{array}
\right.
\label{eq:11q}
\end{equation}
and the triple superproduct $[\hspace{0.02cm}...\hspace{0.02cm},\hspace{0.02cm}...
\hspace{0.02cm},\hspace{0.02cm}...\hspace{0.02cm}]$ is defined as a trilinear mapping
\[
[\hspace{0.02cm}...\hspace{0.02cm},\hspace{0.02cm}...
\hspace{0.02cm},\hspace{0.02cm}...\hspace{0.02cm}];\;V\otimes V\otimes V \rightarrow V.
\]
The triple superproduct is subject to three conditions, which can be found in \cite{okubo_1994}. If $V_f=0$, i.e. $V=V_B$, the triple superproduct $[x, y, z]$ reduces to the standard Lie triple system.\\
\indent
Besides, it is supposed that the underlying vector superspace $V$ always possesses a bilinear form $\langle x|\,y\rangle$ satisfying
\begin{equation}
\begin{array}{ll}
&\langle\hspace{0.02cm} x\hspace{0.02cm}|\hspace{0.04cm}y\rangle = (-1)^{\sigma(x)\sigma(y)} \langle\hspace{0.02cm} y\hspace{0.02cm}|\hspace{0.04cm}x\rangle,
\\[1ex]
&\langle\hspace{0.02cm} x\hspace{0.02cm}|\hspace{0.04cm}y\rangle = 0,\; \mbox{if}\; \sigma(x)\neq\sigma(y).
\end{array}
\label{eq:11w}
\end{equation}
\indent
We now give the formulation of Example 3 from \cite{okubo_1994}. Let $P\!:V\rightarrow V$ be a grade-preserving linear map in $V$, i.e.
\begin{equation}
\sigma(Px) = \sigma(x),\; \mbox{for any}\; x\in V
\label{eq:11e}
\end{equation}
and we assume the validity of
\begin{align}
&P^{\hspace{0.03cm}2} = \lambda\hspace{0.02cm}I, \label{eq:11r}\\[1ex]
&\langle\hspace{0.02cm} x\hspace{0.02cm}|\hspace{0.02cm}Py\rangle
=
-\langle\hspace{0.02cm} Px\hspace{0.01cm}|\hspace{0.03cm}y\hspace{0.02cm}\rangle,
\label{eq:11t}
\end{align}
where $I$ is the identity mapping in $V$ and $\lambda$ is nonzero constant. The following expression for the triple product:
\begin{equation}
[x,y,z] = \langle\hspace{0.02cm} y\hspace{0.02cm}|\hspace{0.03cm}Pz\hspace{0.02cm}\rangle Px
-
(-1)^{\sigma(x)\sigma(y)}\langle\hspace{0.02cm} x\hspace{0.02cm}|Pz\hspace{0.02cm}\rangle Py
-
2\hspace{0.02cm}\langle\hspace{0.02cm}x\hspace{0.02cm}|\hspace{0.03cm}Py\hspace{0.02cm}
\rangle Pz
+
\lambda\hspace{0.02cm}\langle y|\hspace{0.03cm}z\hspace{0.02cm}\rangle\hspace{0.02cm} x
-
(-1)^{\sigma(x)\sigma(y)\!}\lambda\hspace{0.02cm}\langle\hspace{0.02cm} x\hspace{0.02cm}|\hspace{0.03cm}z\hspace{0.02cm}\rangle\hspace{0.02cm} y
\label{eq:11y}
\end{equation}
transforms the superspace $V$ into a Lie-supertriple system with this triple product. Note that the same constant $\lambda$ enters in the condition (\ref{eq:11r}) and in the definition of the triple product (\ref{eq:11y}).\\
\indent
Let us show that the Govorkov trilinear relations (\ref{eq:3q})\,--\,(\ref{eq:3e}) represent particular cases of the general formula (\ref{eq:11y}). In addition, the triple product contains also the standard trilinear relations for the single field $\phi_{a}$ (and $\phi_{b}$),
Eqs.\,(\ref{eq:2q})\,--\,(\ref{eq:2e}). Our first step is to fix two sets of the operators
\[
(a^{\phantom{\dagger}\!}_{k},a^{\dagger}_{k})\;\; \mbox{and}\;\; (b^{\phantom{\dagger}\!}_{k},b^{\dagger}_{k}),\quad k = 1,\ldots,M
\]
between which we specify a map $P$ by the rule (cp. with (\ref{eq:5a}))
\begin{equation}
\begin{array}{lll}
&Pa_{k} = b_{k}, \quad\quad &P\hspace{0.01cm}b_{k} = -\hspace{0.02cm}a_{k},
\\[1ex]
&Pa^{\dagger}_{k} = b^{\dagger}_{k}, \quad\quad &P\hspace{0.01cm}b^{\dagger}_{k} = -\hspace{0.02cm}a^{\dagger}_{k}.
\end{array}
\label{eq:11u}
\end{equation}
It immediately follows that
\[
P^{2}a_{k} = -\hspace{0.02cm}a_{k}, \quad\quad P^{2}b_{k} = -\hspace{0.02cm}b_{k},
\]
and thus, by virtue of the condition (\ref{eq:11r}), the constant $\lambda$ is uniquely fixed:
\begin{equation}
\lambda = -1.
\label{eq:11i}
\end{equation}
Let us consider the second condition (\ref{eq:11t}). We set $x = a_{k}^{\dagger}$ and $y = b_{m}$, then, due to (\ref{eq:11u}), the condition for the bilinear form $\langle\cdot|\,\cdot\rangle $ reduces to
\begin{equation}
\langle\hspace{0.02cm} a^{\dagger}_{k}\hspace{0.02cm}|\hspace{0.04cm}a_{m}\hspace{0.02cm}\rangle
=
\langle\hspace{0.03cm} b^{\dagger}_{k}\hspace{0.02cm}|\hspace{0.04cm}b_{m}\hspace{0.02cm}\rangle.
\label{eq:11o}
\end{equation}
We fix the grade
\[
\sigma(a_{k}) = \sigma(a^{\dagger}_{k}) = 0,\qquad \sigma(b_{m}) = \sigma(b^{\dagger}_{m}) = 0,
\]
then the first condition in (\ref{eq:11w}) gives us
\[
\langle\hspace{0.02cm} x\hspace{0.02cm}|\hspace{0.04cm}y\hspace{0.02cm}\rangle
=
\langle\hspace{0.02cm} y\hspace{0.02cm}|\hspace{0.04cm}x\hspace{0.02cm}\rangle\;
\mbox{for any}\; x,y\in V.
\]
Thus, $V_{F} = 0$ and $V = V_{B}$. We choose the bilinear form
$\langle x|\hspace{0.04cm}y\rangle$ to satisfy
\begin{equation}
\begin{array}{llll}
&\langle\hspace{0.02cm} a^{\dagger}_{k}\hspace{0.02cm}|\hspace{0.03cm}a^{\phantom{\dagger}\!}_{m}\rangle
=
\langle\hspace{0.02cm} a^{\phantom{\dagger}\!}_{m}\hspace{0.02cm} |\hspace{0.03cm}a^{\dagger}_{k}\hspace{0.02cm}\rangle
=
-\hspace{0.02cm}2\hspace{0.02cm}\delta_{km},
\quad\quad
&\langle\hspace{0.03cm} b^{\dagger}_{k}|\hspace{0.04cm}b^{\phantom{\dagger}\!}_{m}\rangle =
\langle\hspace{0.03cm} b^{\phantom{\dagger}\!}_{m} |\hspace{0.04cm}b^{\dagger}_{k}\rangle =
-\hspace{0.02cm}2\hspace{0.02cm}\delta_{km},
\\[1ex]
&\langle\hspace{0.02cm} a^{\dagger}_{k}\hspace{0.02cm}|\hspace{0.03cm}a^{\dagger}_{m}\hspace{0.02cm}\rangle
=
\langle\hspace{0.02cm} a_{k}\hspace{0.02cm} |\hspace{0.04cm}a_{m}\hspace{0.02cm}\rangle = 0,
\quad\quad
&\langle\hspace{0.03cm} b^{\dagger}_{k}\hspace{0.02cm}|\hspace{0.03cm}b^{\dagger}_{m}\hspace{0.02cm}\rangle
=
\langle\hspace{0.03cm} b_{k}\hspace{0.02cm} |\hspace{0.03cm}b_{m}\hspace{0.02cm}\rangle
= 0,\\[1ex]
&\langle\hspace{0.02cm} a^{\dagger}_{k}\hspace{0.02cm}|\hspace{0.03cm}b^{\phantom{\dagger}\!}_{m}
\hspace{0.02cm}\rangle
=
\langle\hspace{0.02cm} a_{k}\hspace{0.02cm} |\hspace{0.03cm}b_{m}\hspace{0.02cm}\rangle = 0,
\quad\quad
&\langle\hspace{0.03cm} b^{\dagger}_{k}\hspace{0.02cm}|\hspace{0.03cm}a^{\phantom{\dagger}\!}_{m}\hspace{0.02cm}
\rangle
=
\langle\hspace{0.03cm} b^{\dagger}_{k}\hspace{0.02cm}|\hspace{0.03cm}a^{\dagger}_{m}\hspace{0.02cm}\rangle = 0.
\end{array}
\label{eq:11p}
\end{equation}
The condition (\ref{eq:11o}) is automatically satisfied.\\
\indent
We now return to the main relation (\ref{eq:11y}) and set $x = a_{k}^{\dagger},\,y = a_{l}$, and
$z = b_{m}$. Then, by virtue of (\ref{eq:11u}), (\ref{eq:11i}) and (\ref{eq:11p}), we have:
\[
[\hspace{0.02cm}a^{\dagger}_{k}, a^{\phantom{\dagger}\!}_{l}, b^{\phantom{\dagger}\!}_{m}] =
-\hspace{0.02cm}\langle\hspace{0.02cm} a^{\phantom{\dagger}\!}_{l}\hspace{0.02cm} |\hspace{0.04cm}a^{\phantom{\dagger}\!}_{m}\rangle\hspace{0.02cm} b^{\dagger}_{k}
+
\langle\hspace{0.02cm} a^{\dagger}_{k}\hspace{0.02cm}|\hspace{0.03cm}
a^{\phantom{\dagger}\!}_{m}\hspace{0.02cm}\rangle\hspace{0.02cm}
b^{\phantom{\dagger}\!}_{l}
+
2\hspace{0.02cm}\langle\hspace{0.02cm} a^{\dagger}_{k}\hspace{0.02cm}|\hspace{0.03cm}
b^{\phantom{\dagger}\!}_{l}\hspace{0.02cm}\rangle\hspace{0.02cm}
a^{\phantom{\dagger}\!}_{m}
-
\langle\hspace{0.02cm} a^{\phantom{\dagger}\!}_{l}\hspace{0.02cm}|\hspace{0.03cm}
b^{\phantom{\dagger}\!}_{m}\hspace{0.02cm}\rangle\hspace{0.03cm}
a^{\dagger}_{k}
+
\langle\hspace{0.02cm} a^{\dagger}_{k}\hspace{0.02cm}|\hspace{0.03cm}
b^{\phantom{\dagger}\!}_{m}\hspace{0.02cm}\rangle\hspace{0.03cm}
a^{\phantom{\dagger}\!}_{l}
=
-\hspace{0.02cm}2\hspace{0.02cm}
\delta^{\phantom{\dagger}\!}_{km}\hspace{0.02cm}b^{\phantom{\dagger}\!}_{l}.
\]
Thus, we reproduce the trilinear relation (\ref{eq:3e}). Further, if we set $x = b_{m},\,y = a_{k}^{\dagger}$, and $z = a_{l}$, then the triple product (\ref{eq:11y}) will take the form
\[
[\hspace{0.03cm}b^{\phantom{\dagger}\!}_{m}, a^{\dagger}_{k}, a^{\phantom{\dagger}\!}_{l}]
=
-\hspace{0.02cm}\langle\hspace{0.02cm} a^{\dagger}_{k}\hspace{0.02cm} |\hspace{0.03cm}b^{\phantom{\dagger}\!}_{l}\hspace{0.02cm}\rangle\hspace{0.02cm} a^{\phantom{\dagger}\!}_{m}
-
\langle\hspace{0.03cm} b^{\phantom{\dagger}\!}_{m}\hspace{0.02cm}|\hspace{0.03cm}
b^{\phantom{\dagger}\!}_{l}\hspace{0.02cm}\rangle
\hspace{0.02cm}b^{\dagger}_{k}
-
2\hspace{0.02cm}\langle\hspace{0.03cm}b^{\phantom{\dagger}\!}_{m}\hspace{0.02cm}|
\hspace{0.04cm}
b^{\dagger}_{k}\hspace{0.02cm}\rangle\hspace{0.02cm}b^{\phantom{\dagger}\!}_{l}
-
\langle\hspace{0.02cm} a^{\dagger}_{k}\hspace{0.02cm}|\hspace{0.03cm}
a^{\phantom{\dagger}\!}_{l}\hspace{0.02cm}\rangle\hspace{0.03cm}
b^{\phantom{\dagger}\!}_{m}
+
\langle\hspace{0.03cm}
b^{\phantom{\dagger}\!}_{m}\hspace{0.02cm}|\hspace{0.03cm}
a^{\phantom{\dagger}\!}_{l}\rangle\hspace{0.03cm}
a^{\dagger}_{k}
\]
\[
=
4\hspace{0.015cm}\delta_{mk}\hspace{0.02cm}b_{l}
+
2\hspace{0.015cm}\delta_{kl}\hspace{0.02cm}b_{m},
\]
that gives us the relation (\ref{eq:3q}). It is not difficult to verify that for $x = a_{l},\, y = b_{m}$, and $z = a_{k}^{\dagger}$ we reproduce (\ref{eq:3w}). One can state that with the rules (\ref{eq:11u}), (\ref{eq:11i}) and (\ref{eq:11p}) all of the Govorkov trilinear relations are contained in an unique formula (\ref{eq:11y}).\\
\indent
To complete our analysis, we consider the triple product for one set of operators, for example, for $(a_{k},\,a_{k}^{\dagger})$. Let $x = a_{k}^{\dagger},\,y = a_{l}$, and $z = a_{m}$, then from (\ref{eq:11y}) we have
\[
[\hspace{0.03cm}a^{\dagger}_{k}, a^{\phantom{\dagger}\!}_{l}, a^{\phantom{\dagger}\!}_{m}\hspace{0.02cm}]
=
-\hspace{0.02cm}\langle\hspace{0.02cm} a^{\phantom{\dagger}\!}_{l}\hspace{0.02cm} |\hspace{0.03cm}a^{\phantom{\dagger}\!}_{m}\hspace{0.02cm}\rangle\hspace{0.02cm} a^{\dagger}_{k}
+
\langle\hspace{0.02cm} a^{\dagger}_{k}\hspace{0.02cm}|\hspace{0.03cm}
a^{\phantom{\dagger}\!}_{m}\rangle\hspace{0.02cm}a^{\phantom{\dagger}\!}_{l}
=
-\hspace{0.02cm}2\hspace{0.02cm}
\delta^{\phantom{\dagger}\!}_{km}\hspace{0.02cm}a^{\phantom{\dagger}\!}_{l}.
\]
We see that the triple product with the rules (\ref{eq:11u}), (\ref{eq:11i}), and (\ref{eq:11p}) correctly reproduces the standard trilinear relations of a para-Fermi statistics. This special case was considered by Okubo \cite{okubo_1994} as Example 2 for the triple product
\begin{equation}
[\hspace{0.02cm}x,y,z\hspace{0.02cm}] = \lambda\hspace{0.02cm}
\bigl\{\langle\hspace{0.02cm} y\hspace{0.02cm}|\hspace{0.03cm}z\hspace{0.02cm}\rangle x
-
(-1)^{\sigma(x)\sigma(y)\!}\lambda\hspace{0.02cm}\langle\hspace{0.02cm} x\hspace{0.02cm}|\hspace{0.03cm}z\hspace{0.02cm}\rangle y\bigr\}.
\label{eq:11a}
\end{equation}
In fact the triple product (\ref{eq:11a}) represents two last terms in (\ref{eq:11y}). However, in our case there are two important distinction from the Okubo case. We have fixed the constant $\lambda$ and the bilinear form as follows:
\[
\lambda = -1,\qquad \langle\hspace{0.02cm}a^{\dagger}_{k}\hspace{0.02cm}|
\hspace{0.03cm}a^{\phantom{\dagger}\!}_{l}\hspace{0.02cm}\rangle
=
\langle\hspace{0.02cm}a^{\phantom{\dagger}\!}_{l}\hspace{0.02cm} |\hspace{0.03cm}a^{\dagger}_{k}\hspace{0.02cm}\rangle
= -\hspace{0.02cm}2\hspace{0.015cm}\delta^{\phantom{\dagger}\!}_{kl},
\]
while Okubo has made this somewhat different:
\[
\lambda = 2,\qquad\;\;\;
\langle\hspace{0.02cm}a^{\dagger}_{k}\hspace{0.02cm}|
\hspace{0.03cm}a^{\phantom{\dagger}\!}_{l}\hspace{0.02cm}\rangle
=
\langle\hspace{0.02cm} a^{\phantom{\dagger}\!}_{l}\hspace{0.02cm} |\hspace{0.03cm}a^{\dagger}_{k}\hspace{0.02cm}\rangle
= \delta^{\phantom{\dagger}\!}_{kl}.
\hspace{0.55cm}
\]
On both cases the triple product (\ref{eq:11a}) correctly reproduces the relations for the para-Fermi statistics, however, in the last case the Govorkov relations are not reproduced.

\section{Connection with the Duffin-Kemmer-Petiau formalism}
\setcounter{equation}{0}
\label{section_12}

In our paper \cite{markov_2015} we have obtained the Fock-Schwinger proper-time representation for the inverse operator $\hat{\cal L}^{-1}$:
\begin{equation}
\frac{1}{\hat{\cal L}} \equiv \frac{\hat{\cal L}^{2}}{\hat{\cal L}^{3}} =
i\!\int\limits_{0}^{\infty}\!d\hspace{0.03cm}T\!
\int\!\frac{d^{\,2}\chi}{T^{2}}\;\hspace{0.02cm}
{\rm e}^{\displaystyle{-\hspace{0.02cm}i\hspace{0.02cm}T\bigl (\hat{H}(z) - i\hspace{0.01cm}\epsilon\hspace{0.02cm}\bigr) +
\frac{1}{2}\,\bigl(\hspace{0.02cm}T\hspace{0.02cm}[\hspace{0.03cm}\chi,\hat{\cal L}\hspace{0.02cm}] + \frac{1}{4}\,T^{2\,}[\hspace{0.03cm}\chi,\hat{\cal L}\hspace{0.02cm}]^{\hspace{0.02cm}2}\hspace{0.03cm}\bigr)}},
\;
\epsilon\rightarrow +\hspace{0.01cm}0,
\label{eq:12q}
\end{equation}
where
\[
\hat{\cal L} \equiv \hat{\cal L}(z,D) =
A\hspace{0.02cm}\biggl(\frac{\!i}{\,\varepsilon^{1/3}(z)}\,\eta_{\mu}(z)
\hspace{0.03cm}D_{\mu} + m\hspace{0.02cm}I\biggr)
\]
and
\[
\hat{H}(z) \equiv \hat{\cal L}^{\hspace{0.01cm}3}(z,D)\]
is the Hamilton operator; $D_{\mu} = \partial_{\mu} + ieA_{\mu}(x)$ is the covariant derivative and $\chi$ is a para-Grassmann variable of order $p = 2$ (i.e. $\chi^{3} = 0$) with the rules of an integration \cite{omote_1979}:
\[
\int\!d^{\,2}\chi  = 0 = \int\!d^{\,2}\chi\,[\hspace{0.03cm}\chi,\hat{\cal L}\hspace{0.03cm}],  \quad
\int\!d^{\,2}\chi\,[\hspace{0.03cm}\chi,\hat{\cal L}\hspace{0.03cm}]^{\,2} =
4\hspace{0.02cm}\hat{\cal L}^{2}.
\]
The operator $\hat{\cal L}(z, D)$ represents the cubic root of some third order wave operator in an external electromagnetic field. Further, the matrices $\eta_{\mu}(z)$ are defined by the matrices $\beta_{\mu}$ of the Duffin-Kemmer-Petiau (DKP) algebra and by the complex deformation parameter $z$ as follows:
\[
\eta_{\mu}(z) \equiv \biggl(1 + \frac{1}{2}\,z\biggr)\beta_{\mu} - z\hspace{0.03cm}\biggl(\frac{\sqrt{3}}{2}\biggr)\hspace{0.02cm}\hspace{0.02cm}\zeta_{\mu},
\]
where
\begin{equation}
\zeta_{\mu} = i\hspace{0.04cm}[\hspace{0.03cm}\beta_{\mu}, \omega\hspace{0.03cm}]
\label{eq:12w}
\end{equation}
and
\begin{equation}
\omega = \frac{\!\!1}{(M!)^{\hspace{0.02cm}2}}\; \epsilon_{\mu_{1}\mu_{2}\ldots\hspace{0.02cm}\mu_{2M}}
\beta_{\mu_{1}}\beta_{\mu_{2}}\ldots\beta_{\mu_{2M}}.
\label{eq:12e}
\end{equation}
At the end of all calculations, it should be necessary to passage to the limit $z \rightarrow q$ ($q$ is a primitive cubic root of unity).\\
\indent
One of the main reasons of appearance of the present work was a hope to develop a convenient mathematical technique, which would enable us within the framework of the DKP approach to construct the path integral representation in parasuperspace for the inverse operator
$\hat{\cal L}^{-1}(z, D)$, Eq.\,(\ref{eq:12q}). Matrix element of the operator $\hat{\cal L}^{-1}(z, D)$ in the corresponding basis of states can be considered as the propagator of a massive vector particle in a background gauge field. Unfortunately, Govorkov's unitary quantization formalism has proved to be unsuitable for this purpose. Below we shall discuss this problem in more detail.\\
\indent
Our first step is to compute the commutator $[\zeta_{\mu}, \omega]$, where $\zeta_{\mu}$ is defined by Eq.\,(\ref{eq:12w}). With the aid of the algebraic relations\footnote{\,All the basic formulae of the $\omega-\beta_{\mu}$ matrix algebra for the spin-1 can be found in Appendix A of our paper \cite{markov_2015}.}
\begin{equation}
\omega^{2\!}\hspace{0.035cm}\beta_{\mu} + \beta_{\mu\,}\omega^2 = \beta_{\mu}
\quad\mbox{and}\quad
\omega\beta_{\mu\,}\omega = 0,
\label{eq:12r}
\end{equation}
it is easily derived that
\begin{equation}
[\hspace{0.03cm}\zeta_{\mu}, \omega\hspace{0.03cm}] = i\hspace{0.01cm}\beta_{\mu}.
\label{eq:12t}
\end{equation}
Confronting the expressions (\ref{eq:12w}) and (\ref{eq:12t}) with (\ref{ap:A17}), one finds that the easiest way to establish a connection between the DKP theory and the unitary quantization scheme is to identify literally the matrices $\beta_{\mu}$ and $\zeta_{\mu}$ from the DKP approach with the quantities $\beta_{\mu}$ and $\zeta_{\mu}$ that appear within uniquantization, Eq.\,(\ref{ap:A6}). It is evident from this identification that
\begin{equation}
\omega \equiv -\frac{1}{2}\,\zeta_{0}.
\label{eq:12y}
\end{equation}
Now it is necessary to verify whether the relations (\ref{ap:A11})\,--\,(\ref{ap:A16}) will be satisfied if we stay only within the framework of the DKP formalism. At first, we consider the bilinear relations (\ref{ap:A15}) and  (\ref{ap:A16}). For the former relation, taking into account (\ref{eq:12r}), we have
\[
[\hspace{0.02cm}\zeta_{\mu},\zeta_{\nu}]
=
- [\hspace{0.02cm}[\hspace{0.03cm}\beta_{\mu}, \omega\hspace{0.03cm}],
[\hspace{0.03cm}\beta_{\nu}, \omega\hspace{0.03cm}]\hspace{0.02cm}]
=
[\hspace{0.03cm}\omega,
[\hspace{0.02cm}\beta_{\mu},\beta_{\nu}\hspace{0.03cm}]\hspace{0.02cm}]
\hspace{0.02cm}\omega
\hspace{0.02cm}+\hspace{0.02cm}
[\hspace{0.02cm}\beta_{\mu},\beta_{\nu}\hspace{0.02cm}].
\]
By virtue of the $\omega-\beta_{\mu}$ algebra the following equality:
\begin{equation}
[\hspace{0.03cm}\omega,
[\hspace{0.02cm}\beta_{\mu},\beta_{\nu}\hspace{0.03cm}]\hspace{0.02cm}] =0
\label{eq:12u}
\end{equation}
takes place and, thus, we arrive at (\ref{ap:A15}). For the bilinear relation (\ref{ap:A16}) with the use of Jacobi's identity and (\ref{eq:12u}), we get
\begin{equation}
[\hspace{0.02cm}\zeta_{\mu},\beta_{\nu}\hspace{0.02cm}] = -\hspace{0.02cm}i\hspace{0.04cm}[\hspace{0.02cm}
[\hspace{0.03cm}\omega,\beta_{\mu}\hspace{0.03cm}],
\beta_{\nu}\hspace{0.02cm}]
=
i\hspace{0.04cm}\bigl(\hspace{0.03cm}[\hspace{0.02cm}
[\hspace{0.02cm}\beta_{\mu},\beta_{\nu}\hspace{0.03cm}],\omega\hspace{0.02cm}]
\hspace{0.02cm}+\hspace{0.02cm}
[\hspace{0.02cm}
[\hspace{0.02cm}\beta_{\nu},\omega\hspace{0.03cm}], \beta_{\mu}\hspace{0.02cm}]\hspace{0.02cm}\bigr)
=
[\hspace{0.02cm}\zeta_{\nu},\beta_{\mu}].
\label{eq:12i}
\end{equation}
where we also observe a perfect coincidence. However, the relation (\ref{eq:12i}) within the DKP theory has actually more ``weak'' form. Indeed, let us consider once more the bilinear relation (\ref{ap:A16}) without resorting to Jacobi's identity at this time. Taking into account the relations
\begin{align}
\omega\hspace{0.015cm}\beta_{\mu}\beta_{\nu}
&+ \beta_{\nu}\beta_{\mu\,}\omega = \omega\hspace{0.02cm}\delta_{\mu\nu}, \notag\\[1ex]
\beta_{\mu}\hspace{0.02cm}\omega\hspace{0.015cm}\beta_{\nu}
&+ \beta_{\nu\,}\omega\beta_{\mu} = 0,
\notag
\end{align}
we have
\[
[\hspace{0.02cm}\zeta_{\mu},\beta_{\nu}\hspace{0.02cm}] = -i\hspace{0.04cm}[\hspace{0.02cm}[\hspace{0.03cm}\omega,\beta_{\mu}\hspace{0.03cm}],
\beta_{\nu}\hspace{0.02cm}]
\equiv
-\hspace{0.02cm}i\hspace{0.03cm}\bigl(\hspace{0.03cm}
\omega\hspace{0.015cm}\beta_{\mu}\beta_{\nu} + \beta_{\nu}\beta_{\mu}\hspace{0.03cm}\omega
-
\beta_{\mu}\hspace{0.03cm}\omega\beta_{\nu} - \beta_{\nu}\hspace{0.03cm}\omega\beta_{\mu}
\hspace{0.02cm}\bigr)
=
-\hspace{0.02cm}i\hspace{0.03cm}\omega\hspace{0.015cm}\delta_{\mu\nu}
\]
and thus, instead of (\ref{ap:A16}), we find
\begin{equation}
[\hspace{0.02cm}\zeta_{\mu},\beta_{\nu}] = [\hspace{0.02cm}\zeta_{\nu},\beta_{\mu}]
=
-\hspace{0.02cm}i\hspace{0.03cm}\omega\hspace{0.02cm}\delta_{\mu\nu}.
\label{eq:12o}
\end{equation}
It is precisely this circumstance that has negative consequence for the trilinear relations which is now under consideration.\\
\indent
The trilinear relation (\ref{ap:A11}) is clearly satisfied by virtue of the DKP algebra
\begin{equation}
\beta_{\mu}\beta_{\nu}\beta_{\lambda} + \beta_{\lambda}\beta_{\nu}\beta_{\mu} =
\delta_{\mu\nu}\beta_{\lambda} + \delta_{\lambda\nu}\beta_{\mu}.
\label{eq:12p}
\end{equation}
The relation (\ref{ap:A12}) also holds, since the same algebra (\ref{eq:12p}) is true for the matrices $\zeta_{\mu}$. Let us now consider the mutual commutation relations between
$\zeta_{\mu}$ and $\beta_{\mu}$. On the strength of (\ref{eq:12o}) and (\ref{eq:12t}), we now
have for (\ref{ap:A13})
\[
[\hspace{0.02cm}\zeta_{\lambda},[\hspace{0.02cm}\zeta_{\mu},\beta_{\nu}]\hspace{0.02cm}]
=
-\hspace{0.02cm}i\hspace{0.03cm}\delta_{\mu\nu}\hspace{0.015cm}
[\hspace{0.03cm}\zeta_{\lambda}, \omega\hspace{0.03cm}]
\equiv
\delta_{\mu\nu}\hspace{0.02cm}\beta_{\lambda},
\]
but there should be
\[
\hspace{0.4cm}
[\hspace{0.02cm}\zeta_{\lambda},[\hspace{0.02cm}\zeta_{\mu},\beta_{\nu}]\hspace{0.02cm}]
=
2\hspace{0.015cm}\delta_{\mu\nu}\beta_{\lambda} +
\delta_{\lambda\nu}\beta_{\mu} + \delta_{\lambda\mu}\beta_{\nu}.
\]
Thus, there is a significant difference between the right-hand sides of these commutators. The trilinear relation (\ref{ap:A13}), just as (\ref{ap:A14}), is not satisfied.\\
\indent
Inconsistence between the unitary quantization scheme and the DKP theory can be also seen if
one considers the relation (\ref{ap:A10}) in which $\zeta_0$ is replaced by $\omega$ in
accordance with the rule (\ref{eq:12y}). Then, taking into account (\ref{eq:12o}), we get
\begin{equation}
\omega = \frac{i}{2\hspace{0.02cm}(2\hspace{0.01cm}M + 1)}\,
\sum^{2M}_{\mu\hspace{0.02cm}=1}\,[\hspace{0.02cm}\zeta_{\mu},\beta_{\mu}\hspace{0.02cm}]
=
\frac{1}{2\hspace{0.02cm}(2\hspace{0.01cm}M + 1)}\;\omega\,
\sum^{2M}_{\mu\hspace{0.02cm}=1}\,\delta_{\mu\mu}
=
\frac{M}{2\hspace{0.01cm}M + 1}\,\omega.
\label{eq:12a}
\end{equation}
Here, we see an apparent contradiction. We can summarize these considerations with the statement that in spite of a close similarity between these two formalisms, the scheme of quantization based on the Duffin-Kemmer-Petiau theory does not embed into the scheme of the unitary quantization suggested by Govorkov.\\
\indent
But, there is one more possibility connected with parafermion quantization in accordance with the Lie algebra of the orthogonal group $SO(2M+2)$. Such a quantization was studied in due time by Geyer \cite{geyer_1968} (see also Fukutome \cite{fukutome_1981}). Here, a very important circumstance is that for the case of the group $SO(2M+2)$ some additional operator denoted in \cite{geyer_1968} as $a_{0}$, arises. This operator should be considered as an analogue of the operator $\zeta_{0}$, Eq.\,(\ref{eq:1i}). Unfortunately, in distinction from the unitary quantization scheme, for the group $SO(2M + 2)$ we have only one set of the operators $(a^{\phantom{\dagger}\!}_{k},\hspace{0.02cm}a_{k}^{\dagger})$ that are associated with the initial quantities $\beta_{\mu}$ by the relations (\ref{ap:A18}). Nevertheless, in such a situation, we can simply introduce by hand the second set of the desired operators $(b^{\phantom{\dagger}\!}_{k},\hspace{0.02cm}b_{k}^{\dagger})$ by
setting\footnote{\,We redefine the operators in \cite{geyer_1968}  for our case as follows:
\[
a_{0}\rightarrow 2\hspace{0.015cm}i\hspace{0.015cm}a_{0},
\quad
a_{k}\rightarrow \sqrt{2}\hspace{0.02cm}a_{k},
\quad
b_{m}\rightarrow 2\hspace{0.015cm}\sqrt{2}\hspace{0.03cm}i\hspace{0.025cm}b_{m}.
\]
}
\[
b_{k}\equiv [\hspace{0.02cm}a_{0}, a_{k}\hspace{0.02cm}],
\quad
b^{\dagger}_{k}\equiv [\hspace{0.02cm}a^{\phantom{\dagger}\!}_{0}, a^{\dagger}_{k}\hspace{0.02cm}],
\quad
a^{\dagger}_{0} = -\hspace{0.02cm}a^{\phantom{\dagger}\!}_{0}.
\]
In this case, the trilinear relations in \cite{geyer_1968} containing the operator $a_0$ take a familiar form
\begin{align}
&[\hspace{0.02cm}a^{\dagger}_{k}, b^{\phantom{\dagger}\!}_{m}\hspace{0.02cm}] = -\hspace{0.02cm}2\hspace{0.015cm}\delta^{\phantom{\dagger}\!}_{km}
\hspace{0.02cm}a^{\phantom{\dagger}\!}_{0},
\quad
[\hspace{0.02cm}b^{\dagger}_{m},a^{\phantom{\dagger}\!}_{k}\hspace{0.02cm}] = 2\hspace{0.015cm}\delta_{mk}\hspace{0.02cm}a^{\phantom{\dagger}\!}_{0},
\notag\\[0.8ex]
&[\hspace{0.02cm}a_{0}, b_{m}\hspace{0.02cm}] = -\hspace{0.025cm}a_{m},
\quad
[\hspace{0.02cm}a^{\phantom{\dagger}\!}_{0}, b^{\dagger}_{m}\hspace{0.02cm}] = -\hspace{0.025cm}a^{\dagger}_{m}, \notag\\[0.8ex]
&[\hspace{0.02cm}a_{k},b_{m}\hspace{0.02cm}] = 0,
\qquad\hspace{0.02cm}
[\hspace{0.02cm}a^{\dagger}_{k},b^{\dagger}_{m}\hspace{0.02cm}] = 0
\notag
\end{align}
and, in particular, we have
\begin{equation}
a_{0} = -\frac{\!1}{4M}\,
\sum^{M}_{k\hspace{0.02cm}=1}\,\Bigl(
[\hspace{0.02cm}a^{\dagger}_{k},b^{\phantom{\dagger}\!}_{k}\hspace{0.02cm}] + [\hspace{0.02cm}b^{\dagger}_{k},a^{\phantom{\dagger}\!}_{k}\hspace{0.02cm}]\Bigr).
\label{eq:12s}
\end{equation}
Furthermore, in the paper \cite{geyer_1968} acting the operator $a_0$ on the vacuum state
(cp. with (\ref{eq:7tt}))
\[
a_{0}|\hspace{0.03cm}0\rangle = \pm\hspace{0.03cm}\frac{i}{2}\,p\hspace{0.03cm}|\hspace{0.03cm}0\rangle
\]
was also defined. Attention should be paid to the fact that the expression on the right-hand side in Eq.\,(\ref{eq:12s}) has a different factor in front of the summation sign in contrast to (\ref{eq:1i}). This enables us to get rid of the contradiction (\ref{eq:12a}), when we identify the operator $a_{0}$ with the operator $\omega$ from the DKP theory. It should be considered that the quantities $\zeta_{\mu}$ are connected with the operators $(b^{\phantom{\dagger}\!}_{k},\hspace{0.02cm}b_{k}^{\dagger})$ through the relations (\ref{ap:A18}). In this case, a direct consequence of the expression (\ref{eq:12w}) is
\[
a_{0}\equiv -\hspace{0.02cm}i\hspace{0.03cm}\omega.
\]
All of these questions related to parafermion quantization based on the orthogonal group $SO(2M+2)$ and an important connection between this quantization scheme and the Duffin-Kemmer-Petiau theory will be considered in detail in our next work. In the same work, on the basis of this connection the construction of the path integral representation for the inverse operator $\hat{\cal L}^{-1}$, Eq.\,(\ref{eq:12q}), will be given.

\section{Conclusion}
\setcounter{equation}{0}
\label{section_10}

In this paper we have considered various aspects of a connection between the unitary quantization and parastatistics. In the analysis of the connection, the primary emphasis has been placed on the use of the Green decomposition of the creation and annihilation operators, and also para-Grassmann numbers. It was found that a system of the commutation relations derived by Govorkov within the framework of uniquantization is very severy, since it has been possible to associate this system only with a particular case of parastatistics, namely, with the para-Fermis statistics of order 2.  However, even so, we needed to introduce a number of additional assumptions and a new operator $\Omega$ (section 3). It should be noted also that in the papers \cite{govorkov_1979, govorkov_1980} the case of an odd number of dimensions, i.e. the unitary group $SU(2M)$, was also considered. Govorkov has shown that the Lie algebra of the unitary group contains the Lie algebra of the symplectic group $Sp\hspace{0.02cm}(2M)$ and also the other operators that complete it to the Lie algebra of the original group $SU(2M)$. As is known \cite{kamefuchi_1962}, the quantization in accordance with the Lie algebra of the symplectic group $Sp\hspace{0.02cm}(2M)$ corresponds to paraboson quantization. By doing so, one can state a similar task of the connection between the unitary quantization scheme based on the Lie algebra of the unitary group $SU(2M)$  and para-Bose statistics.\\
\indent
In this concluding section, however, we would like to discuss in a little more detail a secondary consequence of the constructions presented in this work (section 6), on which very little light was shed because of the large number of the formulae. It deals with the para-Fermi statistics of order 2 itself and is not specific to the unitary quantization scheme. It turns out that certain trilinear relations containing both the operators $a_{k}$ (or $b_{m}$) and the para-Grassmann numbers $\xi_{k}$ have another equivalent (dual?) representation. This can be seen on the example of the relations of the type
\begin{equation}
[\hspace{0.03cm}\xi_{k}, [\hspace{0.02cm}\xi_{l},a_{m}\hspace{0.02cm}]\hspace{0.02cm}] = 0
\quad \mbox{and} \quad
[\hspace{0.03cm}a^{\phantom{\dagger}\!}_{k}, [\hspace{0.02cm}a^{\dagger}_{l},\xi^{\phantom{\dagger}\!}_{m}\hspace{0.02cm}]\hspace{0.02cm}]
=
2\hspace{0.01cm}\delta^{\phantom{\dagger}\!}_{kl}\hspace{0.02cm}\xi^{\phantom{\dagger}\!}_{m},
\label{eq:13q}
\end{equation}
for which the dual representations have the following form:
\begin{equation}
\{\hspace{0.01cm}\xi_{k}, \{\hspace{0.02cm}\xi_{l}, a_{m}\}\} = 0
\quad \mbox{and} \quad
\{\hspace{0.02cm}a^{\phantom{\dagger}\!}_{k}, \{\hspace{0.02cm}a^{\dagger}_{l},\xi^{\phantom{\dagger}\!}_{m}\}\!\hspace{0.02cm}\}
=
2\hspace{0.01cm}\delta^{\phantom{\dagger}\!}_{kl}\hspace{0.02cm}\xi^{\phantom{\dagger}\!}_{m}.
\label{eq:13w}
\end{equation}
All these relations turn into identity with the help of a {\it commonly} used commutation rules for the Green components of the operators $a_k$ and para-Grassmann numbers $\xi_{k}$, Eqs.\,(\ref{eq:2p}) and (\ref{eq:5w}).\\
\indent
Further, in section 8 we have shown that a consequence of Eqs.\,(\ref{eq:13q}) and (\ref{eq:13w})
is an existence of two alternative definitions of the parafermion coherent state, namely,
\[
|\hspace{0.02cm}(\xi)_{2}\hspace{0.02cm}\rangle = \exp\Bigl(-\frac{1}{2}\sum^{M}_{l\hspace{0.02cm} =
\hspace{0.02cm}1}\,
[\hspace{0.02cm}\xi^{\phantom{\dagger}\!}_{l}, a^{\dagger}_{l}\hspace{0.02cm}]\Bigr)|\hspace{0.03cm}0\rangle
\]
and
\[
|\hspace{0.02cm}(\xi)_{2}\hspace{0.02cm}\rangle = \exp\Bigl(\hspace{0.02cm}\frac{1}{2}\sum^{M}_{l\hspace{0.02cm} =
\hspace{0.02cm}1}\,
\{\hspace{0.02cm}\xi^{\phantom{\dagger}\!}_{l}, a^{\dagger}_{l}\hspace{0.02cm}\}\Bigr)|\hspace{0.03cm}0\rangle.
\]
In both cases the main property of the coherent state is fulfilled
\[
a_{k}|\hspace{0.04cm}(\xi)_{2}\hspace{0.02cm}\rangle = \xi_{k}|\hspace{0.04cm}(\xi)_{2}\hspace{0.02cm}\rangle
\]
and, besides, in both cases the overlap function has its usual form
\[
\langle\hspace{0.025cm}(\bar{\xi}^{\,\prime})_{2}\hspace{0.02cm}
|\hspace{0.02cm}(\xi)_{2}\hspace{0.02cm}\rangle
=
{\rm e}^{\hspace{0.025cm}\textstyle\frac{1}{2}\hspace{0.03cm}
[\hspace{0.02cm}
\bar{\xi}^{\,\prime}_{l}\hspace{0.02cm},\xi^{\phantom{\prime}\!}_{l}
\hspace{0.02cm}\hspace{0.02cm}]}.
\]
The precise meaning of appearance of such ``twins'' remains unclear for us. Perhaps one reason of a purely algebraic nature is the fact that of the four basic identities (\ref{ap:B1})\,--\,(\ref{ap:B4}), only two ones are independent, namely (\ref{ap:B2}) and (\ref{ap:B3}). This circumstance and its consequence were analyzed in detail in the paper by Lavrov {\it et al.} \cite{lavrov_2014}. In particular, the Jacobi identity (\ref{ap:B1}) is a consequence of the generalized identity (\ref{ap:B2}). The latter contains double anticommutators on the right-hand side as in (\ref{eq:13w}). This hints that one of the relations (\ref{eq:13q}) and (\ref{eq:13w}) is a consequence of each other for $p=2$. In any case we may state that the para-Fermi statistics of order 2 is a very special case of parastatistics (as well as ordinary Fermi-statistics), since it possesses the properties that are completely absent for higher
$(p \geq 3)$ para-Fermi statistics.


\section*{\bf Acknowledgments}

The work of D.M.G. has been supported by the Russian Foundation for Basic Research
(Project No. 18-02-00149)), as well as by the S\~ao Paulo Research Foundation
(FAPESP, Grant No. 2016/03319-6), the Brazilian National Council for Scientific and Technological Development (CNPq), and by the Tomsk State University competitiveness improvement programme.


\begin{appendices}
\numberwithin{equation}{section}

\section{Lie algebra $su(2M + 1)$}

\label{appendix_A}
The Lie algebra of the unitary group $SU(2M + 1)$ has the form \cite{okubo_1975}
\begin{equation}
\begin{split}
&[\hspace{0.02cm}X_{\mu\nu},X_{\sigma\lambda}] = \delta_{\nu\sigma}X_{\mu\lambda} - \delta_{\mu\lambda}X_{\sigma\nu},\\[1ex]
&\sum^{2M}_{\mu\hspace{0.02cm}=\hspace{0.02cm}0}X_{\mu\mu} = 0,
\label{ap:A1}
\end{split}
\end{equation}
where the indices $\mu,\nu,\ldots$ run values $0, 1, 2,\ldots 2M$. By introducing a new set of operators
\begin{align}
&F_{\mu\nu} = X_{\mu\nu} - X_{\nu\mu}, \quad F_{\mu\nu} = - F_{\nu\mu}
\notag \\[1ex]
&\widetilde{F}_{\mu\nu} = X_{\mu\nu} + X_{\nu\mu}, \quad \widetilde{F}_{\mu\nu} = + \widetilde{F}_{\nu\mu},  \notag
\end{align}
the Lie algebra (\ref{ap:A1}) becomes
\begin{align}
&[\hspace{0.02cm}F_{\mu\nu},F_{\sigma\lambda}] =
\delta_{\nu\sigma}F_{\mu\lambda} + \delta_{\mu\lambda}F_{\nu\sigma}
-
\delta_{\mu\sigma}F_{\nu\lambda} - \delta_{\nu\lambda}F_{\mu\sigma},
\label{ap:A2} \\[1ex]
&[\hspace{0.02cm}\widetilde{F}_{\mu\nu},\widetilde{F}_{\sigma\lambda}] =
\delta_{\nu\sigma}F_{\mu\lambda} + \delta_{\mu\lambda}F_{\nu\sigma}
+
\delta_{\mu\sigma}F_{\nu\lambda} + \delta_{\nu\lambda}F_{\mu\sigma},
\label{ap:A3}  \\[1ex]
&[\hspace{0.02cm}F_{\mu\nu},\widetilde{F}_{\sigma\lambda}] =
\delta_{\nu\sigma}\widetilde{F}_{\mu\lambda} - \delta_{\mu\lambda}\widetilde{F}_{\nu\sigma}
- \delta_{\mu\sigma}\widetilde{F}_{\nu\lambda} + \delta_{\nu\lambda}\widetilde{F}_{\mu\sigma}
\label{ap:A4}
\end{align}
and the condition of speciality turns into
\begin{equation}
\sum^{2M}_{\mu\hspace{0.02cm}=\hspace{0.02cm}0}\widetilde{F}_{\mu\mu} = 0.
\label{ap:A5}
\end{equation}
The operators $F_{\mu\nu}$ form the Lie algebra of the orthogonal group $SO(2M + 1)$ and the
operators $\widetilde{F}_{\mu\nu}$ complete this algebra to the algebra of the unitary group
$SU(2M + 1)$.\\
\indent
The unitary quantization procedure is based on the choice of the Lie algebra of the group $SO(2M+1)$ as the basis algebra. Further, Govorkov \cite{govorkov_1979} has introduced the following quantities:
\begin{equation}
\begin{split}
&\beta_{\mu}\equiv iF_{\mu 0},\quad
\beta_{0} = iF_{00} = 0,\\[1ex]
&\zeta_{\mu}\equiv \widetilde{F}_{\mu 0},\quad\;
\zeta_{0} = \widetilde{F}_{00} \neq 0.
\label{ap:A6}
\end{split}
\end{equation}
The next relations
\begin{align}
&F_{\mu\nu} = [\hspace{0.02cm}\beta_{\mu},\beta_{\nu}] \hspace{0.02cm}-\hspace{0.01cm}
i\hspace{0.01cm}\bigl(\delta_{0\nu}\beta_{\mu} - \delta_{0\mu}\beta_{\nu}\bigr),
\label{ap:A7} \\[1ex]
&[\hspace{0.02cm}\zeta_{\mu},\zeta_{\nu}] =
[\hspace{0.02cm}\beta_{\mu},\beta_{\nu}] \hspace{0.02cm}-\hspace{0.01cm}
2\hspace{0.01cm}i\hspace{0.01cm}\bigl(\delta_{0\nu}\beta_{\mu} - \delta_{0\mu}\beta_{\nu}\bigr),
\label{ap:A8}  \\[1ex]
&[\hspace{0.02cm}\zeta_{\mu},\beta_{\nu}] =
-i\widetilde{F}_{\mu\nu} + i\hspace{0.015cm}\delta_{\mu\nu\,}\zeta_{0}
+
i\hspace{0.01cm}\bigl(\delta_{0\nu}\zeta_{\mu} - \delta_{0\mu}\zeta_{\nu}\bigr)
\label{ap:A9}
\end{align}
are a particular consequence of the algebra (\ref{ap:A2})\,--\,(\ref{ap:A4}).
By virtue of the equality $\widetilde{F}_{\mu \nu}= \widetilde{F}_{\nu \mu}$ it follows from Eq.\,(\ref{ap:A9}) that
$$
[\hspace{0.02cm}\zeta_{\mu},\beta_{\nu}\hspace{0.02cm}] \hspace{0.02cm}-\hspace{0.02cm} [\hspace{0.02cm}\zeta_{\nu},\beta_{\mu}] =
2\hspace{0.01cm}i\hspace{0.01cm}\bigl(\delta_{0\nu}\zeta_{\mu} - \delta_{0\mu}\zeta_{\nu}\bigr).
$$
This relation defines antisymmetric part of the commutator $[\zeta_{\mu}, \beta_{\nu}]$. In the paper \cite{govorkov_1979} in formula (\ref{ap:A7}) all terms in parentheses are absent, in formula (\ref{ap:A8}) there is no the multiplier 2 on the right-hand side, and in (\ref{ap:A9}) next to the last term is missed out. All these lost terms and the factor 2 are important when we check a consistency of the concrete expressions (see below).\\
\indent
Further, setting $\mu = \nu$ in (\ref{ap:A9}) and summing over $\mu$, in view of (\ref{ap:A5}), we derive another important relation
\begin{equation}
\zeta_{0} = -\frac{i}{2\hspace{0.01cm}M + 1}\,
\sum^{2M}_{\mu\hspace{0.02cm}=1}\,[\hspace{0.02cm}\zeta_{\mu},\beta_{\mu}\hspace{0.02cm}],
\label{ap:A10}
\end{equation}
i.e. the operator $\zeta_0$ is not independent one but it is determined by the other operators.\\
\indent
In terms of the variables (\ref{ap:A6}) one can rewrite the algebra (\ref{ap:A2})\,--\,(\ref{ap:A4}) in an equivalent form of the trilinear relations
\begin{align}
&[\hspace{0.02cm}\beta_{\lambda},[\hspace{0.02cm}\beta_{\mu},\beta_{\nu}]\hspace{0.02cm}]
=
\delta_{\lambda\mu}\beta_{\nu} - \delta_{\lambda\nu}\beta_{\mu},
\label{ap:A11} \\[1ex]
&[\hspace{0.02cm}\zeta_{\lambda},[\hspace{0.02cm}\zeta_{\mu},\zeta_{\nu}]\hspace{0.02cm}]
=
\delta_{\lambda\mu\,}\zeta_{\nu} - \delta_{\lambda\nu\,}\zeta_{\mu},
\label{ap:A12}  \\[1ex]
&[\hspace{0.02cm}\zeta_{\lambda},[\hspace{0.02cm}\zeta_{\mu},\beta_{\nu}]\hspace{0.02cm}]
=
2\hspace{0.015cm}\delta_{\mu\nu}\beta_{\lambda} +
\delta_{\lambda\nu}\beta_{\mu} + \delta_{\lambda\mu}\beta_{\nu},
\label{ap:A13}  \\[1ex]
&[\hspace{0.02cm}\beta_{\lambda},[\hspace{0.02cm}\zeta_{\mu},\beta_{\nu}]\hspace{0.02cm}]
=
-2\hspace{0.015cm}\delta_{\mu\nu}\zeta_{\lambda} -
\delta_{\lambda\nu}\zeta_{\mu} - \delta_{\lambda\mu}\zeta_{\nu},
\label{ap:A14}
\end{align}
and the bilinear ones
\begin{align}
&[\hspace{0.02cm}\beta_{\mu},\beta_{\nu}]
=
[\hspace{0.02cm}\zeta_{\mu},\zeta_{\nu}],
\label{ap:A15} \\[1ex]
&[\hspace{0.02cm}\zeta_{\mu},\beta_{\nu}]
=
[\hspace{0.02cm}\zeta_{\nu},\beta_{\mu}].
\label{ap:A16}
\end{align}
Here, the indices run values $1, 2, \ldots , 2M$. In the review \cite{govorkov_1983} Govorkov, however, states that the relations (\ref{ap:A11})\,--\,(\ref{ap:A16}) {\it `` ... are satisfied for $\zeta_0$ by itself by virtue of the fulfilment for the other $\zeta_{\mu}$. Therefore the indices $\mu,\hspace{0.02cm} \nu,\hspace{0.02cm} \lambda$ in these relations can be considered running values $\,0\hspace{0.02cm},\hspace{0.02cm}1,\hspace{0.02cm}2,\hspace{0.02cm}
\ldots\hspace{0.02cm},2M$''}. One can see that this does not actually seem to be the case from a comparison, for example, of the bilinear relations (\ref{ap:A15}), (\ref{ap:A16}) with the relations (\ref{ap:A8}), (\ref{ap:A9}). In the former case, for $\nu = 0$ we have (recall that $\beta_0 = 0$)
$$
[\hspace{0.02cm}\zeta_{\mu},\zeta_{0}] = 0, \quad [\hspace{0.02cm}\beta_{\mu},\zeta_{0}] = 0,
$$
while in the latter case the commutators take the form
\begin{equation}
[\hspace{0.02cm}\zeta_{\mu},\zeta_{0}] = -2\hspace{0.015cm}i\hspace{0.02cm}\beta_{\mu},
\quad
[\hspace{0.02cm}\beta_{\mu},\zeta_{0}] = 2\hspace{0.015cm}i\hspace{0.02cm}\zeta_{\mu}.
\label{ap:A17}
\end{equation}
Further, the generalization of the trilinear relations (\ref{ap:A11}), (\ref{ap:A12}) valid for any values of indices takes the form, correspondingly,
\begin{align}
&[\hspace{0.01cm}\beta_{\lambda},[\hspace{0.01cm}\beta_{\mu},\beta_{\nu}]\hspace{0.01cm}]
=
\delta_{\lambda\mu}\beta_{\nu} - \delta_{\lambda\nu}\beta_{\mu}
+
\delta_{0\nu}\bigl(\delta_{0\lambda}\beta_{\mu} - \delta_{0\mu}\beta_{\lambda}\bigr)
-
\delta_{0\mu}\bigl(\delta_{0\lambda}\beta_{\nu} - \delta_{0\nu}\beta_{\lambda}\bigr),
\notag \\[1ex]
&[\hspace{0.03cm}\zeta_{\lambda}\hspace{0.02cm},
[\hspace{0.03cm}\zeta_{\mu},\zeta_{\nu}\hspace{0.02cm}]\hspace{0.02cm}]
=
\delta_{\lambda\mu\,}\zeta_{\nu} - \delta_{\lambda\nu\,}\zeta_{\mu}
+
\delta_{0\nu}\bigl(\delta_{0\lambda}\zeta_{\mu}  - \delta_{0\mu}\zeta_{\lambda} - \delta_{\mu\lambda}\zeta_{0}\bigr)
-
\delta_{0\mu}\bigl(\delta_{0\lambda}\zeta_{\nu}  - \delta_{0\nu}\zeta_{\lambda} - \delta_{\nu\lambda}\zeta_{0}\bigr)
\notag \\[1ex]
&+2\hspace{0.015cm}i\hspace{0.01cm}\bigl(\delta_{0\nu}
[\hspace{0.02cm}\beta_{\mu},\zeta_{\lambda}] - \delta_{0\mu}[\hspace{0.02cm}\beta_{\nu},\zeta_{\lambda}]\hspace{0.02cm}\bigr).
\notag
\end{align}
A distinguishing feature of the last expression is appearing the terms, which are {\it bilinear} in $\beta$ and $\zeta$ operators. These terms cannot be eliminated by any means.\\
\indent
Finally, a more general relation for (\ref{ap:A13}) has the form
\begin{align}
&[\hspace{0.02cm}\zeta_{\lambda},[\hspace{0.02cm}\zeta_{\mu},\beta_{\nu}]\hspace{0.02cm}]
=
2\hspace{0.02cm}\delta_{\mu\nu}\beta_{\lambda} +
\delta_{\lambda\nu}\beta_{\mu} + \delta_{\lambda\mu}\beta_{\nu}
\notag \\[1ex]
-\
\delta_{0\nu}\bigl(\delta_{0\lambda}&\beta_{\mu} - \delta_{0\mu}\beta_{\lambda}\bigr)
-
\delta_{0\mu}\bigl(\delta_{0\lambda}\beta_{\nu} - \delta_{0\nu}\beta_{\lambda}\bigr)
+
2\hspace{0.01cm}i\hspace{0.01cm}\delta_{0\mu}\hspace{0.02cm}
[\hspace{0.02cm}\zeta_{\nu},\zeta_{\lambda}].
\notag
\end{align}
Here, the right-hand side also contains the term bilinear in $\zeta$.\\
\indent
For the unitary representations of the algebra under consideration, the quantities $\beta_{\mu}$ and $\zeta_{\mu}$ are Hermitian:
$$
\beta^{\dagger}_{\mu} = \beta^{\phantom{\dagger}\!}_{\mu}, \quad \zeta^{\dagger}_{\mu} = \zeta^{\phantom{\dagger}\!}_{\mu}.
$$
This circumstance enables us to introduce the Hermitian conjugate operators
\begin{equation}
\begin{split}
&a_{k} = \beta_{2k - 1} - i\hspace{0.02cm}\beta_{2k}, \quad\;
b_{k} = \zeta_{2k - 1} - i\hspace{0.02cm}\zeta_{2k}, \\[1ex]
&a^{\dagger}_{k} = \beta^{\phantom{\dagger}\!}_{2k - 1} + i\hspace{0.01cm}\beta^{\phantom{\dagger}\!}_{2k}, \quad\;
b^{\dagger}_{k} = \zeta^{\phantom{\dagger}\!}_{2k - 1} + i\hspace{0.02cm}\zeta^{\phantom{\dagger}\!}_{2k},
\end{split}
\label{ap:A18}
\end{equation}
where $k = 1,2,\ldots,M$. The algebra (\ref{eq:1q})\,--\,(\ref{eq:1y}) and (\ref{eq:1o}) for the operators $a_{k},\,b_{k}$ and $\zeta_{0}$ is a direct corollary of (\ref{ap:A11})\,--\,(\ref{ap:A16})
and (\ref{ap:A17}).

\section{Operator identities}
\label{appendix_B}

In this Appendix we give a number of the operator identities, which we use extensively throughout in the text:
\begin{align}
&[\hspace{0.02cm}A,[\hspace{0.02cm}B,C\hspace{0.03cm}]\hspace{0.02cm}] = - [\hspace{0.02cm}B,[\hspace{0.02cm}C,A\hspace{0.02cm}]\hspace{0.02cm}] - [\hspace{0.03cm}C,[A,B\hspace{0.02cm}]\hspace{0.03cm}],
\label{ap:B1} \\[1ex]
&[A,[B,C\hspace{0.03cm}]\hspace{0.03cm}] =  \{C,\{A,B\hspace{0.02cm}\}\} - \{B,\{A,C\hspace{0.02cm}\}\!\},  \label{ap:B2}  \\[1ex]
&\{A,[B,C\hspace{0.03cm}]\hspace{0.02cm}\} =  \{B,[\hspace{0.02cm}C,A\hspace{0.02cm}]\} - [\hspace{0.03cm}C,\{A,B\hspace{0.02cm}\}],   \label{ap:B3} \\[1ex]
&[A,\{B,C\hspace{0.02cm}\}\hspace{0.02cm}] = - [\hspace{0.02cm}B,\{C,A\hspace{0.02cm}\}] - [\hspace{0.03cm}C,\{A,B\hspace{0.02cm}\}],
\label{ap:B4}
\end{align}
where $[\,,]$ and $\{\,,\}$ designate commutator and anticommutator, respectively. In addition to
(\ref{ap:B1})\,--\,(\ref{ap:B4}), the following simple relations are rather useful:
\begin{align}
&[\hspace{0.02cm}A, B\hspace{0.02cm}C\hspace{0.02cm}] = \{A, B\}\hspace{0.04cm}C - B\hspace{0.02cm}\{A, C\hspace{0.02cm}\}
=
B\hspace{0.03cm}[A, C\hspace{0.02cm}] + [A, B\hspace{0.02cm}]\hspace{0.04cm}C,
\label{ap:B5} \\[0.8ex]
&\{A, B\hspace{0.02cm}C\} = \{A, B\}\hspace{0.02cm}C - B\hspace{0.02cm}[\hspace{0.02cm}A, C\hspace{0.02cm}]
=
B\hspace{0.02cm}\{A, C\} + [\hspace{0.02cm}A, B\hspace{0.02cm}]\hspace{0.04cm}C.
\label{ap:B6}
\end{align}
Finally, the operator identities involving exponential functions have the form \cite{mendas_1989, mendas_1990, mendas_2010, pain_2013}
\begin{align}
&{\rm e}^{X}Y{\rm e}^{-X} = Y + [X,Y\hspace{0.02cm}] + \frac{\!1}{2!}\hspace{0.03cm}[X,[X,Y\hspace{0.02cm}]\hspace{0.02cm}]
+ \frac{\!1}{3!}\hspace{0.03cm}[X,[X,[X,Y\hspace{0.02cm}]\hspace{0.02cm}]\hspace{0.02cm}] + \ldots\,,
\label{ap:B7} \\[0.8ex]
&{\rm e}^{X}Y{\rm e}^{X} = Y + \{X, Y\} + \frac{\!1}{2!}\hspace{0.03cm}\{X,\{X,Y\hspace{0.02cm}\}\hspace{0.01cm}\!\}
+ \frac{\!1}{3!}\hspace{0.03cm}\{X\{X,\{X,Y\hspace{0.02cm}\}\hspace{0.01cm}\!\}\hspace{0.01cm}\!\} + \ldots\,,
\label{ap:B8}\\[0.8ex]
&{\rm e}^{X}{\rm e}^{Y}{\rm e}^{-X} = \exp\bigl({\rm e}^{X}Y{\rm e}^{-X}\bigr).
\label{ap:B9}
\end{align}


\section{\bf Commutation relations with the operator ${\rm e}^{\alpha\hspace{0.02cm}i \widetilde{N}}$}
\label{appendix_C}

Here, we bring together all the (anti)commutation relations that involve the operator
${\rm e}^{\alpha\hspace{0.02cm}i \widetilde{N}}$, which have arisen on various occasions in sections 6 and 8. At value of the parameter $\alpha = \pm\hspace{0.025cm}\pi$, we have:
$$
\{{\rm e}^{\pm\hspace{0.015cm}\pi\hspace{0.01cm}i \widetilde{N}}\!,a_{k}\} = 0,
\quad \{{\rm e}^{\pm\hspace{0.015cm}\pi\hspace{0.01cm}i \widetilde{N}}\!,b_{m}\} = 0.
$$
Further, at value of the parameter $\alpha = \pm\hspace{0.025cm}\pi/2$, we have
\begin{align}
&a_{k}\hspace{0.04cm}
{\rm e}^{\pm\hspace{0.015cm}\pi\hspace{0.01cm}i \widetilde{N}/2}
=
\mp\,{\rm e}^{\pm\hspace{0.015cm}\pi\hspace{0.01cm}i \widetilde{N}/2}\hspace{0,015cm}b_{k},
\quad
b_{m}\hspace{0.04cm}
{\rm e}^{\pm\hspace{0.015cm}\pi\hspace{0.01cm}i \widetilde{N}/2}
=
\pm\,{\rm e}^{\pm\hspace{0.015cm}\pi\hspace{0.01cm}i \widetilde{N}/2}\hspace{0,015cm}a_{m},
\notag \\[0.8ex]
&{\rm e}^{\pm\hspace{0.015cm}\pi\hspace{0.01cm}i \widetilde{N}/2}\hspace{0,015cm}a_{k} =
\pm\,b_{k}\hspace{0.04cm}
{\rm e}^{\pm\hspace{0.015cm}\pi\hspace{0.01cm}i \widetilde{N}/2}\hspace{0,015cm},
\,\quad
{\rm e}^{\pm\hspace{0.015cm}\pi\hspace{0.01cm}i \widetilde{N}/2}\hspace{0,015cm}b_{m} =
\mp\,a_{m}\hspace{0.04cm}
{\rm e}^{\pm\hspace{0.015cm}\pi\hspace{0.01cm}i \widetilde{N}/2}\hspace{0,015cm}
\notag
\end{align}
or in the equivalent form
\begin{align}
&a_{k} =
\mp\,{\rm e}^{\pm\hspace{0.015cm}\pi\hspace{0.01cm}i \widetilde{N}/2}\hspace{0,015cm}b_{k}
\hspace{0.04cm}
{\rm e}^{\mp\hspace{0.015cm}\pi\hspace{0.01cm}i \widetilde{N}/2}\hspace{0,015cm},
\quad
b_{m} =
\pm\,{\rm e}^{\pm\hspace{0.015cm}\pi\hspace{0.01cm}i \widetilde{N}/2}\hspace{0,015cm}a_{m}
\hspace{0.04cm}
{\rm e}^{\mp\hspace{0.015cm}\pi\hspace{0.01cm}i \widetilde{N}/2}\hspace{0,015cm},
\notag \\[0.8ex]
&a_{k} =
\pm\,{\rm e}^{\mp\hspace{0.015cm}\pi\hspace{0.01cm}i \widetilde{N}/2}\hspace{0,015cm}b_{k}
\hspace{0.04cm}
{\rm e}^{\pm\hspace{0.015cm}\pi\hspace{0.01cm}i \widetilde{N}/2}\hspace{0,015cm},
\quad
b_{m} =
\mp\,{\rm e}^{\mp\hspace{0.015cm}\pi\hspace{0.01cm}i \widetilde{N}/2}\hspace{0,015cm}a_{m}
\hspace{0.04cm}
{\rm e}^{\pm\hspace{0.015cm}\pi\hspace{0.01cm}i \widetilde{N}/2}\hspace{0,015cm}.
\notag
\end{align}
The commutation relations involving the para-Grassmann numbers $\xi_{k}$ for the $\alpha = \pm\hspace{0.025cm}\pi$ case are
\begin{align}
&{\rm e}^{\pm\hspace{0.015cm}\pi\hspace{0.01cm}i \widetilde{N}}[\hspace{0.03cm}\xi_{l}, a_{k}\hspace{0.02cm}] =
-\hspace{0.025cm}{\rm e}^{\pm\hspace{0.015cm}\pi\widetilde{\Lambda}}\hspace{0.02cm}
[\hspace{0.03cm}\xi_{l}, a_{k}\hspace{0.02cm}]\hspace{0.04cm}
{\rm e}^{\mp\hspace{0.015cm}\pi\hspace{0.01cm}i \widetilde{N}},
\quad\;\;
{\rm e}^{\pm\hspace{0.015cm}\pi\hspace{0.01cm}i \widetilde{N}}\{\hspace{0.03cm}\xi_{l}, a_{k}\hspace{0.02cm}\} =
-\hspace{0.025cm}{\rm e}^{\pm\hspace{0.015cm}\pi\widetilde{\Lambda}}\hspace{0.02cm}
\{\hspace{0.03cm}\xi_{l}, a_{k}\hspace{0.02cm}\}\hspace{0.04cm}
{\rm e}^{\mp\hspace{0.015cm}\pi\hspace{0.01cm}i \widetilde{N}},
\notag \\[1ex]
&{\rm e}^{\pm\hspace{0.015cm}\pi\hspace{0.01cm}i \widetilde{N}}[\hspace{0.03cm}\xi_{l}, b_{m}\hspace{0.02cm}] =
-\hspace{0.025cm}{\rm e}^{\pm\hspace{0.015cm}\pi\widetilde{\Lambda}}\hspace{0.02cm}
[\hspace{0.03cm}\xi_{l}, b_{m}\hspace{0.02cm}]\hspace{0.04cm}
{\rm e}^{\mp\hspace{0.015cm}\pi\hspace{0.01cm}i \widetilde{N}},
\quad\;
{\rm e}^{\pm\hspace{0.015cm}\pi\hspace{0.01cm}i \widetilde{N}}\{\hspace{0.03cm}\xi_{l}, b_{m}\hspace{0.02cm}\} =
-\hspace{0.025cm}{\rm e}^{\pm\hspace{0.015cm}\pi\widetilde{\Lambda}}\hspace{0.02cm}
\{\hspace{0.03cm}\xi_{l}, b_{m}\hspace{0.02cm}\}\hspace{0.04cm}
{\rm e}^{\mp\hspace{0.015cm}\pi\hspace{0.01cm}i \widetilde{N}},
\notag
\end{align}
and for the $\alpha = \pm\hspace{0.015cm}\pi/2$ case, we have
\begin{align}
&{\rm e}^{\pm\hspace{0.015cm}\pi\hspace{0.01cm}i \widetilde{N}/2}\hspace{0,015cm}[\hspace{0.03cm}\xi_{l}, a_{k}\hspace{0.02cm}] =
\mp\,{\rm e}^{\pm\hspace{0.015cm}\pi\widetilde{\Lambda}/2}\hspace{0.02cm}
\{\hspace{0.03cm}\xi_{l}, b_{k}\hspace{0.02cm}\}\hspace{0.04cm}
{\rm e}^{\mp\hspace{0.015cm}\pi\hspace{0.01cm}i \widetilde{N}/2}\hspace{0,015cm}, \notag \\[0.8ex]
&{\rm e}^{\pm\hspace{0.015cm}\pi\hspace{0.01cm}i \widetilde{N}/2}\hspace{0,015cm}[\hspace{0.03cm}\xi_{l}, b_{m}\hspace{0.02cm}] =
\pm\,{\rm e}^{\pm\hspace{0.015cm}\pi\widetilde{\Lambda}/2}\hspace{0.02cm}
\{\hspace{0.03cm}\xi_{l}, a_{m}\hspace{0.02cm}\}\hspace{0.04cm}
{\rm e}^{\mp\hspace{0.015cm}\pi\hspace{0.01cm}i \widetilde{N}/2}\hspace{0,015cm}.
\notag
\end{align}

\end{appendices}

\end{document}